\documentclass[aps,prx,article,twocolumn,preprintnumbers,amsmath,amssymb,superscriptaddress]{revtex4-1}

\date{\today}
\usepackage{epsfig}
\usepackage{subfigure}
\usepackage{graphicx}
\usepackage{dcolumn}
\usepackage{bm}
\usepackage[colorlinks,citecolor=blue,linkcolor=blue,hyperindex,CJKbookmarks]{hyperref}
\usepackage{float}
\usepackage{hyperref}
\usepackage{comment}
\newcommand{\real}      {\mathrm{Re}}

\newcommand{\Ham}   {{\mathcal{H}}}

\newcommand{\kbf}      {\textbf{k}}

\newcommand{\kprbf}      {\textbf{k}^\prime}
\newcommand{\kpr}      {k^\prime}
\newcommand{\qbf}      {\textbf{q}}

\newcommand{\ibf}      {\textbf{i}}
\newcommand{\jbf}      {\textbf{j}}
\newcommand{\rbf}      {\textbf{r}}

\newcommand{\Schrdg} {{Schr\"{o}dinger}}

\newcommand{\Ugs}{U_{\rm GS}}
\newcommand{\Ub}{U_{\rm GS}}
\newcommand{\Upl}{U_{\rm plrn}}
\newcommand{\Ur}{U_{\rm rot}}
\newcommand{\Rq}{R_\qbf}
\newcommand{\Rqd}{R_\qbf^\dagger}
\newcommand{\dt}{\partial_t}

\newcommand{\invrN}{\frac1{\sqrt{N}}}
\newcommand{\gred}{g_\qbf}
\newcommand{\gredZero}{g_0}

\begin{document}

\title{Fluctuating Nature of Light-Enhanced $d$-Wave Superconductivity:\\
A Time-Dependent Variational Non-Gaussian Exact Diagonalization Study}
\author{Yao Wang}
 \email[All correspondence should be addressed to Y.W.(\href{mailto:yaowang@g.clemson.edu}{yaowang@g.clemson.edu}), T.S. (\href{tshi@itp.ac.cn}{tshi@itp.ac.cn}), and C.C.C. (\href{chencc@uab.edu}{chencc@uab.edu}) 
]{}
\affiliation{Department of Physics and Astronomy, Clemson University, Clemson, South Carolina 29631, USA}
\author{Tao Shi}
\affiliation{CAS Key Laboratory of Theoretical Physics, Institute of Theoretical Physics, Chinese Academy of Sciences, Beijing 100190, China}
\affiliation{CAS Center for Excellence in Topological Quantum Computation, University of Chinese Academy of Sciences, Beijing 100049, China}
\author{Cheng-Chien Chen}
 \email[All correspondence should be addressed to Y.W.(\href{mailto:yaowang@g.clemson.edu}{yaowang@g.clemson.edu}), T.S. (\href{tshi@itp.ac.cn}{tshi@itp.ac.cn}), and C.C.C. (\href{chencc@uab.edu}{chencc@uab.edu}) 
]{}
\affiliation{Department of Physics, University of Alabama at Birmingham, Birmingham, Alabama 35294, USA}

\date{\today}
\begin{abstract}
Engineering quantum phases using light is a novel route to designing functional materials, where light-induced superconductivity is a successful example. Although this phenomenon has been realized experimentally, especially for the high-$T_c$ cuprates, the underlying mechanism remains mysterious. Using the recently developed variational non-Gaussian exact diagonalization method, we investigate a particular type of photoenhanced superconductivity by suppressing a competing charge order in a strongly correlated electron-electron and electron-phonon system. We find that the $d$-wave superconductivity pairing correlation can be enhanced by a pulsed laser, consistent with recent experiments based on gap characterizations. However, we also find that the pairing correlation length is heavily suppressed by the pump pulse, indicating that light-enhanced superconductivity may be of fluctuating nature. Our findings also imply a general behavior of nonequilibrium states with competing orders, beyond the description of a mean-field framework.
\end{abstract}
\maketitle

\section{Introduction}
\begin{figure*}[!t]
\begin{center}
\includegraphics[width=16cm]{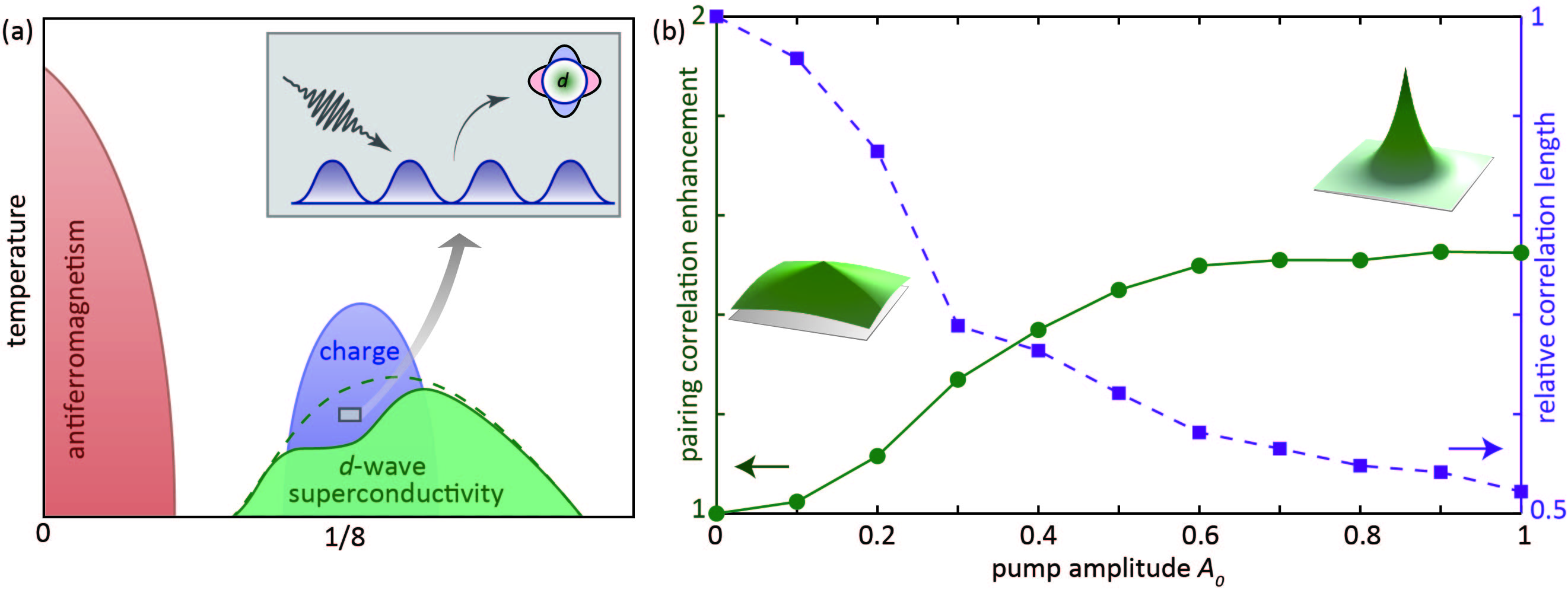}
\vspace{-3mm}\caption{\label{fig:cartoon} \textbf{Light-induced $d$-wave superconductivity experiment and theory.} (a) Sketch of existing light-induced or enhanced superconductivity experiments in the charge-dominant phase in cuprates. (b) Summary of the simulation results in this work. The green circles denote the enhancement of the pairing correlation and the purple squares denote the suppression of the relative correlation length, averaged between time $t=10-30\,t_h^{-1}$. The data are extracted from Figs.~\ref{fig:pumpDynm} and \ref{fig:correlations}. The left and right insets in (b) sketch real-space distribution of pairing correlations, respectively, before and after pump.
}
\end{center}
\end{figure*}

Understanding, controlling, and designing functional quantum phases are major goals and challenges in modern condensed matter physics\,\cite{basov2017towards, wang2018theoretical}. Among a few successful examples, light-induced superconductivity, above its original transition temperature, has been a great surprise and is believed to be a promising route to room-temperature superconductors. Although this novel phenomenon has been realized experimentally in various materials including cuprates, fullerides, and organic salts\,\cite{mankowsky2014nonlinear, hu2014optically,kaiser2014optically,mitrano2016possible, buzzi2020photomolecular}, its mechanism remains controversial.
The underlying physics is partly attractive yet mysterious for the enhanced $d$-wave superconductivity observed in pumped charge-ordered high-$T_c$ cuprates La$_{1.675}$Eu$_{0.2}$Sr$_{0.125}$CuO$_4$ and La$_{2-x}$Ba$_x$CuO$_4$ near 1/8 doping\,\cite{fausti2011light,nicoletti2014optically, forst2014melting, casandruc2015wavelength, khanna2016restoring,nicoletti2018magnetic,cremin2019photoenhanced}, partially due to their original high critical temperatures at equilibrium. As illustrated in Fig.~\ref{fig:cartoon}(a), the occurrence of the Cooper pairs above $T_c$ (reflected by the Josephson plasma resonance in experiments), as a signature of light-induced superconductivity, is observed when these materials are stimulated by a near-infrared pulse laser. 

Motivated by the experiments, various theories and numerical simulations have been conducted to explain the observed light-induced phenomena. In the context of conventional BCS superconductivity, simulations with both mean-field and many-body models have demonstrated the feasibility to manipulate and enhance local Cooper pairs (i.e.~$s$-wave superconductivity)\,\cite{sentef2016theory,sentef2017theory, coulthard2017enhancement,babadi2017theory,murakami2017nonequilibrium,dasari2018transient,bittner2019possible, li2020manipulating, tindall2020dynamical, tindall2020analytical}.
The understanding becomes more challenging for the unconventional $d$-wave superconductivity in cuprates, due to the two-dimensional geometry and the strongly correlated nature. Insightful theoretical perspectives have been proposed in the context of phenomenological or steady-state theory, including the suppression of competing charge order\,\cite{patel2016light}, Floquet engineering of the Fermi surface\,\cite{kennes2019light}, and (for the terahertz pump) parametric amplification\,\cite{michael2020parametric}. In contrast to the studies of $s$-wave superconductivity, rigorous nonequilibrium simulations for nonlocal $d$-wave pairing instability in microscopic models have been limited to undoped systems with truncated phonon modes\,\cite{wang2018light}, distinct from the conditions of existing cuprate-based experiments. The extension to a doped system has been hindered by the difficulty of treating both strong electronic correlation and electron-phonon coupling in a quantum many-body simulation. In particular, the spatial fluctuations of phonons and bosonic excitations in a 1/8-doped cuprates are expected to be important due to the lack of a nesting momentum. Therefore, the microscopic theory for light-induced superconductivity in a doped cuprate remains an open question.

Moreover, recent experiments have revealed the presence of fluctuating superconductivity above the transition temperature $T_c$ in overdoped cuprates and FeSe\,\cite{kondo2015point, he2020fluctuating, faeth2020incoherent, xu2020spectroscopic}. In this regime, the superconductivity gap remains open, but long-range order and zero resistance are absent due to the reduction of correlation length. Recent ultrafast experiments also showed that the pump light can destroy the coherence of Cooper pairs and fluctuate an existing superconductor\,\cite{boschini2018collapse,yang2018terahertz, mootz2020lightwave}. Thus, the resonance and gap in the transient optical conductivity may cause a misassignment of the long-range superconductivity\,\cite{chiriaco2018transient,lemonik2019transport,sun2020transient}. Both observations raise the necessity to further investigate the coherence of superconductivity induced by light in a quantum many-body model.

For this purpose, we study the photoinduced dynamics and superconductivity in a light-driven Hubbard-Holstein model relevant to cuprates. To overcome the numerical difficulties of simulating many-body dynamics with strong electronic correlations and electron-phonon coupling, we construct on top of the recently developed variational non-Gaussian exact diagonalization (NGSED) and develop the time-dependent NGSED (see Sec.~\ref{sec:method}). This hybrid method leverages the merits of both the numerical many-body solver and the variational non-Gaussian solver: The former is necessary to unbiasedly tackle strong electronic correlations, while the latter avoids the phonon's unbounded Hilbert-space problem and has been demonstrated efficient in describing the ground and excited states of systems with the Fr\"{o}hlich-type electron-phonon coupling\,\cite{shi2018variational,shi2020variational, shi2019ultrafast, shi2021spin}. As an extension of the equilibrium NGSED\,\cite{wang2019zero}, this time-dependent method provides an accurate description of far-from-equilibrium states through the Krylov-subspace method and the K\"{a}hlerization of the solvers. These advances of the numerical method allow the simulation of light-induced dynamics in quantum materials with both electronic correlations and electron-phonon couplings. 

As summarized in Fig.~\ref{fig:cartoon}(b), our results suggest that the $d$-wave pairing correlation can be dramatically enhanced by a pulsed laser when the system is charge dominant and close to a phase boundary, consistent with optical experiments. However, we also find that the pairing correlation length is heavily suppressed by the pump pulse. Therefore, light-enhanced $d$-wave superconductivity may be of fluctuating nature. In addition to existing ultrafast reflectivity measurements, our theoretical findings also predict various observations verifiable by future transport and photoemission experiments on pumped La$_{1.675}$Eu$_{0.2}$Sr$_{0.125}$CuO$_4$ and La$_{2-x}$Ba$_x$CuO$_4$.

The organization of this paper is as follows. We introduce our microscopic model for the cuprate system and light-matter interaction in Sec.~\ref{sec:model}. Next, we briefly introduce the assumptions and framework of the time-dependent NGSED method in Sec.~\ref{sec:method}, while the detailed derivations and benchmarks are shown in Appendix \ref{app:NGSED} and \ref{sec:krylov}. The main results about light-enhanced $d$-wave superconductivity and the fluctuating nature are presented in Secs.~\ref{sec:lightInduceddWave} and \ref{sec:correlationLen}. We then discuss the frequency dependence of these observations in Sec.~\ref{sec:freq}. Finally, we conclude our paper and discuss relevant experimental predictions in Sec.~\ref{sec:conclusion}.

\section{Prototypical Models}\label{sec:model}

Unconventional superconductivity is believed to emerge from the intertwined orders of strongly correlated systems, where both spin and charge instabilities exist\,\cite{keimer2015quantum, davis2013concepts}. It was widely believed that spin fluctuation, induced by electron correlation, is a viable candidate to provide the pairing glue for $d$-wave superconductivity\,\cite{scalapino1986d,gros1987superconducting,kotliar1988superexchange,tsuei2000pairing}. The minimal model to represent this strong correlation is the single-band Hubbard model\,\cite{zhang1988effective}. Based on rigorous numerical simulations, many important experimental discoveries have been reproduced using this model, such as antiferromagnetism\,\cite{singh2002spin}, stripe phases\,\cite{zheng2017stripe, huang2017numerical,huang2018stripe,ponsioen2019period}, strange metallicity\,\cite{kokalj2017bad, huang2019strange, cha2020slope}, and superconductivity\,\cite{zheng2016ground,ido2018competition,jiang2019superconductivity}. However, increasing experimental evidence reveals the significant role of phonons in high-$T_c$ superconductors\,\cite{shen2004missing, lanzara2001evidence, reznik2006electron, devereaux2007inelastic,he2018rapid}, in addition to the strong electronic correlation. Together with the crucial lattice effects observed in pump-probe experiments\,\cite{mankowsky2014nonlinear,hu2014optically, kaiser2014optically}, we believe that the minimal model to describe the light-induced $d$-wave superconductivity must involve both interactions.

The Hubbard-Holstein model is the prototypical model for describing correlated quantum materials with both electron-electron interaction and electron-phonon coupling\,\cite{hubbard1963week,holstein1959studies}. Its Hamiltonian is written as
\begin{eqnarray} \label{eq:HHM}
\Ham&=&-t_h\sum_{\langle\ibf,\jbf\rangle,\sigma} \left(c^\dagger_{\ibf \sigma} c_{\jbf\sigma} +H.c. \right) + U\sum_{\ibf }n_{\ibf \uparrow}n_{\ibf \downarrow}\nonumber\\
&&+g\sum_{\ibf\sigma}  (a_\ibf  + a_\ibf^\dagger) n_{\ibf\sigma}+\omega_0\sum_\ibf a_\ibf^\dagger a_\ibf \,.
\end{eqnarray}
Here, $c_{\ibf\sigma}$ ($c_{\ibf\sigma}^\dagger$) annihilates (creates) an electron at site $\ibf$ with spin $\sigma$, and $a_{\ibf}$ ($a_{\ibf}^\dagger$) annihilates (creates) a phonon at site $\ibf$. To reduce the number of model parameters, we restrict the hopping integral $t_h$ to the nearest neighbor, electron-electron interaction to the on-site Coulomb (Hubbard) repulsion $U$, electron-phonon coupling to the on-site electrostatic coupling (Holstein) $g$, and phonon energy to the dispersionless $\omega_0$. The electron-electron interaction and electron-phonon coupling can be depicted by the dimensionless parameters $u=U/t_h$ and $\lambda = g^2/\omega_0 t_h$, respectively. Here, we set $\omega_0=t_h$ in accord with common choices\,\cite{clay2005intermediate, nowadnick2012competition, nowadnick2015renormalization, johnston2013determinant, wang2016using, mendl2017doping}. Throughout the paper, we focus on 12.5\% hole doping simulated on an $N=4\times 4$ cluster, corresponding to the La$_{1.675}$Eu$_{0.2}$Sr$_{0.125}$CuO$_4$ and La$_{1.875}$Ba$_{0.125}$CuO$_4$ experiments\,\cite{fausti2011light,nicoletti2014optically, forst2014melting, casandruc2015wavelength, khanna2016restoring,nicoletti2018magnetic,zhang2018light,cremin2019photoenhanced}.

In an undoped system with a well-defined nesting momentum, one can restrict the phonons to only the $\qbf=(\pi,\pi)$ mode\,\cite{wang2016using, wang2018light}. However, in a doped Hubbard-Holstein model without commensurability, phonon modes at all momenta should be considered. As shown in Ref.~\onlinecite{wang2019zero}, the phase diagram for a doped Hubbard-Holstein model is dominated by a regime with strong charge susceptibility and a regime with strong spin susceptibility, although both are not ordered like the half-filled case. Unlike the striped order demonstrated in the Hubbard model at the thermodynamic limit\,\cite{zheng2017stripe, huang2017numerical,huang2018stripe,ponsioen2019period}, our system size does not support such a period-8 instability. Therefore, to increase the charge correlation and mimic the charge-dominant cuprate system, we exploit relatively strong electron-phonon couplings.

The light-driven physics is described (on the microscopic level) through the Peierls substitution $t_hc^\dagger_{\ibf \sigma} c_{\jbf\sigma}  \rightarrow t_h e^{i\int_{\rbf_\ibf}^{\rbf_\jbf} \mathbf{A}(t)\cdot d\rbf} c^\dagger_{\ibf \sigma} c_{\jbf\sigma}$, where the vector potential $\mathbf{A}(t)$ of the external light pulse affects the many-body Hamiltonian Eq.~\eqref{eq:HHM}. In this paper, we simulate the pump pulse with an oscillatory Gaussian vector potential:
\begin{equation}
    \mathbf{A}(t)=A_0\hat{e}_{\rm pol} \exp\left[-\frac{(t-t_0)^2}{2\sigma^2}\right]\cos(\Omega t)\,,
\end{equation}
and fix the polarization as diagonal $\hat{e}_{\rm pol}=\frac1{\sqrt{2}}(\hat x + \hat y)$ and the pump frequency as $\Omega = 4t_h$ (close to the 800-nm laser for $t_h=350$meV). In a strongly correlated model like the Hubbard model or the extended Hubbard model, the ultrafast pump pulse is coupled to electrons and may manipulate the delicate balance between different competing orders\,\cite{lu2012enhanced,al2013wave,rincon2014photoexcitation,lu2015photoinduced}. When a finite-frequency phonon is involved, the photomanipulated competition of phases also relies on the retardation effect of the phonons.

Throughout this paper, we restrict ourselves to the two-dimensional Hubbard-Holstein model and transient equal-time correlation functions. The relation between the in-plane Cooper pairs and the $c$-axis Josephson plasmon resonance has been studied through semiclassical simulations\,\cite{denny2015proposed,raines2015enhancement,okamoto2017transiently}. Our simulation does not consider specific experimental conditions including the aforementioned probe schemes, material-specific matrix elements, and finite temperature, nevertheless it helps explain the observed phenomena, in general. Therefore, our correlation function analysis in Secs.~\ref{sec:lightInduceddWave}-\ref{sec:freq} aims to address a matter-of-principle question when the balance between charge-density wave (CDW) and superconductivity, in a strongly correlated material, is altered by a pulsed laser.

\section{Time-Dependent Variational Non-Gaussian Exact Diagonalization Method}\label{sec:method}
The simulation of nonequilibrium quantum many-body systems requires either Green's function or wavefunction methods. The Green's function methods, constructed on the Keldysh formalism and represented by the dynamical mean-field theory\,\cite{freericks2006nonequilibrium,aoki2014nonequilibrium}, have achieved great success in solving correlated materials and pump-probe spectroscopies in the thermodynamic limit\,\cite{werner2007efficient,eckstein2010interaction}. However, when multiple instabilities compete in systems with strong electronic correlations and electron-phonon couplings, the accuracy cannot be guaranteed through perturbations which may lead to a biased solution. On the other hand, wavefunction methods, such as exact diagonalization (ED) and density-matrix renormalization group, have well-controlled numerical error, but are restricted to small systems or low dimensions\,\cite{dagotto1994correlated,white1992density}. This issue becomes even more severe when the electron-phonon coupling is non-negligible due to the unbounded phonon Hilbert space\,\cite{zhang1998density,brockt2015matrix}.

To tackle the strong and dynamical electron-phonon coupling and overcome the issue of phonon Hilbert space, we develop a time-dependent extension of the NGSED method. The idea of this method is based on the following observations for electrons and phonons, respectively. Electrons have a complicated form of interactions (four-fermion terms) and intertwined instabilities, which thereby have to be treated by an accurate solver. However, with the Pauli exclusion principle, the electronic Hilbert-space dimension is relatively small, which allows an exact solution for finite-size or low-dimensional systems. In contrast, the phonon-phonon interaction (anharmonicity) is usually weak and electron-phonon coupling has a linear form, with the complexity coming instead from the unbounded local phonon Hilbert space. Therefore, if one can find an entangler transformation involving a general form of entanglements between electrons and phonons, the solution of the correlated Hamiltonian can be mapped (after the transformation) to a factorizable wavefunction as a product state consisting of electrons and phonons separately. Using this trick, we can take advantage of the distinct properties mentioned above and solve the two subsystems using different techniques.

In equilibrium, such an efficient entangler transformation has been found and benchmarked, in terms of the generalized polaron transformation\,\cite{shi2018variational}. This leads to the wavefunction ansatz\,\cite{wang2019zero}
\begin{eqnarray}\label{eq:wvfuncansatz}
	\big|\Psi_{\rm G}\big\rangle\! =\!\Upl |\psi_{\rm ph}\rangle \!\otimes\!|\psi_{\rm e}\rangle \!= \!e^{\frac{i}{\sqrt{N}}\sum_\qbf \lambda_\qbf p_{-\qbf} \rho_\qbf} |\psi_{\rm ph}\rangle \!\otimes\!|\psi_{\rm e}\rangle\,,
\end{eqnarray}
with the electron density operator $\rho_{\qbf}=\sum_{\ibf\sigma} n_{\ibf\sigma}e^{-i\qbf\cdot \rbf_\ibf}$, the phonon momentum $p_{\qbf}=i\sum_{\ibf}(   a_{\ibf}^\dagger-  a_{\ibf})e^{-i\qbf\cdot \rbf_\ibf}/\sqrt{N}$, and the phonon displacement $x_{\qbf}=\sum_{\ibf}  (a_{\ibf}+ a_{\ibf}^\dagger)e^{-i\qbf\cdot \rbf_\ibf}/\sqrt{N}$, where $N$ is the number of lattice sites in the calculation. Here, the right-hand side is a direct product of electron and phonon states: The electronic wavefunction $|\psi_{\rm e}\rangle$ is treated as a full many-body state, while the phonon wavefunction $ |\psi_{\rm ph}\rangle$ is treated as coherent Gaussian state (see Appendix.~\ref{app:NGSED} for details). The entangler transformation involves momentum-dependent variational parameters $\lambda_\qbf$, describing the polaronic dressing. Physically, a larger dressing $\lambda_\qbf\sim g/\omega_0$ accurately describes the phonon and coupling energies, known as the Lang-Firsov transformation\,\cite{lang1962}, while a smaller dressing allows a precise solution for the electron energy. Thus, the variational parameters $\lambda_\qbf$s are optimized numerically as a balance between these two effects. Note that all $\lambda_\qbf$s are independent real numbers and entangle the phonon momentum with electron density. This entangler transformation is demonstrated to be sufficient in equilibrium\,\cite{shi2018variational,shi2021spin}. For systems with both strong electron-phonon coupling and electronic interactions, the electronic part of wavefunction $|\psi_{\rm e}\rangle$ can be solved by ED and the above framework becomes the NGSED method. The application of the NGSED to the equilibrium 2D Hubbard-Holstein model successfully reveals the novel intermediate phases with superconducting instability\,\cite{wang2019zero,karakuzu2017superconductivity, ohgoe2017competition}, which is consistent with the recent DQMC-AFQMC simulations at the thermodynamic limit\,\cite{costa2020phase}.

In the real-time evolution, the electron density is driven by the pump laser, the consequence of which is the varying force acting on the lattice and the finite momentum of phonon. To characterize the electron-density-dependent phonon momentum, we introduce an extra cubic coupling between the position and the phonon density in the non-Gaussian transformation
\begin{equation} \label{eq:NGStransformation}
	U_{\rm NGS}(t) = e^{i\frac{1}{\sqrt{N}}\sum_{\qbf}\lambda_\qbf(t) (p_{-\qbf}\cos \varphi_{\qbf}(t)\,-x_{-\qbf}\sin \varphi_{\qbf}(t))\rho _{\qbf}}
\end{equation}
to construct the variational ansatz
\begin{equation}\label{eq:NGSEDwavefunction}
	\big|\Psi(t)\big\rangle  = U_{\rm NGS}(t) |\psi_{\rm ph}(t)\rangle \otimes|\psi_{\rm e}(t)\rangle\,.
\end{equation}
Here, the additional phase parameter $\varphi_\qbf$ controls the ratio of position and momentum displacements. The time dependence of $\lambda_\qbf$ and $\varphi_{\qbf}$ allows dynamical fluctuation of the polaronic dressing effect. At the same time, these two sets of parameters, as a whole, disentangle the electrons and phonons with the Fr\"{o}hlich-type coupling and allow a relatively accurate description of the transformed wavefunction via the direct-product state in the rightmost of Eq.~\eqref{eq:NGSEDwavefunction}. The price one pays for the transformation is the complication of the effective electronic Hamiltonian [see Eq.~\eqref{eq:realH} in Appendix \ref{app:NGSED}], which is solved by the time-dependent ED with well-controlled numerical errors [see discussions in Appendix \ref{sec:krylov}]. As a minimal extension to the equilibrium wavefunction ansatz Eq.~\eqref{eq:wvfuncansatz}, the structure of Eq.~\eqref{eq:NGStransformation} involves an implicit K\"{a}hlerization, which guarantees the minimization of errors throughout the evolution (see discussions in Appendix \ref{app:NGSED}). In each numerically small step of the time evolution, i.e. $|\Psi(t)\rangle\rightarrow|\Psi(t+\delta t)\rangle$, we simultaneously evolve both the variational parameters and the full electronic wavefunction $|\psi_e(t)\rangle$, as explained in Appendixes \ref{app:NGSED} and \ref{sec:krylov}, respectively.

\begin{figure*}[!th]
\begin{center}
\includegraphics[width=18cm]{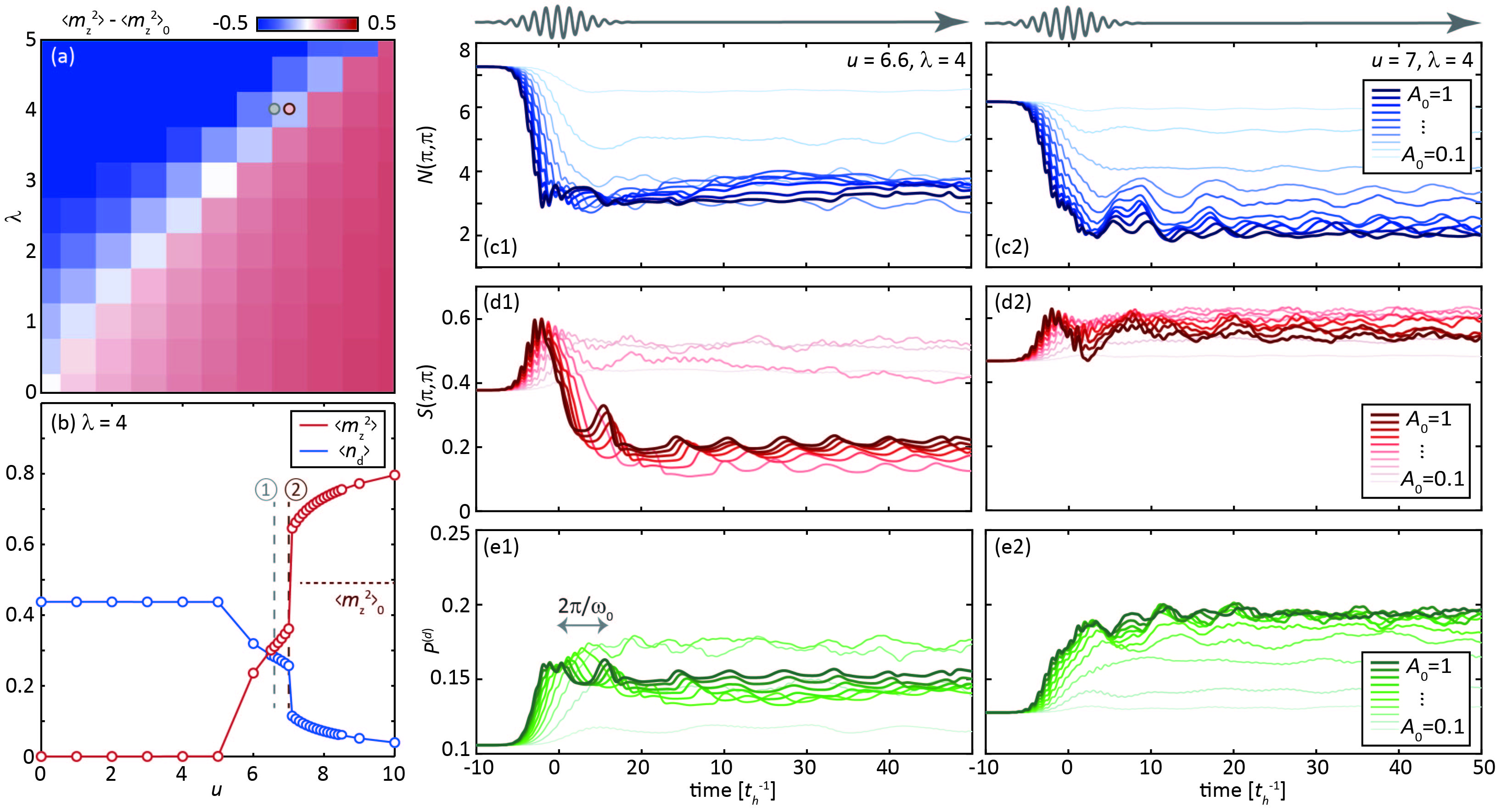}
\vspace{-6mm}\caption{\label{fig:pumpDynm} \textbf{Light-driven dynamics of charge, spin, and pairing correlations.} (a). The square of local moment $\langle m_z^2\rangle$ for different model parameters, subtracting off that for noninteracting electrons $\langle m_z^2\rangle_0$. (b). The square of local moment $\langle m_z^2\rangle$ (red) and the double occupation $\langle n_d\rangle$ as a function of $u$ for a fixed $\lambda=4$, where $\langle m_z^2\rangle_0$ is indicated by a horizontal dotted line. The two parameter sets used on right are marked as the dots in (a) and the dashed lines in (b). (c1-c2) The evolution of charge correlation $N(\pi,\pi)$ after a linear pump of various strengths $A_0$ ranging from 0.1 to 1 (denoted by different colors) with $\lambda=4$ and (c1) $u=6.6$ and (c2) $u=7$, respectively. (d1-e2) The same as (c1-c2) but for (d1-d2) the spin correlation $S(\pi,\pi)$ and (e1-e2) the $d$-wave pairing susceptibility $P^{(d)}$. The pump pulse is sketched above.\vspace{-5mm}
}
\end{center}
\end{figure*}

\section{Light-Enhanced Pairing Correlations}\label{sec:lightInduceddWave}
Figure~\ref{fig:pumpDynm}(a) shows the square of local moment $\langle m_z^2\rangle = \sum_i\langle (n_{i\uparrow} - n_{i\downarrow})^2\rangle/N$ calculated for different $u$ and $\lambda$ of the 12.5\% doped Hubbard-Holstein model, where $N$ is the system size. $\langle m_z^2\rangle$ provides an indication of the overall spin moment and spin correlation. An abrupt change of $\langle m_z^2\rangle$ with varying $u$ and/or $\lambda$ can signal a phase transition. At the large-$u$ regime, the system is dominated by the spin correlations inherent from the Hubbard model. In this regime, $d$-wave superconductivity is identified in a quasi-1D system\,\cite{jiang2019superconductivity,chung2020plaquette}, although inthe 2D thermodynamic limit its existence is not fully established\,\cite{maier2005systematic}. In contrast, at the large-$\lambda$ regime, the electrons are bound with phonons as bipolarons, exhibiting a local moment lower than that of noninteracting electrons. This is also reflected in the average double occupation $\langle n_d\rangle =\sum_i \langle n_{i\uparrow} n_{i\downarrow} \rangle/N$ presented in Fig.~\ref{fig:pumpDynm}(b). There is an intermediate regime between these two limits, where the charge correlation slightly dominates (reflected by the comparison to non-interacting-electron local moment $\langle m_z^2\rangle_0$), while the spin correlations remain finite. Based on recent studies using NGSED and QMC methods, this intermediate regime persists at the thermodynamics limit and reflects the realistic phases in a competing-order system\,\cite{wang2019zero, costa2020phase}.  We focus on this intermediate regime, which we believe corresponds to the situation in cuprate experiments\,\cite{fausti2011light,nicoletti2014optically, forst2014melting, casandruc2015wavelength, khanna2016restoring,nicoletti2018magnetic,cremin2019photoenhanced, wang2018light}.

\begin{figure*}[!th]
\begin{center}
\includegraphics[width=18cm]{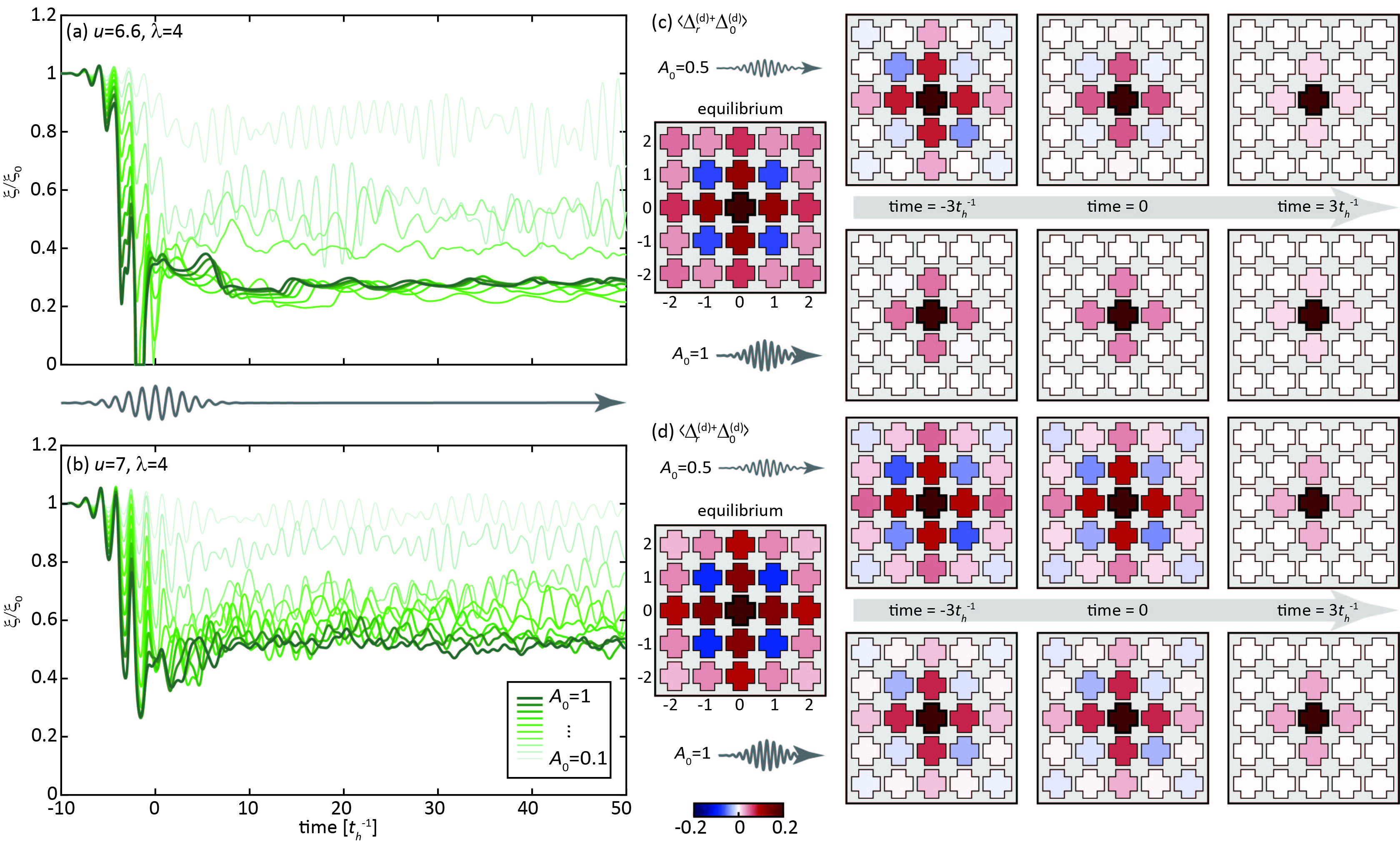}\vspace{-2mm}
\caption{\label{fig:correlations} \textbf{Time-dependent correlation length and spatial distribution of Cooper pairs.} (a,b) Evolution of the correlation length $\xi$ of $d$-wave Cooper pairs renormalized by its equilibrium value $\xi_0$, obtained for (a) $u=6.6$ and (b) $u=7$, respectively. (c) The real-space distribution of $\langle\Delta_{\rbf}^{(d)\dagger}\Delta_{\bf 0}^{(d)}\rangle$ for different $\rbf$ (considering translational invariance) evaluated at equilibrium, $t=-3t_h^{-1}$, 0, and $3t_h^{-1}$, respectively. The upper and lower present the dynamics induced by two different pump strengths $A_0=0.5$ and $A_0=1$, respectively. The simulations are obtained for the $u=6.6$ and $\lambda=4$ system. (d) The same as (c) but for the $u=7$ and $\lambda=4$ system.\vspace{-4mm}
}
\end{center}
\end{figure*}

We select two sets of model parameters inside this intermediate regime. Set 1 ($u=6.6$ and $\lambda=4$) lies in the center of the intermediate regime, while Set 2 ($u=7$ and $\lambda=4$) lies close to the boundary. We first focus on Set 1 and examine the light-induced dynamics for the charge structure factor $N(\qbf) = \langle \rho_{-\qbf}\rho_\qbf\rangle /N$ and spin structure factor $S(\qbf) = \langle \rho^{s}_{-\qbf}\rho^{s}_\qbf\rangle /N$, where $\rho^{s}_\qbf=\sum_{\ibf} (n_{\ibf\uparrow}-n_{\ibf\downarrow})e^{-i\qbf\cdot \rbf_\ibf}$. We focus on the momentum $\qbf=(\pi,\pi)$, due to the important role of spin fluctuations at this momentum on superconductivity; data at other momenta are presented in the Appendix \ref{app:othermomentum}. Figures \ref{fig:pumpDynm} (c1) and (d1) show the pump dynamics of $N({\pi,\pi})$ and $S({\pi,\pi})$ for Set 1. The same as the half-filled case\,\cite{wang2018light}, the pump pulse suppresses the charge correlations and enhances spin fluctuations, which possibly serve as the pairing glue, at short time ($t<0$). This leads to the rise of $d$-wave superconductivity instability [see Fig.~\ref{fig:pumpDynm}(e1)], characterized by the pairing susceptibility
\begin{eqnarray}
P^{(d)} = \frac1N\left\langle \Delta^{(d)\dagger}_0 \Delta^{(d)}_0\right\rangle,
\end{eqnarray}
where the $d$-wave pairing operator reads
\begin{eqnarray}
	 \Delta^{(d)}_\kbf &= &\sum_{\kprbf} c_{\kbf-\kprbf\downarrow} c_{\kprbf\uparrow} \left[\cos \kpr_x - \cos  \kpr_y\right]\,.
\end{eqnarray}
As Set 1 is relatively far from the spin-dominant regime, spin fluctuation is unstable, manifest as the drop of $S({\pi,\pi})$ for strong pump ($A_0>0.5$) at longer timescales. As a consequence, the $d$-wave pairing susceptibility stops increasing and starts to decrease for these strong pump conditions. This nonmonotonic behavior reflects that the transient $d$-wave pairing emerges from the dedicated balance among charge, spin, and electronic itineracy. While the latter two can be induced by light and enhance the $d$-wave pairing, their internal competition may reduce the enhancement if the pump strength keeps increasing. Noticeably, when the nonequilibrium state finally melts into such a ``metallic'' state, lattice distortions are released. In these cases, the dynamics exhibit a period $\sim 2\pi/\omega_0$. This is reflected in the corresponding Fourier spectra (see also the Appendixes).

The situation becomes different if one drives the system with Set 2 ($u=7$ and $\lambda=4$), where the system resides in proximity to a phase boundary. Because of the existing strong magnetic fluctuations, we find that the $d$-wave pairing correlation is easier to enhance. As shown in Figs.~\ref{fig:pumpDynm}(c2-e2), the pump pulse increasingly enhances $S({\pi,\pi})$ and $P^{\rm(d)}$ after suppressing the charge fluctuations. Although the enhancement of $S({\pi,\pi})$ saturates at large pump strengths, spin fluctuations are more robust against the increase of pump, due to proximity to a quantum phase boundary. Therefore, they never drop below the equilibrium values. This change of spin fluctuations causes the saturation of $d$-wave pairing correlations in Fig.~\ref{fig:pumpDynm}(e2). Similarly, this enhanced $d$-wave pairing concurs with a melting of  the competing CDW order, which leads to an oscillation of phonon energy. Since the CDW order is already very weak near the phase boundary, the intensity of the amplitude mode is no longer visible in the dynamics.

\section{Quantum Fluctuations and Correlation Length}\label{sec:correlationLen}
The increase of pairing correlation reflects the formation of Cooper pairs induced by the pump pulse. The experimental reflection of this correlation is the opening of a gap (or Josephson plasma) characterized by ultrafast optical reflectivity\,\cite{fausti2011light, nicoletti2014optically, forst2014melting, casandruc2015wavelength, khanna2016restoring, nicoletti2018magnetic, cremin2019photoenhanced}. However, we emphasize that the opening of a gap does not necessarily reflect the onset of superconductivity. As recently shown in cuprates and FeSe experiments, above $T_c$ there exists a fluctuating superconductivity phase that exhibits a superconducting gap but also a finite resistance, reflecting preformed Cooper pairs\,\cite{kondo2015point, he2020fluctuating, faeth2020incoherent, xu2020spectroscopic}. This is a signature of strong correlations in quantum material distinct from the mean-field notions, due to strong thermal or quantum fluctuations. For the nonequilibrium superconductivity in this paper, we focus on the zero-temperature dynamics and fluctuations may arise from quantum instead of thermal origins. 

\begin{figure*}[!th]
\begin{center}
\includegraphics[width=16.5cm]{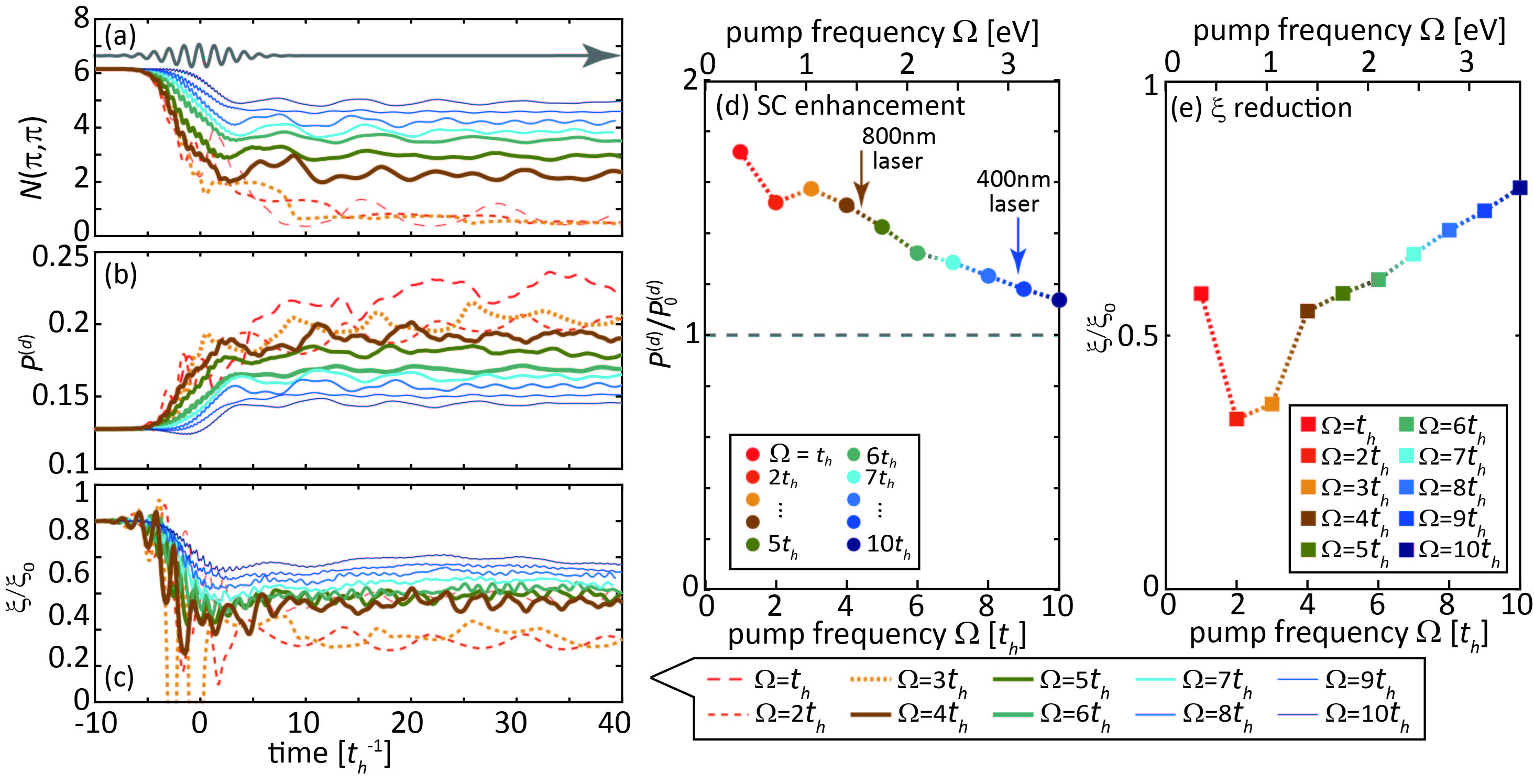}\vspace{-2mm}
\caption{\label{fig:freqDep} \textbf{Frequency dependence of the light-induced dynamics and superconductivity.} (a-c) Evolution of the (a) charge correlation $N(\pi,\pi)$, (b) the $d$-wave pairing susceptibility $P^{(d)}$, and (c) the correlation length $\xi$, at fixed pump strength $A_0=0.8$ but with different pump frequencies $\Omega$. (d,e) The postpump (d) pairing correlation $P^{(d)}$ and (e) correlation length $\xi$ averaged between time $t=10-30\,t_h^{-1}$ for different pump frequencies, relative to their equilibrium values.\vspace{-4mm}
}
\end{center}
\end{figure*}

To better clarify the nature of this photoinduced many-body state, we further consider the spatial fluctuation through the FFLO pairing correlation with finite momentum $P^{(d)}_\kbf = \frac1N\left\langle \Delta^{(d)\dagger}_\kbf \Delta^{(d)}_\kbf\right\rangle$\,\cite{fulde1964superconductivity,larkin1965nonuniform}. For $\kbf=0$, it recovers to the BCS pairing correlation, reflecting the total number of Cooper pairs. The correlation length $\xi$ can be estimated through
\begin{equation}
    \xi =\lim_{|\kbf|\rightarrow 0}\frac1{|\kbf|}\sqrt{\frac{P^{(d)}_0}{P^{(d)}_\kbf} - 1}\,,
\end{equation}
based on the definition of correlation length satisfying $\langle \Delta_\ibf^{(d)\dagger} \Delta_\jbf^{(d)}\rangle \sim e^{-|\rbf_i-\rbf_j|/\xi}$. Figures \ref{fig:correlations}(a) and (b) present the evolution of $\xi$, estimated using the smallest momentum accessible in our cluster $\kbf=(\pi/2,0)$ and normalized by the equilibrium correlation length $\xi_0$. For the $u=6.6$ system where charge order is robust [\ref{fig:correlations} (a)], the correlation length is dramatically suppressed for $A_0<0.5$, where the $P^{(d)}$ is enhanced in Fig.~\ref{fig:pumpDynm}(e1). When the pump strength increases to $A_0 \ge 0.5$, $\xi$ even drops to zero at the center of the pump, indicating the destruction of local Cooper pairs and consistent with the drop of $P^{(d)}$ in Fig.~\ref{fig:pumpDynm}(e1). This overall drop of correlation length can be visualized in the real-space distribution of $\langle \Delta_\rbf^{(d)\dagger} \Delta_0^{(d)}\rangle$ shown in Fig.~\ref{fig:correlations}(c): The Cooper pairs become strongly localized and spatially decoherent after the pump.

What is more interesting is the system near the phase boundary ($u=7$), where the BCS pairing correlation $P^{(d)}$ always increases after the pump, as discussed in Fig.~\ref{fig:pumpDynm}(e2). However, this enhancement of total Cooper pairs is always accompanied by a drop of the correlation length, as shown in Fig.~\ref{fig:correlations}(b). This means that the photoinduced Cooper pairing is more fluctuating than the equilibrium one. Different from the $u=6.6$ system, here the decrease of $\xi$ saturates at a moderate value, about half of the equilibrium $\xi_0$. To filter out the oscillatory dynamics, we extract the average enhancement of pairing correlation $P_d(t)/P_d(-\infty)$ and the suppression of $\xi(t)/\xi(-\infty)$ during the time window $10t_h^{-1} < t< 30t_h^{-1}$, long after the pump pulse disappears. As already shown in Fig.~\ref{fig:cartoon}(b), the relative changes (suppression or enhancement) in these two quantities are comparable for all pump strengths. Considering that long-range correlations asymptotically approach $\langle \Delta_\ibf^{(d)\dagger} \Delta_\jbf^{(d)}\rangle\sim P^{(d)} e^{-|\rbf_i-\rbf_j|/\xi}$, our simulation reflects an enhancement of short-range superconductivity but a suppression of the long-range one\,\cite{definitionOfSC}.

The difference between a short-range enhancement and long-range suppression can be visualized via the spatial distribution obtained by the Fourier transform of $P^{(d)}_\kbf$. As shown in Fig.~\ref{fig:correlations}(c), the local pairing correlation increases and the range where the correlation is visible remains finite after the pump. Both facts suggest that this type of short-range superconductivity is visible in pump-probe spectroscopies, albeit with a finite broadening. Thus, the fluctuating nature of light-enhanced superconductivity does not conflict with existing optical experiments in La$_{1.675}$Eu$_{0.2}$Sr$_{0.125}$CuO$_4$ and La$_{1.875}$Ba$_{0.125}$CuO$_4$\,\cite{fausti2011light, nicoletti2014optically, forst2014melting, casandruc2015wavelength, khanna2016restoring, nicoletti2018magnetic, cremin2019photoenhanced}. Such a suppression of the coherence is associated with the nonlocal nature of $d$-wave Cooper pairs, which is in contrast to the light-enhanced strength and coherence of local pairing\,\cite{sentef2017theory, coulthard2017enhancement, li2020manipulating, tindall2020dynamical}. This anticorrelation between the strength and coherence is also distinct from previous light-induced competing orders\,\cite{lu2012enhanced,al2013wave,rincon2014photoexcitation,lu2015photoinduced}, possibly resulting from the fact that superconductivity arises as a third instability emergent from the competition.

\section{Frequency Dependence}\label{sec:freq}

We further investigate the impact of the pump frequencies on superconductivity and the competing orders. With a fixed pump amplitude $A_0=0.8$, Figs.~\ref{fig:freqDep}(a-c) present the dynamics of various quantities discussed above for different frequencies. As $\Omega$ increases from $t_h$ (350\,meV) to $10t_h$ (3.5\,eV), the suppression of the charge correlations is weaker. This can be attributed to the loss of resonances to the phonon dynamics, which plays the crucial role in the melting of the charge order. Interestingly, these suppressions of CDW do not always translate into a monotonic increase of the $d$-wave superconductivity. As shown in Fig.~\ref{fig:freqDep}(b), the low-frequency pump with $\Omega=2 t_h$ leads to a slightly weaker enhancement and a dramatic decrease of the correlation length $\xi$. The transient correlation length is suppressed to zero in the center of the pump [see Fig.~\ref{fig:freqDep}(c)], indicating the loss of superconductivity. We tentatively attribute this anomaly to the fact that the parametric driving condition $\Omega\sim 2\omega_0$ is reached\,\cite{knap2016dynamical, babadi2017theory}, leading to strong fluctuations of phonons and overwhelming the $d$-wave pairing\,\cite{macridin2006synergistic,mendl2017doping}. 

Excepting the two frequencies which correspond to parametric driving, the enhancement of the $d$-wave pairing susceptibility drops monotonically with the pump frequency, consistent with the melting of the charge correlations. We stress that such a phenomenon reflects that the dynamics are driven by light-induced quantum fluctuations instead of a thermal effect, as the fluence increases (instead of decreases) with the pump frequency. Although some nonequilibrium effects are simply attributed to the sudden heating of the electronic states caused by the pump pulse, this is not the situation we found in the light-enhanced $d$-wave superconductivity. More rigorous discrimination of the thermal effects relies on the calculation of pump-probe spectroscopies and a fitting with the thermal distribution function\,\cite{wang2017producing,shvaika2018interpreting,wang2018theory, matveev2019stroboscopic, mootz2020lightwave}, which is beyond the scope of this paper. 

Figures~\ref{fig:freqDep}(d) and (e) summarize the frequency dependence of the pairing susceptibility and the correlation length, averaged between time $t=10-30\,t_h^{-1}$ after the pump, similar to the pump strength dependence shown in Fig.~\ref{fig:cartoon}. Except for the situation close to phonon parametric resonance, we find that the maximal superconductivity enhancement is reached at approximately $1.4$\,eV and rapidly decreases for higher frequencies. As this frequency is close to the Mott gap resonance (approximately $4t_h\sim 1.4\,$eV) for $U=8t_h$, this enhancement can be attributed to the generation of spin fluctuations across the Mott gap. This observation is consistent with the wavelength-dependence study in LBCO, where the 800-nm laser is found to enhance superconductivity move efficiently than a 400-nm laser\,\cite{casandruc2015wavelength}. Although other mechanisms beyond the single-band model have been proposed to address this wavelength dependence, such a consistency strengthens the connection between our microscopic simulation and the real experiments.

\section{Summary and Outlook}\label{sec:conclusion}

Altogether, our study provides a novel perspective of interpreting the light-induced $d$-wave pairing in the context of strong correlation and electron-phonon coupling. Although it does not rule out other interpretations, our result reflects that the light-induced Cooper pairs may be spatially local. Therefore, the Josephson plasma resonance and the optical gap may coexist with a small Drude weight, which reflects the superfluid density and cannot be directly resolved in ultrafast reflectivity experiments. Fortunately, this mystery was recently answered by a state-of-the-art ultrafast transport measurement, in another type of light-induced superconductor (K$_3$C$_{60}$)\,\cite{budden2021evidence}. Such a transport measurement directly provides the transient resistivity of the material, which reflects the superconductivity long-range order. Therefore, a similar transport measurement in La$_{1.875}$Ba$_{0.125}$CuO$_4$ will help to distinguish fluctuations from ordered $d$-wave superconductivity. 

In addition to optical and transport properties, the single-particle electronic structure measurable by photoemission is employed frequently to characterize the $d$-wave superconductors. In the single-particle context, the equilibrium fluctuating superconductivity translates into the (single-particle) gap opening above the transition temperature $T_c$\,\cite{he2020fluctuating, faeth2020incoherent, xu2020spectroscopic}, while the distinction from long-range ordering may hide in the quantitative spectral shape (e.g., quasiparticle width and temperature dependence of the Fermi-surface spectral weight) and is still an ongoing research. Identifying this fluctuation would be even harder, but also promising, for a nonequilibrium state measured by the time-resolved ARPES. The findings of our simulations indicate that the light-driven La$_{1.875}$Ba$_{0.125}$CuO$_4$ should exhibit a (single-particle) superconducting gap, but its quasiparticle peak should be damped by the fluctuations. Complementary to regular angle-resolved photoemission spectroscopy (ARPES), future developments on two-electron ARPES measurement\,\cite{su2020coincidence, mahmood2021distinguishing}, at ultrafast timescales, are promising directions to distinguish long-range superconductivity from a fluctuating one.

Alternatively, a recent attempt has investigated the inverse process by measuring the coherence of CDW states using x-ray scattering\,\cite{wandel2020light}. Light-induced CDW also has been recently observed in LaTe$_3$ with two competing CDW orders\,\cite{kogar2020light}. Here, we investigate specifically light-driven cuprates, but appropriate modification\,\cite{demler1995theory,demler1998pi} of our study and microscopic model can provide a general platform to examine light-driven fluctuations for intrinsically competing orders, as pointed out in a recent phenomenological study~\cite{sun2020transient}. The fluctuating nature of the enhanced quantum phase suggests that the nonequilibrium states of competing-order systems are beyond the description of a mean-field framework.

To tackle the dynamics of systems with strong electronic correlation and electron-phonon coupling, we develop a new technique by generalizing the variational non-Gaussian exact diagonalization framework out of equilibrium. This method overcomes the issues of unbounded phonon Hilbert space, while keeping the numerical accuracy through the exact diagonalization of the (transformed) electronic problem. Although we restrict the application into the Hubbard-Holstein model relevant for the cuprate $d$-wave superconductivity, our derivation is rather general and can be applied to dispersive phonons, extended electronic interactions, and more complicated electron-phonon coupling. The formalism also can be applied to multiple phonon branches as long as the phonon-phonon interaction can be ignored, but the generalization to multiband electrons is restricted by the Hilbert-space size issue in exact diagonalization. Future application of this method also includes the evolution of pump-probe spectroscopies\,\cite{freericks2009theoretical,lenarvcivc2014optical,shimizu2011sum,wang2017producing,wang2018theory} by tracking the time-dependent gauges and evaluating the off-diagonal overlap of Gaussian states.

\section*{Acknowledgement}
The authors thank J.I. Cirac, E. Demler, J.-F. He, M. Mitrano, M. A. Sentef, and I. E. Perakis for insightful discussions. Y.W. acknowledges support from National Science Foundation (NSF) Grant No.~DMR-2038011. C.-C.C. acknowledges support from NSF Grant No.~OIA-1738698 and OAC-2031563. The calculations were performed on the Frontera computing system at the Texas Advanced Computing Center. Frontera is made possible by NSF NSF Grant No.~OAC-1818253.

\appendix

\section{Detailed Derivation of the Time-Dependent NGSED Method}\label{app:NGSED}

To obtain the equations of motion for variational parameters in Eq.~\eqref{eq:NGSEDwavefunction}, where the phonon  state is
\begin{eqnarray}
 	|\psi_{\rm ph}\rangle = e^{-\frac12 R_0^T \sigma_y \Delta_R} e^{-i\frac14 \sum_\qbf \Rqd \xi_\qbf  \Rq} |0\rangle = \Ub |0\rangle\,,
\end{eqnarray}
we notice that the generalization from $\Upl$ to $U_{\rm NGS}$ involves a rotation of phonon operators in the phase space. Therefore, the $U_{\rm NGS}$ can be expressed by a composite transformation, leading to an equivalent form of Eq.~\eqref{eq:NGSEDwavefunction}:
\begin{eqnarray}
    \big|\Psi(t)\big\rangle  = \Ur(t) \Upl(t) |\psi_{\rm ph}(t)\rangle \otimes|\psi_{\rm e}(t)\rangle\,,
\end{eqnarray}
where the rotation is reflected in the unitary (Gaussian) transformation
\begin{eqnarray}
	\Ur(t) = e^{-i\frac14 \sum_\qbf \Rqd \eta_\qbf  \Rq},
\end{eqnarray}
and $\eta_\qbf$ is a variational matrix.
In the derivation of variational equations of motion, it is usually convenient to employ the linearized form $\bar{S}_{\qbf}$ to parametrize the rotational transformation, i.e.~$\Ur^\dagger \Rq \Ur = \bar{S}_{\qbf}\Rq$. Thus, ignoring the redundant degrees of freedom, which can be absorbed into $|\psi_{\rm ph}\rangle$, we obtain the following relation between $\varphi_\qbf$ and $\bar{S}_{\qbf}$
\begin{eqnarray}\label{eq:fixedGaugeSbarq}
\bar{S}_{\qbf}(t) = e^{i\sigma_y\eta_\qbf} =\left(\begin{array}{cc}
    \cos \varphi_\qbf(t)  & -\sin\varphi_\qbf(t)\\
    \sin\varphi_\qbf(t) & \cos\varphi_\qbf(t)
\end{array}\right).
\end{eqnarray}
In the above wavefunction prototype, $\Delta_R$ (vector), $\xi_\qbf$ (matrix), $\lambda_\qbf$ and $\varphi_\qbf$ (scalar) are variational parameters, which are allowed to evolve during the dynamics.

To evaluate the nonequilibrium dynamics within the hybrid wavefunction ansatz, we project the \Schrdg\ equation $i\dt\big|\Psi(t)\big\rangle  = \Ham(t) \big|\Psi(t)\big\rangle$ onto the tangential space in the variational manifold, which gives equations of motion for both the variational parameters and the electronic wavefunction $|\psi_e(t)\rangle$. In contrast to an energy minimization (imaginary-time evolution), the algebra of this projection in a real-time evolution may be ill defined. For the general time-dependent variational principle, the real and imaginary parts in the projection of Schr\"{o}dinger equation on the tangential space 
\begin{equation}\label{eq:TDVP}i\langle \xi_\alpha|\dt\big |{\Psi}(t)\big\rangle = \langle \xi_\alpha|\Ham(t) \big |{\Psi}(t)\big\rangle,
\end{equation}
with $|\xi_\alpha\rangle$ denoting any tangential vector, may give rise to two sets of incompatible equations of motion, which correspond to the Dirac-Frankel and McLachlan variational principles, respectively\,\cite{hackl2020geometry}. If and only if the variational manifold is K\"{a}hler manifold, the two principles result in the same equation of motion. By designing the wavefunction ansatz in Eqs.~\eqref{eq:NGSEDwavefunction} and \eqref{eq:NGStransformation}, we K\"{a}hlerize the variational manifold such that the two principles are compatible and the derivations below are self-consistent. Physically,  the K\"{a}hlerity guarantees that the variational principle Eq.~\eqref{eq:TDVP} minimizes errors with respect to the realistic time-dependent wavefunctions.

Taking advantage of the composite representation of $U_{\rm NGS}$, it is convenient to define
\begin{eqnarray}
	\big|\bar{\Psi}(t)\big\rangle  = \Ur^\dagger \big|\Psi(t)\big\rangle = \Upl(t) |\psi_{\rm ph}\rangle \otimes|\psi_{\rm e}\rangle\,,
\end{eqnarray}
and solve the \Schrdg\ equation in a rotating frame $i\dt\big|\bar{\Psi}(t)\big\rangle  = \bar{\Ham} \big|\bar{\Psi}(t)\big\rangle$.
Thus, the rotated Hamiltonian becomes
\begin{eqnarray}
\bar{\Ham} &=& \Ur^\dagger \Ham \Ur-i\Ur^\dagger\dt \Ur\nonumber\\
&=&  \sum_{\kbf\sigma} (\varepsilon_\kbf +\mu)c^\dagger_{\kbf\sigma}c_{\kbf\sigma}  + \frac14\sum_\qbf \Rqd (\omega_\qbf + \dt\varphi_\qbf)\Rq\nonumber\\
&&  + \frac1{\sqrt{N}}\sum_\qbf \gred\Rqd \bar{S}_\qbf^\dagger e_1 \rho_{\qbf} + \Ham_{e-e}.
\end{eqnarray}
Here, we denote the general electron-electron interaction terms as $\Ham_{e-e}$ and assume it to commute with the electronic density operator $n_\ibf$, which is satisfied in the Hubbard-Holstein model Eq.~\eqref{eq:HHM}. This assumption substantially simplifies the derivation and leads to the following Eq.~\eqref{eq:realH}.

On the left-hand side of the rotated \Schrdg\ equation, we obtain
\begin{eqnarray}\label{eq:idtNGS}
i\dt\big |\bar{\Psi}\big\rangle \!
&=&\! \Upl \Ub \Big[ -i \frac14\Delta_R^T\sigma_y\dt \Delta_R -i  \frac12    R_0^TS_0^T \sigma_y\dt \Delta_R\nonumber\\
&&-i \frac14\sum_\qbf\Rqd S_\qbf^\dagger \sigma_y\dt S_\qbf\Rq-\frac1{\sqrt{N}}\sum_\qbf \Rqd S_\qbf^\dagger e_2 \rho_{\qbf}\dt\lambda_\qbf\nonumber\\
&&-\frac1{\sqrt{N}} \Delta_R^T e_2 \rho_{0}\dt\lambda_0 \Big]|0\rangle_{\rm ph}\otimes|\psi_e\rangle\nonumber\\
&& + \Upl |\psi_{\rm ph}\rangle \otimes( i\dt\big |\psi_e\big\rangle)\,.
\end{eqnarray}
Here, $S_{\qbf} = e^{i\sigma_y\xi_\qbf}$ is the linearized transformation matrix of the $\Ugs$
and the effective Hamiltonian becomes
\begin{eqnarray}\label{eq:realH}
\Ham_{\rm  eff}&=&\Upl^\dagger\bar{\Ham}\Upl\nonumber\\
&=&-t_h\sum_{\jbf\sigma\alpha\delta_\alpha} e^{i\frac1{\sqrt{N}}\sum_\qbf \lambda_\qbf\Rqd  e_2 e^{-i\qbf\cdot\jbf}\left(1-e^{-i\qbf\cdot\delta_\alpha}\right) } c_{\jbf+\delta_\alpha,\sigma}^\dagger c_{\jbf\sigma}\nonumber\\
&& -\frac1{2N}\sum_\qbf\sum_{\kbf,\kbf^\prime\atop \sigma, \sigma^\prime} V_\qbf\, c_{\kbf+\qbf,\sigma}^\dagger c_{\kbf\sigma} c_{\kbf^\prime-\qbf, \sigma^\prime}^\dagger c_{\kbf^\prime \sigma^\prime}+ \Ham_{e-e} \nonumber\\
&&-\invrN\!\sum_\qbf\Rqd  \!\left[\lambda_\qbf\!\left(\omega_0 \!+\!\dt\varphi_\qbf \right)\!-\! g\bar{S}_\qbf^\dagger \right]\!e_1\rho_\qbf\nonumber\\
&& +\frac14\sum_\qbf(\omega_0 \!+ \!\dt\varphi_\qbf) \Rqd \Rq\,,
\end{eqnarray}
for the product state $|\psi_{\rm ph}\rangle \otimes\big |\psi_e\big\rangle$.
Here, $\alpha$ takes the $x$ and $y$ directions, while $\delta_\alpha$ denotes the vector pointing to the nearest neighbors along the corresponding directions. 

After transforming the Hamiltonian to the basis of product states, the electron-phonon dressing effects are reflected in both the interaction and the kinetic energy. For the effective electronic interaction, the dressing of phonons mediates an additional $V_\qbf$ on top of the original $H_{e-e}$, whose variational expression is 
\begin{eqnarray}\label{eq:vqexplicit}
V_\qbf = 4g \cos \varphi_\qbf \real[ \lambda_\qbf] -2|\lambda_\qbf|^2(\omega_0+\dt\varphi_\qbf).
\end{eqnarray}
(Note, that we employed the assumption that $H_{e-e}$ commutes with $n_\ibf$ and, therefore, $\Upl$.) For the kinetic energy, the phonon-dressed effective hopping term can be reformulated in a closed form
\begin{eqnarray}\label{eq:taexpr}
    t_\alpha =t_h e^{-\sum_\qbf\frac{|\lambda_\qbf|^2}{N}(1-\cos q_\alpha) e_2^T \Gamma_\qbf e_2 }\,,
\end{eqnarray}
when taking the expectation with respect to the phonon Gaussian state $|\psi\rangle_{\rm ph}$.

Now with both the time derivative and the effective Hamiltonian transformed into the product-state basis, we can project Eqs.~\eqref{eq:idtNGS} and \eqref{eq:realH} onto various tangential vectors sequentially. For the zeroth-order tangential vector, which is proportional to the phonon vacuum state $|0\rangle_{\rm ph}$, we obtain the equation of motion for the electronic state:
\begin{eqnarray}
i\dt\big |\psi_e\big\rangle &=& \Big[\langle \psi_{\rm ph}|\Ham_{\rm  eff}(t)|\psi_{\rm ph}\rangle +i \frac14\Delta_R^T\sigma_y\dt \Delta_R \nonumber\\
&&+\frac1{\sqrt{N}} \Delta_R^T e_2 \rho_{0}\dt\lambda_0-\frac12\sum_\qbf \dt \varphi_\qbf   \Big]|\psi_e\big\rangle.\nonumber\\
\end{eqnarray}
The (phonon-traced) $\Ham_{\rm eff}$ together with the scalar terms in the square brackets govern the time evolution of the electronic wavefunction $|\psi_e\rangle$. In the NGSED framework, we treat $|\psi_e\rangle$ as a full many-body state and evaluate this evolution stepwisely by the Krylov-subspace method (see Appendix \ref{sec:krylov}).

Moreover, the equations of motion for (the variational parameters of) $|\psi_{\rm ph}\rangle$ can be obtained by projecting Eqs.~\eqref{eq:idtNGS} and \eqref{eq:realH} onto the first-order tangential vector ($R_0^T S_0^T|0\rangle_{\rm ph}\otimes|\psi_e\rangle$) and the second-order tangential vector ($\Rq^\dagger \Rq |0\rangle_{\rm ph}\otimes|\psi_e\rangle$). After some algebra, these two equations are
\begin{eqnarray}\label{eq:EOMDel}
\dt \Delta_R&=& i\sigma_y\Big\{\left(\omega_0+\dt\varphi_0\right) \Delta_R -2\sqrt N\rho_{\rm f} \Big[\left(\lambda_0\omega_0+\lambda_0\dt\varphi_0\right.\nonumber\\
&&\left.-\gredZero \bar{S}_0^\dagger\right)e_1 - \dt \lambda_0 e_2\Big]\Big\},
\end{eqnarray}
and
\begin{eqnarray}\label{eq:secondBosonDynm}
-iS^\dagger_\qbf\sigma_y\dt S_\qbf =S_\qbf^\dagger \tilde{\Omega}_\qbf(t) S_\qbf,
\end{eqnarray}
respectively.
Here, $\rho_{\rm f} = N_e/N$ is the average electron density, and the renormalized phonon energy matrix is
\begin{eqnarray}\label{eq:renormalizedPhononFreq}
\tilde{\Omega}_\qbf(t)&= &\frac{8|\lambda_\qbf|^2}{N}\sum_{k\alpha}t_\alpha \big[1-\cos q_\alpha\big]\langle n_\kbf\rangle\cos k_\alpha e_2 e_2^T\nonumber\\
&&+\omega_\qbf + \dt\varphi_\qbf.
\end{eqnarray}
As we are interested in the equal-time measurements (see Sec.~\ref{sec:lightInduceddWave}), the gauge degrees of freedom are redundant. Therefore, we employ a gauge-invariant covariance matrix $\Gamma_\qbf=S_\qbf S_\qbf^\dagger$ and track the evolution of $\Gamma_\qbf$ instead of $S_\qbf$. Equation \eqref{eq:secondBosonDynm} becomes
\begin{eqnarray}\label{eq:bosonQuadraticTerm3}
 \dt \Gamma_\qbf = i\sigma_y\tilde{\Omega}_\qbf(t)\Gamma_\qbf - i\Gamma_\qbf\tilde{\Omega}_\qbf(t)\sigma_y.
\end{eqnarray}

\begin{figure}[!t]
\begin{center}
\includegraphics[width=8.5cm]{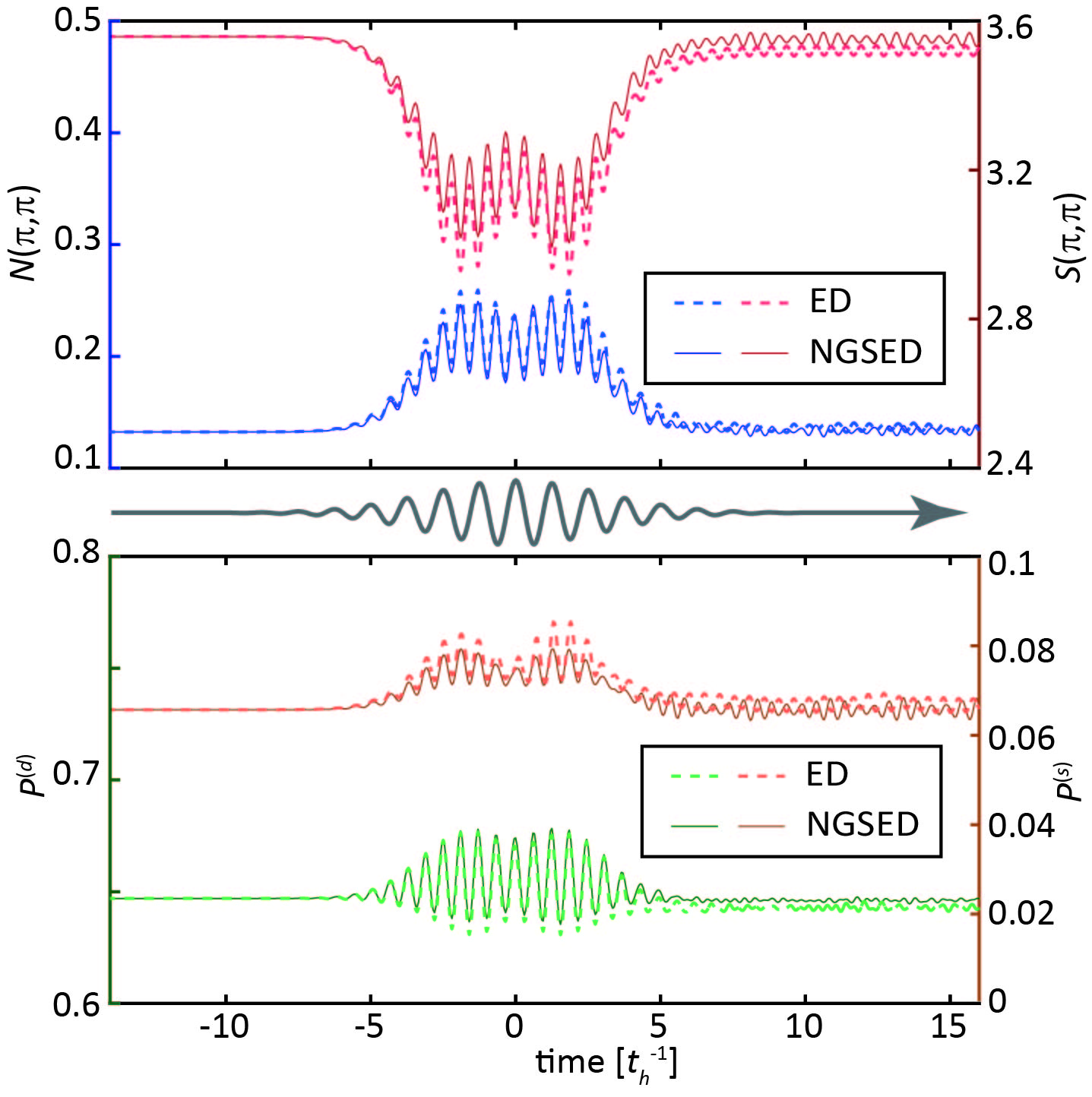}
\caption{\label{fig:EDbenchmark} Time evolution of the charge structure factor (blue curve), spin structure factor (red curve), $d$-wave pairing susceptibility (green curve), and $s$-wave pairing susceptibility (orange curve), for a pump pulse with $A_0 = 0.4$ and frequency $\Omega=5t_h$. The solid and dotted lines denote the results obtained by NGSED and the standard ED (with maximal phonon number $M=10$), respectively. The inset guides the eye for the pump pulse.
}
\end{center}
\end{figure}

Finally, the equations of motion of $\lambda_\qbf$ and $\varphi_\qbf$, which appear in $U_{\rm NGS}$ and entangle phonons and electrons, can be obtained by projecting to the third-order tangential vector $a_\qbf^\dagger\rho_{\qbf}|0\rangle_{\rm ph}\otimes|\psi_e\rangle$. These two equations are
\begin{eqnarray}
\dt\lambda_\qbf = g_\qbf\sin\varphi_\qbf + 2\lambda_\qbf\frac{\Pi_\qbf}{C_\qbf}e_1^T \Gamma_\qbf e_2,
\end{eqnarray}
and
\begin{eqnarray}
\dt\varphi_\qbf = \frac{g_\qbf\cos \varphi_\qbf}{\lambda_\qbf} -\omega_\qbf +2\frac{\Pi_\qbf}{C_\qbf}e_2^T\Gamma_\qbf e_2,
\end{eqnarray}
respectively, where the modulated electronic correlation is
\begin{eqnarray}
\Pi_\qbf = \sum_{k\sigma\alpha } \!t_\alpha [\cos k_\alpha \!-\! \cos(k_\alpha\!+\!q_\alpha)]    \Big\langle \!{\rho_{-\qbf}}{c}_{\kbf\sigma}^\dagger {c}_{\kbf\!+\!\qbf,\sigma} \!\Big\rangle \,,
\end{eqnarray}
 and the density correlation is $C_\qbf =\langle{\rho_{-\qbf}}{\rho_{\qbf}} \rangle$.

With the variational wavefunction Eq.~\eqref{eq:NGSEDwavefunction}, we can express all (equal-time) observables using analytical expressions formed by the variational parameters and the correlations of fermionic operated evaluated in the electronic wavefunction $|\psi_e\rangle$\,\cite{wang2019zero}. The most important observable in this manuscript is $d$-wave pairing correlation, which can be explicitly written as
\begin{widetext}
\begin{eqnarray}
P^{(d)}_\kbf\!  &=&  \frac1N\sum_{\kprbf} \sum_{\kbf_1,\kbf_2}\xi(\kprbf+\kbf,\kbf_1,\kbf_2) \left[\zeta_{x}(\kprbf)\cos\left(\frac{\kpr_x}{2} - k_{1x}\right) \cos\left(\frac{\kpr_x}{2} - k_{2x}\right)+ \zeta_{y}(\kprbf)\cos\left(\frac{\kpr_y}{2} - k_{1y}\right) \cos\left(\frac{\kpr_y}{2} - k_{2y}\right)\right.\nonumber\\
	 &&\left.-\zeta_{xy}(\kprbf)\cos\left(\frac{\kpr_x}{2}-k_{2x}\right) \cos\left(\frac{\kpr_y}{2}-k_{1y}\right)-\zeta_{xy}(\kprbf)\cos\left(\frac{\kpr_y}{2}-k_{2y}\right) \cos\left(\frac{\kpr_x}{2}-k_{1x}\right) \right]\,.
\end{eqnarray}
\end{widetext}
This expression contains the pairing correlation in the electronic part of wavefunction
\begin{eqnarray}
	 \xi(\kprbf,\kbf_1,\kbf_2)=\langle\psi_e|  c_{\kbf_2\downarrow}^\dagger c_{\kprbf-\kbf_2\uparrow}^\dagger   c_{\kprbf-\kbf_1\uparrow} c_{\kbf_1\downarrow} |\psi_e\rangle
\end{eqnarray}
and analytical dressing factors
\begin{eqnarray}
	\zeta_{x}(\kprbf) &=& \sum_{\rbf} e^{-i\kprbf\cdot\rbf}e^{-\sum_\qbf\left[(1+\cos q_x)(1-e^{i\qbf\cdot\rbf})\right]  \frac{2|\lambda_\qbf|^2}{N}e_2^T \Gamma_\qbf e_2}\nonumber\\
	  \zeta_{y}(\kprbf) &=& \sum_{\rbf} e^{-i\kprbf\cdot\rbf}e^{-\sum_\qbf\left[(1+\cos q_y)(1-e^{i\qbf\cdot\rbf})\right]  \frac{2|\lambda_\qbf|^2}{N}e_2^T \Gamma_\qbf e_2}\nonumber\\
	  \zeta_{xy}(\kprbf) &=& \sum_{\bar{\rbf}}e^{-i\kprbf\cdot\bar{\rbf}}\exp\bigg\{-\sum_\qbf\Big[(2+\cos q_x+\cos q_y)\nonumber\\
	  &&-4e^{i\qbf\cdot\bar{\rbf}}\cos\frac{q_x}{2}\cos\frac{q_y}{2}\Big]  \frac{|\lambda_\qbf|^2}{N}e_2^T \Gamma_\qbf e_2 \bigg\}\,.
\end{eqnarray}
Here, the $\bar{\rbf}$ in the last summation denotes the half-unit-cell-shifted coordinates $\bar{\rbf} = \rbf + \hat{x}/2 - \hat{y}/2$.

Thus, we have now extended the NGSED framework into nonequilibrium dynamics. Each step of the time evolution is achieved by the evaluation of above coupled differential equations, as well as large-scale Krylov-subspace method (for the electronic wavefunction). To benchmark the accuracy, we compare the simulation results with the ED in a 2D Hubbard-Holstein model with $g=t_h$ and $U=8t_h$. The ED simulation is exact, except for the finite phonon occupation, which is truncated at $M=10$ in our calculation. To allow the ED simulation with acceptable computational complexity and accuracy, we choose a small eight-site cluster and a relatively high phonon frequency $\omega_0=5t_h$. For a medium pump strength ($A_0=0.4$) with frequency $\Omega=5t_h$, the relative errors for two dominant correlations, i.e., the spin structure factor $S(\pi,\pi)$ and the $d$-wave pairing susceptibility $P^{(d)}$ are both below 1\% even in the center of the pump pulse (see Fig.~\ref{fig:EDbenchmark}). At the same time, the relative error for the $s$-wave pairing susceptibility $P^{(s)}$ reaches approximately $5\%$. This relatively larger deviation mainly originates from the fact that the variational wavefunction tends to primarily capture the dominant correlations with limited parameters, leaving a slightly larger deviation for other correlations. In addition, the truncation error for the phonon Hilbert space in ED simulations is also more prominent out of equilibrium, which also contributes to the numerical error shown in Fig.~\ref{fig:EDbenchmark}.

\section{Krylov-Subspace Method for Time-Evolution}\label{sec:krylov}
With instantaneous variational parameters, the Hamiltonian can be reduced to an effective electronic one 
\begin{eqnarray}
\Ham_{\rm eff}(t) = \langle \phi_{\rm ph}(t)| U_{\rm NGS}(t)^\dagger \Ham(t)U_{\rm NGS}(t)| \phi_{\rm ph}(t)\rangle.
\end{eqnarray}
Then the single-step electronic state evolution becomes
\begin{eqnarray}
| \psi_e(t+\delta t)\rangle  \approx e^{-i\mathcal{H}(t)\delta t}| \psi_e(t)\rangle.
\end{eqnarray}
Here, the large Hilbert-space dimension ($> 10^8$) requires stable and efficient evaluation of the wavefunction, which in this study is based on the Krylov-subspace method. The Krylov subspace for an instantaneous wavefunction $|\psi_e(t)\rangle$ and Hamiltonian $\Ham_{\rm eff}(t)$ is defined as
\begin{eqnarray*}
    \mathcal{K}_n(t) =\textrm{span}\{|\psi_e(t)\rangle,\Ham_{\rm eff}(t) |\psi_e(t)\rangle,\cdots,\Ham_{\rm eff}(t)^{n-1} |\psi_e(t)\rangle\}. 
\end{eqnarray*}
A widely used Krylov-subspace generator is the Lanczos algorithm, where the wavefunction can be approximated by\,\cite{manmana2007strongly, balzer2012krylov, hochbruck1997krylov,innerberger2020electron}
\begin{eqnarray}\label{eq:eqKrylovTime2}
| \psi(t+\delta t)\rangle &\approx& \exp\left[-iU_n(t) T_n(t) U_n(t)^\dagger\delta t\right] | \psi(t)\rangle.
\end{eqnarray}
Here, $U_n(t)$ is the basis matrix of the Krylov subspace $\mathcal{K}_n(t)$, and the $T_n(t)$ is the tridiagonal matrix generated by the Lanczos algorithm.
Since the projected matrix is with dimension $n$, much smaller than the Hilbert-space dimension, the evaluation of Eq.~\eqref{eq:eqKrylovTime2} is much cheaper. The error of evaluating the vector propagation is well controlled by $\epsilon_n\leq 12 e^{-{(\rho\,\delta t)^2}/{16n}}\left({e\rho \, \delta t}/{4n}\right)^n$ given the Krylov dimension $n\geq \rho\,\delta t/2$ and spectral radius $\rho=|E_{\max}-E_{\min}|$\,\cite{hochbruck1997krylov}. Depending on the simulated pump strengths, we adopt $n=50-100$ in this paper.

\begin{figure*}[!ht]
\begin{center}
\includegraphics[width=18cm]{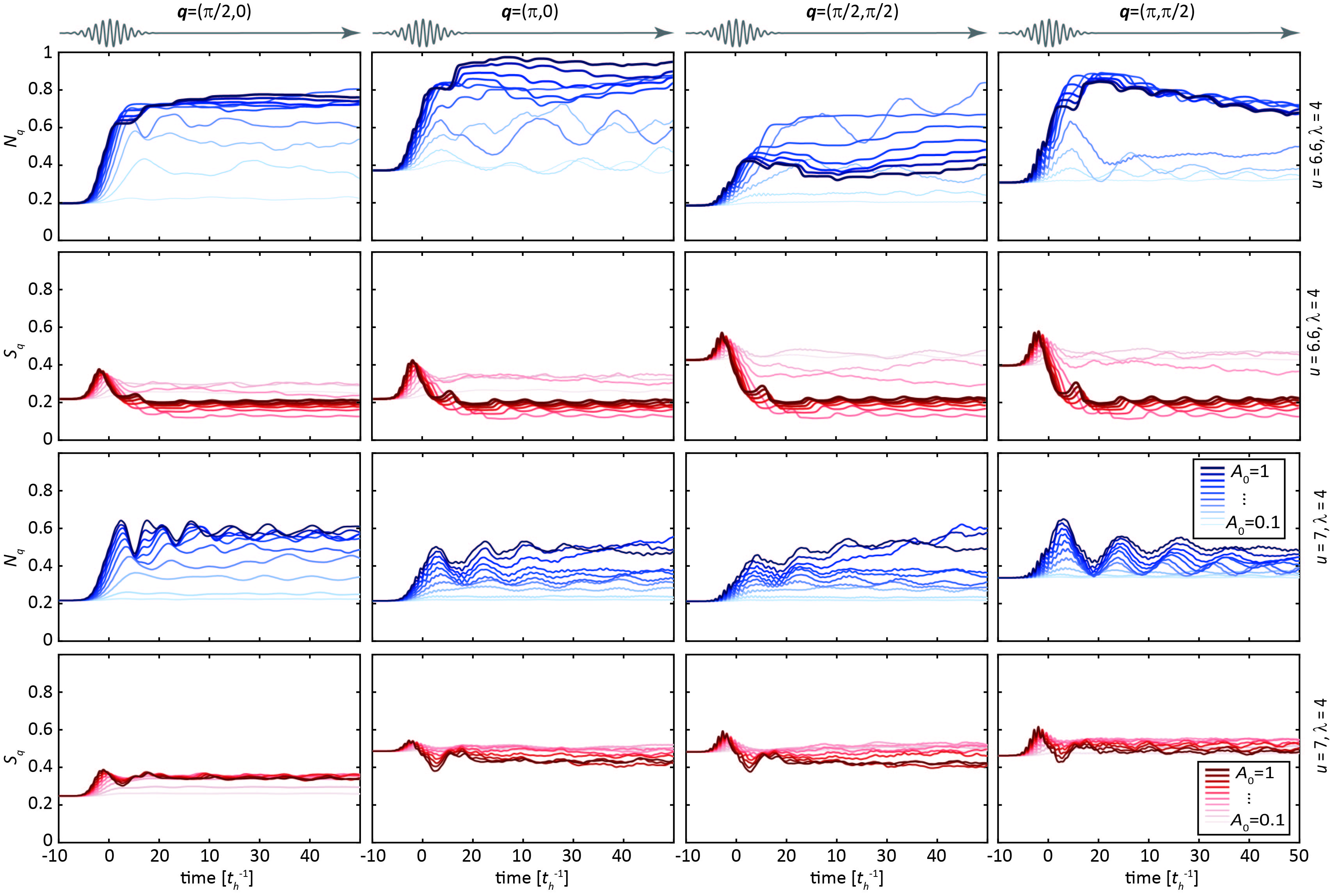}
\caption{\label{fig:allMomenta} Dynamics of $N(\qbf,t)$ and $S(\qbf,t)$ for other momenta [left to right: $(\pi/2,0)$, $(\pi,0)$, $(\pi/2,\pi/2)$, and $(\pi,\pi/2)$] not shown in the main text. The upper two rows are obtained for parameter Set 1, while the lower two rows are for parameter Set 2.
}
\end{center}
\end{figure*}

\section{Spin and Charge Dynamics for Other Momenta}\label{app:othermomentum}
In Fig.~2 in the main text, we show the time evolution for the charge and spin structure factors for only a single momentum $\qbf=(\pi,\pi)$, which plays the dominant role in $d$-wave superconductivity. In this section, we also present the calculations for other momenta.

The upper two rows in Fig.~\ref{fig:allMomenta} show the dynamics for $u=6.6$ and $\lambda=4$ (Set 1). In contrast to the suppression of $N(\pi,\pi)$, the charge structure factors are always enhanced for other momenta, due to the light-melting of the dominant order. The dynamics for spin structure factors are more complicated: They are enhanced during the pump, but for large pump strengths ($A_0 \ge 0.5$) the structure factors decrease rapidly after the pump pulse. As discussed in the main text, this means that a strong pump further suppresses the spin fluctuations and is, thereby, unfavorable for $d$-wave superconductivity.

In contrast to the dynamics with parameter Set 1, the evolution with parameter Set 2 (lower rows in Fig.~\ref{fig:allMomenta}) suggests unchanged spin correlations after the pump. Together with the dominant $(\pi,\pi)$ momentum in the main text, these spin fluctuations give rise to the $d$-wave superconductivity after melting the charge-ordered state.

\begin{figure}[!t]
\begin{center}
\includegraphics[width=8.5cm]{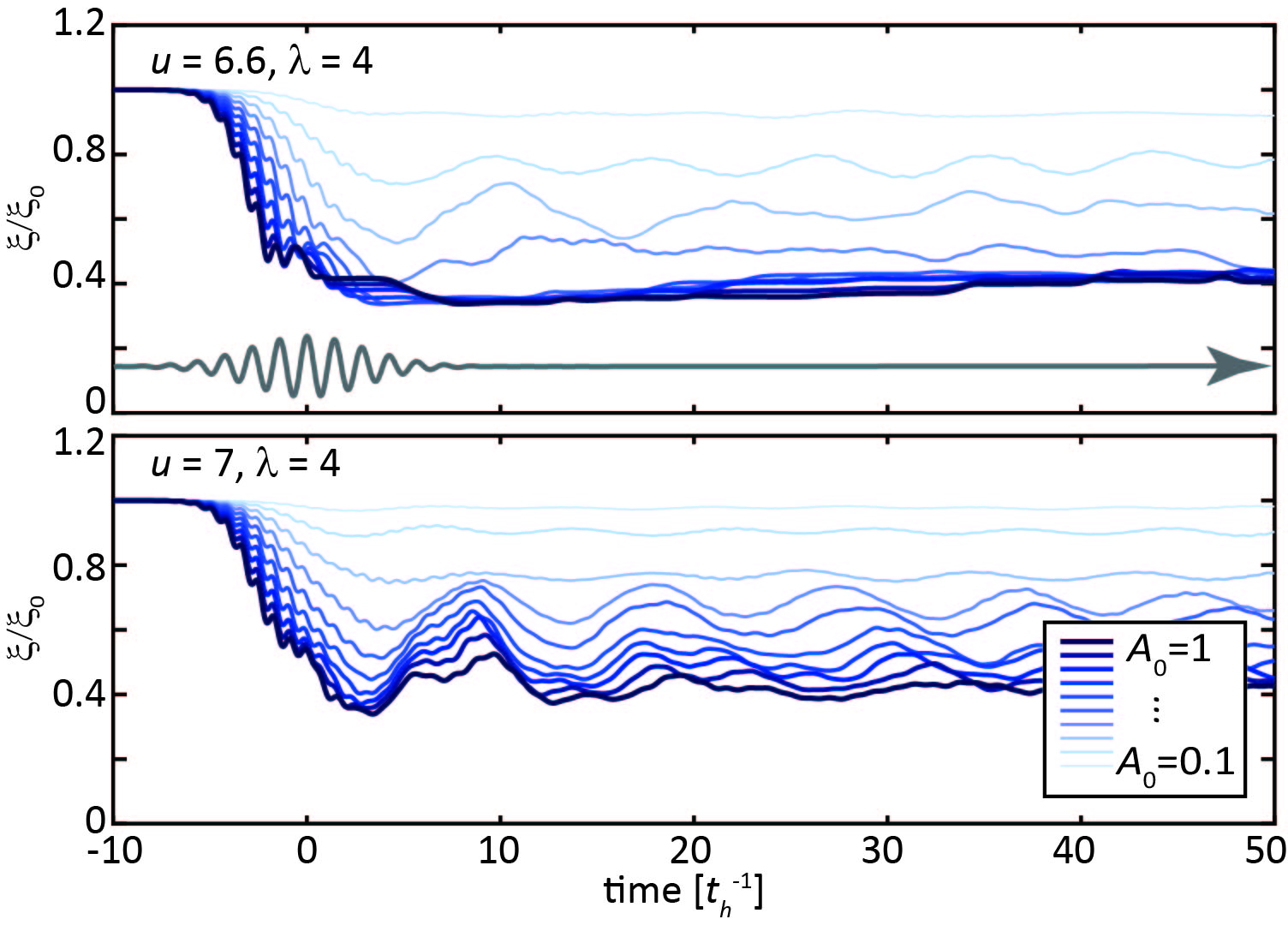}
\caption{\label{fig:chargeCoherence} Evolution of the correlation length $\xi$ of the charge structure factor renormalized by its equilibrium value $\xi_0$, obtained for (upper) $u=6.6$ and (lower) $u=7$, respectively.
}
\end{center}
\end{figure}

Using the momentum distribution of the spin and charge structure factors, we can also obtain the evolution of the correlation length. As shown in Fig.~\ref{fig:chargeCoherence}, the charge correlation length decreases rapidly after the pump pulse enters. Together with the reduction of $N(\pi,\pi)$ shown in Fig.~\ref{fig:pumpDynm}, it reflects the light melting of the charge instability, otherwise dominant at equilibrium. This trend is distinct from the light-driven dynamics of the $d$-wave pairing correlation, whose intensity is enhanced but the correlation length is reduced.

\section{Frequency Distribution of Dynamics}

\begin{figure}[!th]
\begin{center}
\includegraphics[width=8cm]{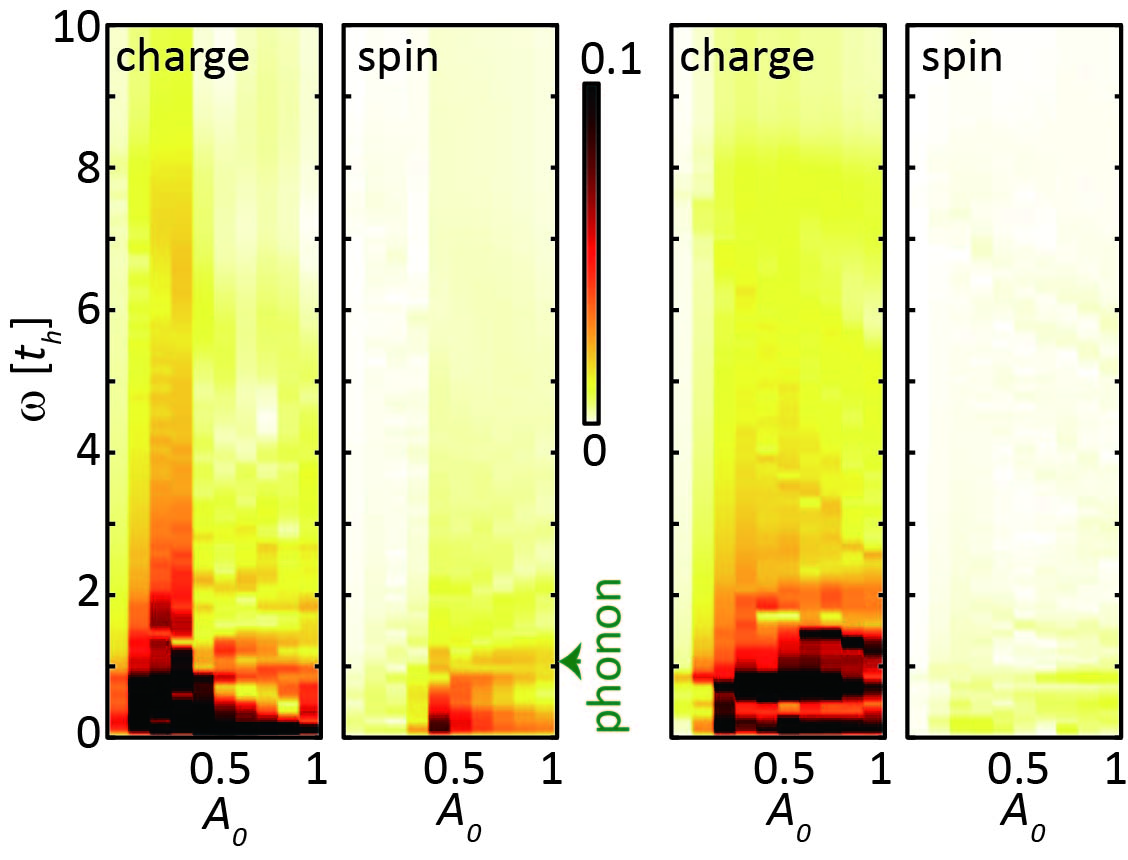}
\caption{\label{fig:fourierSpec} Fourier spectra for the pump-induced evolution of $N(\pi,\pi)$ and $S(\pi,\pi)$ for (left) $u=6.6$, (right) $u=7$, and different pump strengths shown in the main text.
}
\end{center}
\end{figure}
Besides the increase and decrease of the correlations, the postpump dynamics reflect some underlying physical quantities. Figure \ref{fig:fourierSpec} presents the Fourier spectra of the charge and spin dynamics shown in Fig.~2 in the main text. For the model parameter Set 1, the dynamics have low-energy periodicity approximately $2\pi/\omega_0$, reflecting the role of phonons in forming and melting a charge-ordered state. For the model parameter Set 2 close to the phase boundary, the phonon energy becomes highly renormalized\,\cite{wang2016using}. Therefore, the low-energy spectrum becomes complicated and dependent on the pump strengths. This strength dependence is also reflected by the dynamics of other momenta in Fig.~\ref{fig:allMomenta}.

\bibliography{paper}

\begin{thebibliography}{120}%
\makeatletter
\providecommand \@ifxundefined [1]{%
 \@ifx{#1\undefined}
}%
\providecommand \@ifnum [1]{%
 \ifnum #1\expandafter \@firstoftwo
 \else \expandafter \@secondoftwo
 \fi
}%
\providecommand \@ifx [1]{%
 \ifx #1\expandafter \@firstoftwo
 \else \expandafter \@secondoftwo
 \fi
}%
\providecommand \natexlab [1]{#1}%
\providecommand \enquote  [1]{``#1''}%
\providecommand \bibnamefont  [1]{#1}%
\providecommand \bibfnamefont [1]{#1}%
\providecommand \citenamefont [1]{#1}%
\providecommand \href@noop [0]{\@secondoftwo}%
\providecommand \href [0]{\begingroup \@sanitize@url \@href}%
\providecommand \@href[1]{\@@startlink{#1}\@@href}%
\providecommand \@@href[1]{\endgroup#1\@@endlink}%
\providecommand \@sanitize@url [0]{\catcode `\\12\catcode `\$12\catcode
  `\&12\catcode `\#12\catcode `\^12\catcode `\_12\catcode `\%12\relax}%
\providecommand \@@startlink[1]{}%
\providecommand \@@endlink[0]{}%
\providecommand \url  [0]{\begingroup\@sanitize@url \@url }%
\providecommand \@url [1]{\endgroup\@href {#1}{\urlprefix }}%
\providecommand \urlprefix  [0]{URL }%
\providecommand \Eprint [0]{\href }%
\providecommand \doibase [0]{http://dx.doi.org/}%
\providecommand \selectlanguage [0]{\@gobble}%
\providecommand \bibinfo  [0]{\@secondoftwo}%
\providecommand \bibfield  [0]{\@secondoftwo}%
\providecommand \translation [1]{[#1]}%
\providecommand \BibitemOpen [0]{}%
\providecommand \bibitemStop [0]{}%
\providecommand \bibitemNoStop [0]{.\EOS\space}%
\providecommand \EOS [0]{\spacefactor3000\relax}%
\providecommand \BibitemShut  [1]{\csname bibitem#1\endcsname}%
\let\auto@bib@innerbib\@empty
\bibitem [{\citenamefont {Basov}\ \emph {et~al.}(2017)\citenamefont {Basov},
  \citenamefont {Averitt},\ and\ \citenamefont {Hsieh}}]{basov2017towards}%
  \BibitemOpen
  \bibfield  {author} {\bibinfo {author} {\bibfnamefont {D.}~\bibnamefont
  {Basov}}, \bibinfo {author} {\bibfnamefont {R.}~\bibnamefont {Averitt}}, \
  and\ \bibinfo {author} {\bibfnamefont {D.}~\bibnamefont {Hsieh}},\
  }\href@noop {} {\bibfield  {journal} {\bibinfo  {journal} {Nat. Mater.}\
  }\textbf {\bibinfo {volume} {16}},\ \bibinfo {pages} {1077} (\bibinfo {year}
  {2017})}\BibitemShut {NoStop}%
\bibitem [{\citenamefont {Wang}\ \emph
  {et~al.}(2018{\natexlab{a}})\citenamefont {Wang}, \citenamefont {Claassen},
  \citenamefont {Pemmaraju}, \citenamefont {Jia}, \citenamefont {Moritz},\ and\
  \citenamefont {Devereaux}}]{wang2018theoretical}%
  \BibitemOpen
  \bibfield  {author} {\bibinfo {author} {\bibfnamefont {Y.}~\bibnamefont
  {Wang}}, \bibinfo {author} {\bibfnamefont {M.}~\bibnamefont {Claassen}},
  \bibinfo {author} {\bibfnamefont {C.~D.}\ \bibnamefont {Pemmaraju}}, \bibinfo
  {author} {\bibfnamefont {C.}~\bibnamefont {Jia}}, \bibinfo {author}
  {\bibfnamefont {B.}~\bibnamefont {Moritz}}, \ and\ \bibinfo {author}
  {\bibfnamefont {T.~P.}\ \bibnamefont {Devereaux}},\ }\href@noop {} {\bibfield
   {journal} {\bibinfo  {journal} {Nat. Rev. Mater.}\ }\textbf {\bibinfo
  {volume} {3}},\ \bibinfo {pages} {312} (\bibinfo {year}
  {2018}{\natexlab{a}})}\BibitemShut {NoStop}%
\bibitem [{\citenamefont {Mankowsky}\ \emph {et~al.}(2014)\citenamefont
  {Mankowsky}, \citenamefont {Subedi}, \citenamefont {F{\"o}rst}, \citenamefont
  {Mariager}, \citenamefont {Chollet}, \citenamefont {Lemke}, \citenamefont
  {Robinson}, \citenamefont {Glownia}, \citenamefont {Minitti}, \citenamefont
  {Frano} \emph {et~al.}}]{mankowsky2014nonlinear}%
  \BibitemOpen
  \bibfield  {author} {\bibinfo {author} {\bibfnamefont {R.}~\bibnamefont
  {Mankowsky}}, \bibinfo {author} {\bibfnamefont {A.}~\bibnamefont {Subedi}},
  \bibinfo {author} {\bibfnamefont {M.}~\bibnamefont {F{\"o}rst}}, \bibinfo
  {author} {\bibfnamefont {S.}~\bibnamefont {Mariager}}, \bibinfo {author}
  {\bibfnamefont {M.}~\bibnamefont {Chollet}}, \bibinfo {author} {\bibfnamefont
  {H.}~\bibnamefont {Lemke}}, \bibinfo {author} {\bibfnamefont
  {J.}~\bibnamefont {Robinson}}, \bibinfo {author} {\bibfnamefont
  {J.}~\bibnamefont {Glownia}}, \bibinfo {author} {\bibfnamefont
  {M.}~\bibnamefont {Minitti}}, \bibinfo {author} {\bibfnamefont
  {A.}~\bibnamefont {Frano}},  \emph {et~al.},\ }\href@noop {} {\bibfield
  {journal} {\bibinfo  {journal} {Nature}\ }\textbf {\bibinfo {volume} {516}},\
  \bibinfo {pages} {71} (\bibinfo {year} {2014})}\BibitemShut {NoStop}%
\bibitem [{\citenamefont {Hu}\ \emph {et~al.}(2014)\citenamefont {Hu},
  \citenamefont {Kaiser}, \citenamefont {Nicoletti}, \citenamefont {Hunt},
  \citenamefont {Gierz}, \citenamefont {Hoffmann}, \citenamefont {Le~Tacon},
  \citenamefont {Loew}, \citenamefont {Keimer},\ and\ \citenamefont
  {Cavalleri}}]{hu2014optically}%
  \BibitemOpen
  \bibfield  {author} {\bibinfo {author} {\bibfnamefont {W.}~\bibnamefont
  {Hu}}, \bibinfo {author} {\bibfnamefont {S.}~\bibnamefont {Kaiser}}, \bibinfo
  {author} {\bibfnamefont {D.}~\bibnamefont {Nicoletti}}, \bibinfo {author}
  {\bibfnamefont {C.~R.}\ \bibnamefont {Hunt}}, \bibinfo {author}
  {\bibfnamefont {I.}~\bibnamefont {Gierz}}, \bibinfo {author} {\bibfnamefont
  {M.~C.}\ \bibnamefont {Hoffmann}}, \bibinfo {author} {\bibfnamefont
  {M.}~\bibnamefont {Le~Tacon}}, \bibinfo {author} {\bibfnamefont
  {T.}~\bibnamefont {Loew}}, \bibinfo {author} {\bibfnamefont {B.}~\bibnamefont
  {Keimer}}, \ and\ \bibinfo {author} {\bibfnamefont {A.}~\bibnamefont
  {Cavalleri}},\ }\href@noop {} {\bibfield  {journal} {\bibinfo  {journal}
  {Nat. Mater.}\ } (\bibinfo {year} {2014})}\BibitemShut {NoStop}%
\bibitem [{\citenamefont {Kaiser}\ \emph {et~al.}(2014)\citenamefont {Kaiser},
  \citenamefont {Hunt}, \citenamefont {Nicoletti}, \citenamefont {Hu},
  \citenamefont {Gierz}, \citenamefont {Liu}, \citenamefont {Le~Tacon},
  \citenamefont {Loew}, \citenamefont {Haug}, \citenamefont {Keimer} \emph
  {et~al.}}]{kaiser2014optically}%
  \BibitemOpen
  \bibfield  {author} {\bibinfo {author} {\bibfnamefont {S.}~\bibnamefont
  {Kaiser}}, \bibinfo {author} {\bibfnamefont {C.}~\bibnamefont {Hunt}},
  \bibinfo {author} {\bibfnamefont {D.}~\bibnamefont {Nicoletti}}, \bibinfo
  {author} {\bibfnamefont {W.}~\bibnamefont {Hu}}, \bibinfo {author}
  {\bibfnamefont {I.}~\bibnamefont {Gierz}}, \bibinfo {author} {\bibfnamefont
  {H.}~\bibnamefont {Liu}}, \bibinfo {author} {\bibfnamefont {M.}~\bibnamefont
  {Le~Tacon}}, \bibinfo {author} {\bibfnamefont {T.}~\bibnamefont {Loew}},
  \bibinfo {author} {\bibfnamefont {D.}~\bibnamefont {Haug}}, \bibinfo {author}
  {\bibfnamefont {B.}~\bibnamefont {Keimer}},  \emph {et~al.},\ }\href@noop {}
  {\bibfield  {journal} {\bibinfo  {journal} {Phys. Rev. B}\ }\textbf {\bibinfo
  {volume} {89}},\ \bibinfo {pages} {184516} (\bibinfo {year}
  {2014})}\BibitemShut {NoStop}%
\bibitem [{\citenamefont {Mitrano}\ \emph {et~al.}(2016)\citenamefont
  {Mitrano}, \citenamefont {Cantaluppi}, \citenamefont {Nicoletti},
  \citenamefont {Kaiser}, \citenamefont {Perucchi}, \citenamefont {Lupi},
  \citenamefont {Di~Pietro}, \citenamefont {Pontiroli}, \citenamefont
  {Ricc{\`o}}, \citenamefont {Clark}, \citenamefont {Jaksch},\ and\
  \citenamefont {Cavalleri}}]{mitrano2016possible}%
  \BibitemOpen
  \bibfield  {author} {\bibinfo {author} {\bibfnamefont {M.}~\bibnamefont
  {Mitrano}}, \bibinfo {author} {\bibfnamefont {A.}~\bibnamefont {Cantaluppi}},
  \bibinfo {author} {\bibfnamefont {D.}~\bibnamefont {Nicoletti}}, \bibinfo
  {author} {\bibfnamefont {S.}~\bibnamefont {Kaiser}}, \bibinfo {author}
  {\bibfnamefont {A.}~\bibnamefont {Perucchi}}, \bibinfo {author}
  {\bibfnamefont {S.}~\bibnamefont {Lupi}}, \bibinfo {author} {\bibfnamefont
  {P.}~\bibnamefont {Di~Pietro}}, \bibinfo {author} {\bibfnamefont
  {D.}~\bibnamefont {Pontiroli}}, \bibinfo {author} {\bibfnamefont
  {M.}~\bibnamefont {Ricc{\`o}}}, \bibinfo {author} {\bibfnamefont {S.~R.}\
  \bibnamefont {Clark}}, \bibinfo {author} {\bibfnamefont {D.}~\bibnamefont
  {Jaksch}}, \ and\ \bibinfo {author} {\bibfnamefont {A.}~\bibnamefont
  {Cavalleri}},\ }\href@noop {} {\bibfield  {journal} {\bibinfo  {journal}
  {Nature}\ }\textbf {\bibinfo {volume} {530}},\ \bibinfo {pages} {461}
  (\bibinfo {year} {2016})}\BibitemShut {NoStop}%
\bibitem [{\citenamefont {Buzzi}\ \emph {et~al.}(2020)\citenamefont {Buzzi},
  \citenamefont {Nicoletti}, \citenamefont {Fechner}, \citenamefont
  {Tancogne-Dejean}, \citenamefont {Sentef}, \citenamefont {Georges},
  \citenamefont {Biesner}, \citenamefont {Uykur}, \citenamefont {Dressel},
  \citenamefont {Henderson} \emph {et~al.}}]{buzzi2020photomolecular}%
  \BibitemOpen
  \bibfield  {author} {\bibinfo {author} {\bibfnamefont {M.}~\bibnamefont
  {Buzzi}}, \bibinfo {author} {\bibfnamefont {D.}~\bibnamefont {Nicoletti}},
  \bibinfo {author} {\bibfnamefont {M.}~\bibnamefont {Fechner}}, \bibinfo
  {author} {\bibfnamefont {N.}~\bibnamefont {Tancogne-Dejean}}, \bibinfo
  {author} {\bibfnamefont {M.}~\bibnamefont {Sentef}}, \bibinfo {author}
  {\bibfnamefont {A.}~\bibnamefont {Georges}}, \bibinfo {author} {\bibfnamefont
  {T.}~\bibnamefont {Biesner}}, \bibinfo {author} {\bibfnamefont
  {E.}~\bibnamefont {Uykur}}, \bibinfo {author} {\bibfnamefont
  {M.}~\bibnamefont {Dressel}}, \bibinfo {author} {\bibfnamefont
  {A.}~\bibnamefont {Henderson}},  \emph {et~al.},\ }\href@noop {} {\bibfield
  {journal} {\bibinfo  {journal} {Phys. Rev. X}\ }\textbf {\bibinfo {volume}
  {10}},\ \bibinfo {pages} {031028} (\bibinfo {year} {2020})}\BibitemShut
  {NoStop}%
\bibitem [{\citenamefont {Fausti}\ \emph {et~al.}(2011)\citenamefont {Fausti},
  \citenamefont {Tobey}, \citenamefont {Dean}, \citenamefont {Kaiser},
  \citenamefont {Dienst}, \citenamefont {Hoffmann}, \citenamefont {Pyon},
  \citenamefont {Takayama}, \citenamefont {Takagi},\ and\ \citenamefont
  {Cavalleri}}]{fausti2011light}%
  \BibitemOpen
  \bibfield  {author} {\bibinfo {author} {\bibfnamefont {D.}~\bibnamefont
  {Fausti}}, \bibinfo {author} {\bibfnamefont {R.}~\bibnamefont {Tobey}},
  \bibinfo {author} {\bibfnamefont {N.}~\bibnamefont {Dean}}, \bibinfo {author}
  {\bibfnamefont {S.}~\bibnamefont {Kaiser}}, \bibinfo {author} {\bibfnamefont
  {A.}~\bibnamefont {Dienst}}, \bibinfo {author} {\bibfnamefont
  {M.}~\bibnamefont {Hoffmann}}, \bibinfo {author} {\bibfnamefont
  {S.}~\bibnamefont {Pyon}}, \bibinfo {author} {\bibfnamefont {T.}~\bibnamefont
  {Takayama}}, \bibinfo {author} {\bibfnamefont {H.}~\bibnamefont {Takagi}}, \
  and\ \bibinfo {author} {\bibfnamefont {A.}~\bibnamefont {Cavalleri}},\
  }\href@noop {} {\bibfield  {journal} {\bibinfo  {journal} {Science}\ }\textbf
  {\bibinfo {volume} {331}},\ \bibinfo {pages} {189} (\bibinfo {year}
  {2011})}\BibitemShut {NoStop}%
\bibitem [{\citenamefont {Nicoletti}\ \emph {et~al.}(2014)\citenamefont
  {Nicoletti}, \citenamefont {Casandruc}, \citenamefont {Laplace},
  \citenamefont {Khanna}, \citenamefont {Hunt}, \citenamefont {Kaiser},
  \citenamefont {Dhesi}, \citenamefont {Gu}, \citenamefont {Hill},\ and\
  \citenamefont {Cavalleri}}]{nicoletti2014optically}%
  \BibitemOpen
  \bibfield  {author} {\bibinfo {author} {\bibfnamefont {D.}~\bibnamefont
  {Nicoletti}}, \bibinfo {author} {\bibfnamefont {E.}~\bibnamefont
  {Casandruc}}, \bibinfo {author} {\bibfnamefont {Y.}~\bibnamefont {Laplace}},
  \bibinfo {author} {\bibfnamefont {V.}~\bibnamefont {Khanna}}, \bibinfo
  {author} {\bibfnamefont {C.~R.}\ \bibnamefont {Hunt}}, \bibinfo {author}
  {\bibfnamefont {S.}~\bibnamefont {Kaiser}}, \bibinfo {author} {\bibfnamefont
  {S.}~\bibnamefont {Dhesi}}, \bibinfo {author} {\bibfnamefont
  {G.}~\bibnamefont {Gu}}, \bibinfo {author} {\bibfnamefont {J.}~\bibnamefont
  {Hill}}, \ and\ \bibinfo {author} {\bibfnamefont {A.}~\bibnamefont
  {Cavalleri}},\ }\href@noop {} {\bibfield  {journal} {\bibinfo  {journal}
  {Phys. Rev. B}\ }\textbf {\bibinfo {volume} {90}},\ \bibinfo {pages} {100503}
  (\bibinfo {year} {2014})}\BibitemShut {NoStop}%
\bibitem [{\citenamefont {F{\"o}rst}\ \emph {et~al.}(2014)\citenamefont
  {F{\"o}rst}, \citenamefont {Tobey}, \citenamefont {Bromberger}, \citenamefont
  {Wilkins}, \citenamefont {Khanna}, \citenamefont {Caviglia}, \citenamefont
  {Chuang}, \citenamefont {Lee}, \citenamefont {Schlotter}, \citenamefont
  {Turner} \emph {et~al.}}]{forst2014melting}%
  \BibitemOpen
  \bibfield  {author} {\bibinfo {author} {\bibfnamefont {M.}~\bibnamefont
  {F{\"o}rst}}, \bibinfo {author} {\bibfnamefont {R.}~\bibnamefont {Tobey}},
  \bibinfo {author} {\bibfnamefont {H.}~\bibnamefont {Bromberger}}, \bibinfo
  {author} {\bibfnamefont {S.}~\bibnamefont {Wilkins}}, \bibinfo {author}
  {\bibfnamefont {V.}~\bibnamefont {Khanna}}, \bibinfo {author} {\bibfnamefont
  {A.}~\bibnamefont {Caviglia}}, \bibinfo {author} {\bibfnamefont {Y.-D.}\
  \bibnamefont {Chuang}}, \bibinfo {author} {\bibfnamefont {W.}~\bibnamefont
  {Lee}}, \bibinfo {author} {\bibfnamefont {W.}~\bibnamefont {Schlotter}},
  \bibinfo {author} {\bibfnamefont {J.}~\bibnamefont {Turner}},  \emph
  {et~al.},\ }\href@noop {} {\bibfield  {journal} {\bibinfo  {journal} {Phys.
  Rev. Lett.}\ }\textbf {\bibinfo {volume} {112}},\ \bibinfo {pages} {157002}
  (\bibinfo {year} {2014})}\BibitemShut {NoStop}%
\bibitem [{\citenamefont {Casandruc}\ \emph {et~al.}(2015)\citenamefont
  {Casandruc}, \citenamefont {Nicoletti}, \citenamefont {Rajasekaran},
  \citenamefont {Laplace}, \citenamefont {Khanna}, \citenamefont {Gu},
  \citenamefont {Hill},\ and\ \citenamefont
  {Cavalleri}}]{casandruc2015wavelength}%
  \BibitemOpen
  \bibfield  {author} {\bibinfo {author} {\bibfnamefont {E.}~\bibnamefont
  {Casandruc}}, \bibinfo {author} {\bibfnamefont {D.}~\bibnamefont
  {Nicoletti}}, \bibinfo {author} {\bibfnamefont {S.}~\bibnamefont
  {Rajasekaran}}, \bibinfo {author} {\bibfnamefont {Y.}~\bibnamefont
  {Laplace}}, \bibinfo {author} {\bibfnamefont {V.}~\bibnamefont {Khanna}},
  \bibinfo {author} {\bibfnamefont {G.}~\bibnamefont {Gu}}, \bibinfo {author}
  {\bibfnamefont {J.}~\bibnamefont {Hill}}, \ and\ \bibinfo {author}
  {\bibfnamefont {A.}~\bibnamefont {Cavalleri}},\ }\href@noop {} {\bibfield
  {journal} {\bibinfo  {journal} {Phys. Rev. B}\ }\textbf {\bibinfo {volume}
  {91}},\ \bibinfo {pages} {174502} (\bibinfo {year} {2015})}\BibitemShut
  {NoStop}%
\bibitem [{\citenamefont {Khanna}\ \emph {et~al.}(2016)\citenamefont {Khanna},
  \citenamefont {Mankowsky}, \citenamefont {Petrich}, \citenamefont
  {Bromberger}, \citenamefont {Cavill}, \citenamefont {M{\"o}hr-Vorobeva},
  \citenamefont {Nicoletti}, \citenamefont {Laplace}, \citenamefont {Gu},
  \citenamefont {Hill} \emph {et~al.}}]{khanna2016restoring}%
  \BibitemOpen
  \bibfield  {author} {\bibinfo {author} {\bibfnamefont {V.}~\bibnamefont
  {Khanna}}, \bibinfo {author} {\bibfnamefont {R.}~\bibnamefont {Mankowsky}},
  \bibinfo {author} {\bibfnamefont {M.}~\bibnamefont {Petrich}}, \bibinfo
  {author} {\bibfnamefont {H.}~\bibnamefont {Bromberger}}, \bibinfo {author}
  {\bibfnamefont {S.}~\bibnamefont {Cavill}}, \bibinfo {author} {\bibfnamefont
  {E.}~\bibnamefont {M{\"o}hr-Vorobeva}}, \bibinfo {author} {\bibfnamefont
  {D.}~\bibnamefont {Nicoletti}}, \bibinfo {author} {\bibfnamefont
  {Y.}~\bibnamefont {Laplace}}, \bibinfo {author} {\bibfnamefont
  {G.}~\bibnamefont {Gu}}, \bibinfo {author} {\bibfnamefont {J.}~\bibnamefont
  {Hill}},  \emph {et~al.},\ }\href@noop {} {\bibfield  {journal} {\bibinfo
  {journal} {Phys. Rev. B}\ }\textbf {\bibinfo {volume} {93}},\ \bibinfo
  {pages} {224522} (\bibinfo {year} {2016})}\BibitemShut {NoStop}%
\bibitem [{\citenamefont {Nicoletti}\ \emph {et~al.}(2018)\citenamefont
  {Nicoletti}, \citenamefont {Fu}, \citenamefont {Mehio}, \citenamefont
  {Moore}, \citenamefont {Disa}, \citenamefont {Gu},\ and\ \citenamefont
  {Cavalleri}}]{nicoletti2018magnetic}%
  \BibitemOpen
  \bibfield  {author} {\bibinfo {author} {\bibfnamefont {D.}~\bibnamefont
  {Nicoletti}}, \bibinfo {author} {\bibfnamefont {D.}~\bibnamefont {Fu}},
  \bibinfo {author} {\bibfnamefont {O.}~\bibnamefont {Mehio}}, \bibinfo
  {author} {\bibfnamefont {S.}~\bibnamefont {Moore}}, \bibinfo {author}
  {\bibfnamefont {A.}~\bibnamefont {Disa}}, \bibinfo {author} {\bibfnamefont
  {G.}~\bibnamefont {Gu}}, \ and\ \bibinfo {author} {\bibfnamefont
  {A.}~\bibnamefont {Cavalleri}},\ }\href@noop {} {\bibfield  {journal}
  {\bibinfo  {journal} {Phys. Rev. Lett.}\ }\textbf {\bibinfo {volume} {121}},\
  \bibinfo {pages} {267003} (\bibinfo {year} {2018})}\BibitemShut {NoStop}%
\bibitem [{\citenamefont {Cremin}\ \emph {et~al.}(2019)\citenamefont {Cremin},
  \citenamefont {Zhang}, \citenamefont {Homes}, \citenamefont {Gu},
  \citenamefont {Sun}, \citenamefont {Fogler}, \citenamefont {Millis},
  \citenamefont {Basov},\ and\ \citenamefont
  {Averitt}}]{cremin2019photoenhanced}%
  \BibitemOpen
  \bibfield  {author} {\bibinfo {author} {\bibfnamefont {K.~A.}\ \bibnamefont
  {Cremin}}, \bibinfo {author} {\bibfnamefont {J.}~\bibnamefont {Zhang}},
  \bibinfo {author} {\bibfnamefont {C.~C.}\ \bibnamefont {Homes}}, \bibinfo
  {author} {\bibfnamefont {G.~D.}\ \bibnamefont {Gu}}, \bibinfo {author}
  {\bibfnamefont {Z.}~\bibnamefont {Sun}}, \bibinfo {author} {\bibfnamefont
  {M.~M.}\ \bibnamefont {Fogler}}, \bibinfo {author} {\bibfnamefont {A.~J.}\
  \bibnamefont {Millis}}, \bibinfo {author} {\bibfnamefont {D.~N.}\
  \bibnamefont {Basov}}, \ and\ \bibinfo {author} {\bibfnamefont {R.~D.}\
  \bibnamefont {Averitt}},\ }\href@noop {} {\bibfield  {journal} {\bibinfo
  {journal} {Proc. Nat. Acad. Sci.}\ }\textbf {\bibinfo {volume} {116}},\
  \bibinfo {pages} {19875} (\bibinfo {year} {2019})}\BibitemShut {NoStop}%
\bibitem [{\citenamefont {Sentef}\ \emph {et~al.}(2016)\citenamefont {Sentef},
  \citenamefont {Kemper}, \citenamefont {Georges},\ and\ \citenamefont
  {Kollath}}]{sentef2016theory}%
  \BibitemOpen
  \bibfield  {author} {\bibinfo {author} {\bibfnamefont {M.~A.}\ \bibnamefont
  {Sentef}}, \bibinfo {author} {\bibfnamefont {A.}~\bibnamefont {Kemper}},
  \bibinfo {author} {\bibfnamefont {A.}~\bibnamefont {Georges}}, \ and\
  \bibinfo {author} {\bibfnamefont {C.}~\bibnamefont {Kollath}},\ }\href@noop
  {} {\bibfield  {journal} {\bibinfo  {journal} {Phys. Rev. B}\ }\textbf
  {\bibinfo {volume} {93}},\ \bibinfo {pages} {144506} (\bibinfo {year}
  {2016})}\BibitemShut {NoStop}%
\bibitem [{\citenamefont {Sentef}\ \emph {et~al.}(2017)\citenamefont {Sentef},
  \citenamefont {Tokuno}, \citenamefont {Georges},\ and\ \citenamefont
  {Kollath}}]{sentef2017theory}%
  \BibitemOpen
  \bibfield  {author} {\bibinfo {author} {\bibfnamefont {M.~A.}\ \bibnamefont
  {Sentef}}, \bibinfo {author} {\bibfnamefont {A.}~\bibnamefont {Tokuno}},
  \bibinfo {author} {\bibfnamefont {A.}~\bibnamefont {Georges}}, \ and\
  \bibinfo {author} {\bibfnamefont {C.}~\bibnamefont {Kollath}},\ }\href@noop
  {} {\bibfield  {journal} {\bibinfo  {journal} {Phys. Rev. Lett.}\ }\textbf
  {\bibinfo {volume} {118}},\ \bibinfo {pages} {087002} (\bibinfo {year}
  {2017})}\BibitemShut {NoStop}%
\bibitem [{\citenamefont {Coulthard}\ \emph {et~al.}(2017)\citenamefont
  {Coulthard}, \citenamefont {Clark}, \citenamefont {Al-Assam}, \citenamefont
  {Cavalleri},\ and\ \citenamefont {Jaksch}}]{coulthard2017enhancement}%
  \BibitemOpen
  \bibfield  {author} {\bibinfo {author} {\bibfnamefont {J.}~\bibnamefont
  {Coulthard}}, \bibinfo {author} {\bibfnamefont {S.~R.}\ \bibnamefont
  {Clark}}, \bibinfo {author} {\bibfnamefont {S.}~\bibnamefont {Al-Assam}},
  \bibinfo {author} {\bibfnamefont {A.}~\bibnamefont {Cavalleri}}, \ and\
  \bibinfo {author} {\bibfnamefont {D.}~\bibnamefont {Jaksch}},\ }\href@noop {}
  {\bibfield  {journal} {\bibinfo  {journal} {Phys. Rev. B}\ }\textbf {\bibinfo
  {volume} {96}},\ \bibinfo {pages} {085104} (\bibinfo {year}
  {2017})}\BibitemShut {NoStop}%
\bibitem [{\citenamefont {Babadi}\ \emph {et~al.}(2017)\citenamefont {Babadi},
  \citenamefont {Knap}, \citenamefont {Martin}, \citenamefont {Refael},\ and\
  \citenamefont {Demler}}]{babadi2017theory}%
  \BibitemOpen
  \bibfield  {author} {\bibinfo {author} {\bibfnamefont {M.}~\bibnamefont
  {Babadi}}, \bibinfo {author} {\bibfnamefont {M.}~\bibnamefont {Knap}},
  \bibinfo {author} {\bibfnamefont {I.}~\bibnamefont {Martin}}, \bibinfo
  {author} {\bibfnamefont {G.}~\bibnamefont {Refael}}, \ and\ \bibinfo {author}
  {\bibfnamefont {E.}~\bibnamefont {Demler}},\ }\href@noop {} {\bibfield
  {journal} {\bibinfo  {journal} {Phys. Rev. B}\ }\textbf {\bibinfo {volume}
  {96}},\ \bibinfo {pages} {014512} (\bibinfo {year} {2017})}\BibitemShut
  {NoStop}%
\bibitem [{\citenamefont {Murakami}\ \emph {et~al.}(2017)\citenamefont
  {Murakami}, \citenamefont {Tsuji}, \citenamefont {Eckstein},\ and\
  \citenamefont {Werner}}]{murakami2017nonequilibrium}%
  \BibitemOpen
  \bibfield  {author} {\bibinfo {author} {\bibfnamefont {Y.}~\bibnamefont
  {Murakami}}, \bibinfo {author} {\bibfnamefont {N.}~\bibnamefont {Tsuji}},
  \bibinfo {author} {\bibfnamefont {M.}~\bibnamefont {Eckstein}}, \ and\
  \bibinfo {author} {\bibfnamefont {P.}~\bibnamefont {Werner}},\ }\href@noop {}
  {\bibfield  {journal} {\bibinfo  {journal} {Phys. Rev. B}\ }\textbf {\bibinfo
  {volume} {96}},\ \bibinfo {pages} {045125} (\bibinfo {year}
  {2017})}\BibitemShut {NoStop}%
\bibitem [{\citenamefont {Dasari}\ and\ \citenamefont
  {Eckstein}(2018)}]{dasari2018transient}%
  \BibitemOpen
  \bibfield  {author} {\bibinfo {author} {\bibfnamefont {N.}~\bibnamefont
  {Dasari}}\ and\ \bibinfo {author} {\bibfnamefont {M.}~\bibnamefont
  {Eckstein}},\ }\href@noop {} {\bibfield  {journal} {\bibinfo  {journal}
  {Phys. Rev. B}\ }\textbf {\bibinfo {volume} {98}},\ \bibinfo {pages} {235149}
  (\bibinfo {year} {2018})}\BibitemShut {NoStop}%
\bibitem [{\citenamefont {Bittner}\ \emph {et~al.}(2019)\citenamefont
  {Bittner}, \citenamefont {Tohyama}, \citenamefont {Kaiser},\ and\
  \citenamefont {Manske}}]{bittner2019possible}%
  \BibitemOpen
  \bibfield  {author} {\bibinfo {author} {\bibfnamefont {N.}~\bibnamefont
  {Bittner}}, \bibinfo {author} {\bibfnamefont {T.}~\bibnamefont {Tohyama}},
  \bibinfo {author} {\bibfnamefont {S.}~\bibnamefont {Kaiser}}, \ and\ \bibinfo
  {author} {\bibfnamefont {D.}~\bibnamefont {Manske}},\ }\href@noop {}
  {\bibfield  {journal} {\bibinfo  {journal} {J. Phys. Soc. Jpn}\ }\textbf
  {\bibinfo {volume} {88}},\ \bibinfo {pages} {044704} (\bibinfo {year}
  {2019})}\BibitemShut {NoStop}%
\bibitem [{\citenamefont {Li}\ and\ \citenamefont
  {Eckstein}(2020)}]{li2020manipulating}%
  \BibitemOpen
  \bibfield  {author} {\bibinfo {author} {\bibfnamefont {J.}~\bibnamefont
  {Li}}\ and\ \bibinfo {author} {\bibfnamefont {M.}~\bibnamefont {Eckstein}},\
  }\href@noop {} {\bibfield  {journal} {\bibinfo  {journal} {Phys. Rev. Lett.}\
  }\textbf {\bibinfo {volume} {125}},\ \bibinfo {pages} {217402} (\bibinfo
  {year} {2020})}\BibitemShut {NoStop}%
\bibitem [{\citenamefont {Tindall}\ \emph
  {et~al.}(2020{\natexlab{a}})\citenamefont {Tindall}, \citenamefont
  {Schlawin}, \citenamefont {Buzzi}, \citenamefont {Nicoletti}, \citenamefont
  {Coulthard}, \citenamefont {Gao}, \citenamefont {Cavalleri}, \citenamefont
  {Sentef},\ and\ \citenamefont {Jaksch}}]{tindall2020dynamical}%
  \BibitemOpen
  \bibfield  {author} {\bibinfo {author} {\bibfnamefont {J.}~\bibnamefont
  {Tindall}}, \bibinfo {author} {\bibfnamefont {F.}~\bibnamefont {Schlawin}},
  \bibinfo {author} {\bibfnamefont {M.}~\bibnamefont {Buzzi}}, \bibinfo
  {author} {\bibfnamefont {D.}~\bibnamefont {Nicoletti}}, \bibinfo {author}
  {\bibfnamefont {J.}~\bibnamefont {Coulthard}}, \bibinfo {author}
  {\bibfnamefont {H.}~\bibnamefont {Gao}}, \bibinfo {author} {\bibfnamefont
  {A.}~\bibnamefont {Cavalleri}}, \bibinfo {author} {\bibfnamefont
  {M.}~\bibnamefont {Sentef}}, \ and\ \bibinfo {author} {\bibfnamefont
  {D.}~\bibnamefont {Jaksch}},\ }\href@noop {} {\bibfield  {journal} {\bibinfo
  {journal} {Phys. Rev. Lett.}\ }\textbf {\bibinfo {volume} {125}},\ \bibinfo
  {pages} {137001} (\bibinfo {year} {2020}{\natexlab{a}})}\BibitemShut
  {NoStop}%
\bibitem [{\citenamefont {Tindall}\ \emph
  {et~al.}(2020{\natexlab{b}})\citenamefont {Tindall}, \citenamefont
  {Schlawin}, \citenamefont {Sentef},\ and\ \citenamefont
  {Jaksch}}]{tindall2020analytical}%
  \BibitemOpen
  \bibfield  {author} {\bibinfo {author} {\bibfnamefont {J.}~\bibnamefont
  {Tindall}}, \bibinfo {author} {\bibfnamefont {F.}~\bibnamefont {Schlawin}},
  \bibinfo {author} {\bibfnamefont {M.~A.}\ \bibnamefont {Sentef}}, \ and\
  \bibinfo {author} {\bibfnamefont {D.}~\bibnamefont {Jaksch}},\ }\href@noop {}
  {\bibfield  {journal} {\bibinfo  {journal} {arXiv:2011.04417}\ } (\bibinfo
  {year} {2020}{\natexlab{b}})}\BibitemShut {NoStop}%
\bibitem [{\citenamefont {Patel}\ and\ \citenamefont
  {Eberlein}(2016)}]{patel2016light}%
  \BibitemOpen
  \bibfield  {author} {\bibinfo {author} {\bibfnamefont {A.~A.}\ \bibnamefont
  {Patel}}\ and\ \bibinfo {author} {\bibfnamefont {A.}~\bibnamefont
  {Eberlein}},\ }\href@noop {} {\bibfield  {journal} {\bibinfo  {journal}
  {Phys. Rev. B}\ }\textbf {\bibinfo {volume} {93}},\ \bibinfo {pages} {195139}
  (\bibinfo {year} {2016})}\BibitemShut {NoStop}%
\bibitem [{\citenamefont {Kennes}\ \emph {et~al.}(2019)\citenamefont {Kennes},
  \citenamefont {Claassen}, \citenamefont {Sentef},\ and\ \citenamefont
  {Karrasch}}]{kennes2019light}%
  \BibitemOpen
  \bibfield  {author} {\bibinfo {author} {\bibfnamefont {D.~M.}\ \bibnamefont
  {Kennes}}, \bibinfo {author} {\bibfnamefont {M.}~\bibnamefont {Claassen}},
  \bibinfo {author} {\bibfnamefont {M.~A.}\ \bibnamefont {Sentef}}, \ and\
  \bibinfo {author} {\bibfnamefont {C.}~\bibnamefont {Karrasch}},\ }\href@noop
  {} {\bibfield  {journal} {\bibinfo  {journal} {Phys. Rev. B}\ }\textbf
  {\bibinfo {volume} {100}},\ \bibinfo {pages} {075115} (\bibinfo {year}
  {2019})}\BibitemShut {NoStop}%
\bibitem [{\citenamefont {Michael}\ \emph {et~al.}(2020)\citenamefont
  {Michael}, \citenamefont {von Hoegen}, \citenamefont {Fechner}, \citenamefont
  {F{\"o}rst}, \citenamefont {Cavalleri},\ and\ \citenamefont
  {Demler}}]{michael2020parametric}%
  \BibitemOpen
  \bibfield  {author} {\bibinfo {author} {\bibfnamefont {M.~H.}\ \bibnamefont
  {Michael}}, \bibinfo {author} {\bibfnamefont {A.}~\bibnamefont {von Hoegen}},
  \bibinfo {author} {\bibfnamefont {M.}~\bibnamefont {Fechner}}, \bibinfo
  {author} {\bibfnamefont {M.}~\bibnamefont {F{\"o}rst}}, \bibinfo {author}
  {\bibfnamefont {A.}~\bibnamefont {Cavalleri}}, \ and\ \bibinfo {author}
  {\bibfnamefont {E.}~\bibnamefont {Demler}},\ }\href@noop {} {\bibfield
  {journal} {\bibinfo  {journal} {Phys. Rev. B}\ }\textbf {\bibinfo {volume}
  {102}},\ \bibinfo {pages} {174505} (\bibinfo {year} {2020})}\BibitemShut
  {NoStop}%
\bibitem [{\citenamefont {Wang}\ \emph
  {et~al.}(2018{\natexlab{b}})\citenamefont {Wang}, \citenamefont {Chen},
  \citenamefont {Moritz},\ and\ \citenamefont {Devereaux}}]{wang2018light}%
  \BibitemOpen
  \bibfield  {author} {\bibinfo {author} {\bibfnamefont {Y.}~\bibnamefont
  {Wang}}, \bibinfo {author} {\bibfnamefont {C.-C.}\ \bibnamefont {Chen}},
  \bibinfo {author} {\bibfnamefont {B.}~\bibnamefont {Moritz}}, \ and\ \bibinfo
  {author} {\bibfnamefont {T.}~\bibnamefont {Devereaux}},\ }\href@noop {}
  {\bibfield  {journal} {\bibinfo  {journal} {Phys. Rev. Lett.}\ }\textbf
  {\bibinfo {volume} {120}},\ \bibinfo {pages} {246402} (\bibinfo {year}
  {2018}{\natexlab{b}})}\BibitemShut {NoStop}%
\bibitem [{\citenamefont {Kondo}\ \emph {et~al.}(2015)\citenamefont {Kondo},
  \citenamefont {Malaeb}, \citenamefont {Ishida}, \citenamefont {Sasagawa},
  \citenamefont {Sakamoto}, \citenamefont {Takeuchi}, \citenamefont {Tohyama},\
  and\ \citenamefont {Shin}}]{kondo2015point}%
  \BibitemOpen
  \bibfield  {author} {\bibinfo {author} {\bibfnamefont {T.}~\bibnamefont
  {Kondo}}, \bibinfo {author} {\bibfnamefont {W.}~\bibnamefont {Malaeb}},
  \bibinfo {author} {\bibfnamefont {Y.}~\bibnamefont {Ishida}}, \bibinfo
  {author} {\bibfnamefont {T.}~\bibnamefont {Sasagawa}}, \bibinfo {author}
  {\bibfnamefont {H.}~\bibnamefont {Sakamoto}}, \bibinfo {author}
  {\bibfnamefont {T.}~\bibnamefont {Takeuchi}}, \bibinfo {author}
  {\bibfnamefont {T.}~\bibnamefont {Tohyama}}, \ and\ \bibinfo {author}
  {\bibfnamefont {S.}~\bibnamefont {Shin}},\ }\href@noop {} {\bibfield
  {journal} {\bibinfo  {journal} {Nat. Commun.}\ }\textbf {\bibinfo {volume}
  {6}},\ \bibinfo {pages} {7699} (\bibinfo {year} {2015})}\BibitemShut
  {NoStop}%
\bibitem [{\citenamefont {He}\ \emph {et~al.}(2021)\citenamefont {He},
  \citenamefont {Chen}, \citenamefont {Li}, \citenamefont {Zhao}, \citenamefont
  {Song}, \citenamefont {Yoshida}, \citenamefont {Eisaki}, \citenamefont {Wu},
  \citenamefont {Chen}, \citenamefont {Lu} \emph {et~al.}}]{he2020fluctuating}%
  \BibitemOpen
  \bibfield  {author} {\bibinfo {author} {\bibfnamefont {Y.}~\bibnamefont
  {He}}, \bibinfo {author} {\bibfnamefont {S.-D.}\ \bibnamefont {Chen}},
  \bibinfo {author} {\bibfnamefont {Z.-X.}\ \bibnamefont {Li}}, \bibinfo
  {author} {\bibfnamefont {D.}~\bibnamefont {Zhao}}, \bibinfo {author}
  {\bibfnamefont {D.}~\bibnamefont {Song}}, \bibinfo {author} {\bibfnamefont
  {Y.}~\bibnamefont {Yoshida}}, \bibinfo {author} {\bibfnamefont
  {H.}~\bibnamefont {Eisaki}}, \bibinfo {author} {\bibfnamefont
  {T.}~\bibnamefont {Wu}}, \bibinfo {author} {\bibfnamefont {X.-H.}\
  \bibnamefont {Chen}}, \bibinfo {author} {\bibfnamefont {D.-H.}\ \bibnamefont
  {Lu}},  \emph {et~al.},\ }\href@noop {} {\bibfield  {journal} {\bibinfo
  {journal} {Phys. Rev. X (in press)}\ } (\bibinfo {year} {2021})}\BibitemShut
  {NoStop}%
\bibitem [{\citenamefont {Faeth}\ \emph {et~al.}(2021)\citenamefont {Faeth},
  \citenamefont {Yang}, \citenamefont {Kawasaki}, \citenamefont {Nelson},
  \citenamefont {Mishra}, \citenamefont {Chen}, \citenamefont {Schlom},\ and\
  \citenamefont {Shen}}]{faeth2020incoherent}%
  \BibitemOpen
  \bibfield  {author} {\bibinfo {author} {\bibfnamefont {B.~D.}\ \bibnamefont
  {Faeth}}, \bibinfo {author} {\bibfnamefont {S.}~\bibnamefont {Yang}},
  \bibinfo {author} {\bibfnamefont {J.~K.}\ \bibnamefont {Kawasaki}}, \bibinfo
  {author} {\bibfnamefont {J.~N.}\ \bibnamefont {Nelson}}, \bibinfo {author}
  {\bibfnamefont {P.}~\bibnamefont {Mishra}}, \bibinfo {author} {\bibfnamefont
  {L.}~\bibnamefont {Chen}}, \bibinfo {author} {\bibfnamefont {D.~G.}\
  \bibnamefont {Schlom}}, \ and\ \bibinfo {author} {\bibfnamefont {K.~M.}\
  \bibnamefont {Shen}},\ }\href@noop {} {\bibfield  {journal} {\bibinfo
  {journal} {Phys. Rev. X}\ }\textbf {\bibinfo {volume} {11}},\ \bibinfo
  {pages} {021054} (\bibinfo {year} {2021})}\BibitemShut {NoStop}%
\bibitem [{\citenamefont {Xu}\ \emph {et~al.}(2021)\citenamefont {Xu},
  \citenamefont {Rong}, \citenamefont {Wang}, \citenamefont {Wu}, \citenamefont
  {Hu}, \citenamefont {Cai}, \citenamefont {Gao}, \citenamefont {Yan},
  \citenamefont {Li}, \citenamefont {Yin} \emph
  {et~al.}}]{xu2020spectroscopic}%
  \BibitemOpen
  \bibfield  {author} {\bibinfo {author} {\bibfnamefont {Y.}~\bibnamefont
  {Xu}}, \bibinfo {author} {\bibfnamefont {H.}~\bibnamefont {Rong}}, \bibinfo
  {author} {\bibfnamefont {Q.}~\bibnamefont {Wang}}, \bibinfo {author}
  {\bibfnamefont {D.}~\bibnamefont {Wu}}, \bibinfo {author} {\bibfnamefont
  {Y.}~\bibnamefont {Hu}}, \bibinfo {author} {\bibfnamefont {Y.}~\bibnamefont
  {Cai}}, \bibinfo {author} {\bibfnamefont {Q.}~\bibnamefont {Gao}}, \bibinfo
  {author} {\bibfnamefont {H.}~\bibnamefont {Yan}}, \bibinfo {author}
  {\bibfnamefont {C.}~\bibnamefont {Li}}, \bibinfo {author} {\bibfnamefont
  {C.}~\bibnamefont {Yin}},  \emph {et~al.},\ }\href@noop {} {\bibfield
  {journal} {\bibinfo  {journal} {Nat. Commun.}\ }\textbf {\bibinfo {volume}
  {12}},\ \bibinfo {pages} {2840} (\bibinfo {year} {2021})}\BibitemShut
  {NoStop}%
\bibitem [{\citenamefont {Boschini}\ \emph {et~al.}(2018)\citenamefont
  {Boschini}, \citenamefont {da~Silva~Neto}, \citenamefont {Razzoli},
  \citenamefont {Zonno}, \citenamefont {Peli}, \citenamefont {Day},
  \citenamefont {Michiardi}, \citenamefont {Schneider}, \citenamefont
  {Zwartsenberg}, \citenamefont {Nigge} \emph {et~al.}}]{boschini2018collapse}%
  \BibitemOpen
  \bibfield  {author} {\bibinfo {author} {\bibfnamefont {F.}~\bibnamefont
  {Boschini}}, \bibinfo {author} {\bibfnamefont {E.}~\bibnamefont
  {da~Silva~Neto}}, \bibinfo {author} {\bibfnamefont {E.}~\bibnamefont
  {Razzoli}}, \bibinfo {author} {\bibfnamefont {M.}~\bibnamefont {Zonno}},
  \bibinfo {author} {\bibfnamefont {S.}~\bibnamefont {Peli}}, \bibinfo {author}
  {\bibfnamefont {R.}~\bibnamefont {Day}}, \bibinfo {author} {\bibfnamefont
  {M.}~\bibnamefont {Michiardi}}, \bibinfo {author} {\bibfnamefont
  {M.}~\bibnamefont {Schneider}}, \bibinfo {author} {\bibfnamefont
  {B.}~\bibnamefont {Zwartsenberg}}, \bibinfo {author} {\bibfnamefont
  {P.}~\bibnamefont {Nigge}},  \emph {et~al.},\ }\href@noop {} {\bibfield
  {journal} {\bibinfo  {journal} {Nat. Mater.}\ }\textbf {\bibinfo {volume}
  {17}},\ \bibinfo {pages} {416} (\bibinfo {year} {2018})}\BibitemShut
  {NoStop}%
\bibitem [{\citenamefont {Yang}\ \emph {et~al.}(2018)\citenamefont {Yang},
  \citenamefont {Vaswani}, \citenamefont {Sundahl}, \citenamefont {Mootz},
  \citenamefont {Gagel}, \citenamefont {Luo}, \citenamefont {Kang},
  \citenamefont {Orth}, \citenamefont {Perakis}, \citenamefont {Eom} \emph
  {et~al.}}]{yang2018terahertz}%
  \BibitemOpen
  \bibfield  {author} {\bibinfo {author} {\bibfnamefont {X.}~\bibnamefont
  {Yang}}, \bibinfo {author} {\bibfnamefont {C.}~\bibnamefont {Vaswani}},
  \bibinfo {author} {\bibfnamefont {C.}~\bibnamefont {Sundahl}}, \bibinfo
  {author} {\bibfnamefont {M.}~\bibnamefont {Mootz}}, \bibinfo {author}
  {\bibfnamefont {P.}~\bibnamefont {Gagel}}, \bibinfo {author} {\bibfnamefont
  {L.}~\bibnamefont {Luo}}, \bibinfo {author} {\bibfnamefont {J.}~\bibnamefont
  {Kang}}, \bibinfo {author} {\bibfnamefont {P.}~\bibnamefont {Orth}}, \bibinfo
  {author} {\bibfnamefont {I.}~\bibnamefont {Perakis}}, \bibinfo {author}
  {\bibfnamefont {C.}~\bibnamefont {Eom}},  \emph {et~al.},\ }\href@noop {}
  {\bibfield  {journal} {\bibinfo  {journal} {Nat. Mater.}\ }\textbf {\bibinfo
  {volume} {17}},\ \bibinfo {pages} {586} (\bibinfo {year} {2018})}\BibitemShut
  {NoStop}%
\bibitem [{\citenamefont {Mootz}\ \emph {et~al.}(2020)\citenamefont {Mootz},
  \citenamefont {Wang},\ and\ \citenamefont {Perakis}}]{mootz2020lightwave}%
  \BibitemOpen
  \bibfield  {author} {\bibinfo {author} {\bibfnamefont {M.}~\bibnamefont
  {Mootz}}, \bibinfo {author} {\bibfnamefont {J.}~\bibnamefont {Wang}}, \ and\
  \bibinfo {author} {\bibfnamefont {I.~E.}\ \bibnamefont {Perakis}},\
  }\href@noop {} {\bibfield  {journal} {\bibinfo  {journal} {Phys. Rev. B}\
  }\textbf {\bibinfo {volume} {102}},\ \bibinfo {pages} {054517} (\bibinfo
  {year} {2020})}\BibitemShut {NoStop}%
\bibitem [{\citenamefont {Chiriac{\`o}}\ \emph {et~al.}(2018)\citenamefont
  {Chiriac{\`o}}, \citenamefont {Millis},\ and\ \citenamefont
  {Aleiner}}]{chiriaco2018transient}%
  \BibitemOpen
  \bibfield  {author} {\bibinfo {author} {\bibfnamefont {G.}~\bibnamefont
  {Chiriac{\`o}}}, \bibinfo {author} {\bibfnamefont {A.~J.}\ \bibnamefont
  {Millis}}, \ and\ \bibinfo {author} {\bibfnamefont {I.~L.}\ \bibnamefont
  {Aleiner}},\ }\href@noop {} {\bibfield  {journal} {\bibinfo  {journal} {Phys.
  Rev. B}\ }\textbf {\bibinfo {volume} {98}},\ \bibinfo {pages} {220510}
  (\bibinfo {year} {2018})}\BibitemShut {NoStop}%
\bibitem [{\citenamefont {Lemonik}\ and\ \citenamefont
  {Mitra}(2019)}]{lemonik2019transport}%
  \BibitemOpen
  \bibfield  {author} {\bibinfo {author} {\bibfnamefont {Y.}~\bibnamefont
  {Lemonik}}\ and\ \bibinfo {author} {\bibfnamefont {A.}~\bibnamefont
  {Mitra}},\ }\href@noop {} {\bibfield  {journal} {\bibinfo  {journal} {Phys.
  Rev. B}\ }\textbf {\bibinfo {volume} {100}},\ \bibinfo {pages} {094503}
  (\bibinfo {year} {2019})}\BibitemShut {NoStop}%
\bibitem [{\citenamefont {Sun}\ and\ \citenamefont
  {Millis}(2020)}]{sun2020transient}%
  \BibitemOpen
  \bibfield  {author} {\bibinfo {author} {\bibfnamefont {Z.}~\bibnamefont
  {Sun}}\ and\ \bibinfo {author} {\bibfnamefont {A.~J.}\ \bibnamefont
  {Millis}},\ }\href@noop {} {\bibfield  {journal} {\bibinfo  {journal} {Phys.
  Rev. X}\ }\textbf {\bibinfo {volume} {10}},\ \bibinfo {pages} {021028}
  (\bibinfo {year} {2020})}\BibitemShut {NoStop}%
\bibitem [{\citenamefont {Shi}\ \emph {et~al.}(2018)\citenamefont {Shi},
  \citenamefont {Demler},\ and\ \citenamefont {Cirac}}]{shi2018variational}%
  \BibitemOpen
  \bibfield  {author} {\bibinfo {author} {\bibfnamefont {T.}~\bibnamefont
  {Shi}}, \bibinfo {author} {\bibfnamefont {E.}~\bibnamefont {Demler}}, \ and\
  \bibinfo {author} {\bibfnamefont {J.~I.}\ \bibnamefont {Cirac}},\ }\href@noop
  {} {\bibfield  {journal} {\bibinfo  {journal} {Ann. Phys.}\ }\textbf
  {\bibinfo {volume} {390}},\ \bibinfo {pages} {245} (\bibinfo {year}
  {2018})}\BibitemShut {NoStop}%
\bibitem [{\citenamefont {Shi}\ \emph {et~al.}(2020)\citenamefont {Shi},
  \citenamefont {Demler},\ and\ \citenamefont {Cirac}}]{shi2020variational}%
  \BibitemOpen
  \bibfield  {author} {\bibinfo {author} {\bibfnamefont {T.}~\bibnamefont
  {Shi}}, \bibinfo {author} {\bibfnamefont {E.}~\bibnamefont {Demler}}, \ and\
  \bibinfo {author} {\bibfnamefont {J.~I.}\ \bibnamefont {Cirac}},\ }\href@noop
  {} {\bibfield  {journal} {\bibinfo  {journal} {Phys. Rev. Lett.}\ }\textbf
  {\bibinfo {volume} {125}},\ \bibinfo {pages} {180602} (\bibinfo {year}
  {2020})}\BibitemShut {NoStop}%
\bibitem [{\citenamefont {Shi}\ \emph {et~al.}(2019)\citenamefont {Shi},
  \citenamefont {Cirac},\ and\ \citenamefont {Demler}}]{shi2019ultrafast}%
  \BibitemOpen
  \bibfield  {author} {\bibinfo {author} {\bibfnamefont {T.}~\bibnamefont
  {Shi}}, \bibinfo {author} {\bibfnamefont {J.~I.}\ \bibnamefont {Cirac}}, \
  and\ \bibinfo {author} {\bibfnamefont {E.}~\bibnamefont {Demler}},\
  }\href@noop {} {\bibfield  {journal} {\bibinfo  {journal} {Phys. Rev.
  Research}\ }\textbf {\bibinfo {volume} {2}},\ \bibinfo {pages} {033379}
  (\bibinfo {year} {2019})}\BibitemShut {NoStop}%
\bibitem [{\citenamefont {Kn\"{o}rzer}\ \emph {et~al.}(2021)\citenamefont
  {Kn\"{o}rzer}, \citenamefont {Shi}, \citenamefont {Demler},\ and\
  \citenamefont {Cirac}}]{shi2021spin}%
  \BibitemOpen
  \bibfield  {author} {\bibinfo {author} {\bibfnamefont {J.}~\bibnamefont
  {Kn\"{o}rzer}}, \bibinfo {author} {\bibfnamefont {T.}~\bibnamefont {Shi}},
  \bibinfo {author} {\bibfnamefont {E.}~\bibnamefont {Demler}}, \ and\ \bibinfo
  {author} {\bibfnamefont {J.}~\bibnamefont {Cirac}},\ }\href@noop {}
  {\bibfield  {journal} {\bibinfo  {journal} {arXiv:2108.13730}\ } (\bibinfo
  {year} {2021})}\BibitemShut {NoStop}%
\bibitem [{\citenamefont {Wang}\ \emph {et~al.}(2019)\citenamefont {Wang},
  \citenamefont {Esterlis}, \citenamefont {Shi}, \citenamefont {Cirac},\ and\
  \citenamefont {Demler}}]{wang2019zero}%
  \BibitemOpen
  \bibfield  {author} {\bibinfo {author} {\bibfnamefont {Y.}~\bibnamefont
  {Wang}}, \bibinfo {author} {\bibfnamefont {I.}~\bibnamefont {Esterlis}},
  \bibinfo {author} {\bibfnamefont {T.}~\bibnamefont {Shi}}, \bibinfo {author}
  {\bibfnamefont {J.~I.}\ \bibnamefont {Cirac}}, \ and\ \bibinfo {author}
  {\bibfnamefont {E.}~\bibnamefont {Demler}},\ }\href@noop {} {\bibfield
  {journal} {\bibinfo  {journal} {Phys. Rev. Research}\ }\textbf {\bibinfo
  {volume} {2}},\ \bibinfo {pages} {043258} (\bibinfo {year}
  {2019})}\BibitemShut {NoStop}%
\bibitem [{\citenamefont {Keimer}\ \emph {et~al.}(2015)\citenamefont {Keimer},
  \citenamefont {Kivelson}, \citenamefont {Norman}, \citenamefont {Uchida},\
  and\ \citenamefont {Zaanen}}]{keimer2015quantum}%
  \BibitemOpen
  \bibfield  {author} {\bibinfo {author} {\bibfnamefont {B.}~\bibnamefont
  {Keimer}}, \bibinfo {author} {\bibfnamefont {S.}~\bibnamefont {Kivelson}},
  \bibinfo {author} {\bibfnamefont {M.}~\bibnamefont {Norman}}, \bibinfo
  {author} {\bibfnamefont {S.}~\bibnamefont {Uchida}}, \ and\ \bibinfo {author}
  {\bibfnamefont {J.}~\bibnamefont {Zaanen}},\ }\href@noop {} {\bibfield
  {journal} {\bibinfo  {journal} {Nature}\ }\textbf {\bibinfo {volume} {518}},\
  \bibinfo {pages} {179} (\bibinfo {year} {2015})}\BibitemShut {NoStop}%
\bibitem [{\citenamefont {Davis}\ and\ \citenamefont
  {Lee}(2013)}]{davis2013concepts}%
  \BibitemOpen
  \bibfield  {author} {\bibinfo {author} {\bibfnamefont {J.~S.}\ \bibnamefont
  {Davis}}\ and\ \bibinfo {author} {\bibfnamefont {D.-H.}\ \bibnamefont
  {Lee}},\ }\href@noop {} {\bibfield  {journal} {\bibinfo  {journal} {Proc.
  Natl. Acad. Sci. U.S.A.}\ }\textbf {\bibinfo {volume} {110}},\ \bibinfo
  {pages} {17623} (\bibinfo {year} {2013})}\BibitemShut {NoStop}%
\bibitem [{\citenamefont {Scalapino}\ \emph {et~al.}(1986)\citenamefont
  {Scalapino}, \citenamefont {Loh~Jr},\ and\ \citenamefont
  {Hirsch}}]{scalapino1986d}%
  \BibitemOpen
  \bibfield  {author} {\bibinfo {author} {\bibfnamefont {D.}~\bibnamefont
  {Scalapino}}, \bibinfo {author} {\bibfnamefont {E.}~\bibnamefont {Loh~Jr}}, \
  and\ \bibinfo {author} {\bibfnamefont {J.}~\bibnamefont {Hirsch}},\
  }\href@noop {} {\bibfield  {journal} {\bibinfo  {journal} {Phys. Rev. B}\
  }\textbf {\bibinfo {volume} {34}},\ \bibinfo {pages} {8190} (\bibinfo {year}
  {1986})}\BibitemShut {NoStop}%
\bibitem [{\citenamefont {Gros}\ \emph {et~al.}(1987)\citenamefont {Gros},
  \citenamefont {Joynt},\ and\ \citenamefont {Rice}}]{gros1987superconducting}%
  \BibitemOpen
  \bibfield  {author} {\bibinfo {author} {\bibfnamefont {C.}~\bibnamefont
  {Gros}}, \bibinfo {author} {\bibfnamefont {R.}~\bibnamefont {Joynt}}, \ and\
  \bibinfo {author} {\bibfnamefont {T.}~\bibnamefont {Rice}},\ }\href@noop {}
  {\bibfield  {journal} {\bibinfo  {journal} {Z. Phys. B Condens. Matter}\
  }\textbf {\bibinfo {volume} {68}},\ \bibinfo {pages} {425} (\bibinfo {year}
  {1987})}\BibitemShut {NoStop}%
\bibitem [{\citenamefont {Kotliar}\ and\ \citenamefont
  {Liu}(1988)}]{kotliar1988superexchange}%
  \BibitemOpen
  \bibfield  {author} {\bibinfo {author} {\bibfnamefont {G.}~\bibnamefont
  {Kotliar}}\ and\ \bibinfo {author} {\bibfnamefont {J.}~\bibnamefont {Liu}},\
  }\href@noop {} {\bibfield  {journal} {\bibinfo  {journal} {Phys. Rev. B}\
  }\textbf {\bibinfo {volume} {38}},\ \bibinfo {pages} {5142} (\bibinfo {year}
  {1988})}\BibitemShut {NoStop}%
\bibitem [{\citenamefont {Tsuei}\ and\ \citenamefont
  {Kirtley}(2000)}]{tsuei2000pairing}%
  \BibitemOpen
  \bibfield  {author} {\bibinfo {author} {\bibfnamefont {C.}~\bibnamefont
  {Tsuei}}\ and\ \bibinfo {author} {\bibfnamefont {J.}~\bibnamefont
  {Kirtley}},\ }\href@noop {} {\bibfield  {journal} {\bibinfo  {journal} {Rev.
  Mod. Phys.}\ }\textbf {\bibinfo {volume} {72}},\ \bibinfo {pages} {969}
  (\bibinfo {year} {2000})}\BibitemShut {NoStop}%
\bibitem [{\citenamefont {Zhang}\ and\ \citenamefont
  {Rice}(1988)}]{zhang1988effective}%
  \BibitemOpen
  \bibfield  {author} {\bibinfo {author} {\bibfnamefont {F.}~\bibnamefont
  {Zhang}}\ and\ \bibinfo {author} {\bibfnamefont {T.}~\bibnamefont {Rice}},\
  }\href@noop {} {\bibfield  {journal} {\bibinfo  {journal} {Phys. Rev. B}\
  }\textbf {\bibinfo {volume} {37}},\ \bibinfo {pages} {3759} (\bibinfo {year}
  {1988})}\BibitemShut {NoStop}%
\bibitem [{\citenamefont {Singh}\ and\ \citenamefont
  {Goswami}(2002)}]{singh2002spin}%
  \BibitemOpen
  \bibfield  {author} {\bibinfo {author} {\bibfnamefont {A.}~\bibnamefont
  {Singh}}\ and\ \bibinfo {author} {\bibfnamefont {P.}~\bibnamefont
  {Goswami}},\ }\href@noop {} {\bibfield  {journal} {\bibinfo  {journal} {Phys.
  Rev. B}\ }\textbf {\bibinfo {volume} {66}},\ \bibinfo {pages} {092402}
  (\bibinfo {year} {2002})}\BibitemShut {NoStop}%
\bibitem [{\citenamefont {Zheng}\ \emph {et~al.}(2017)\citenamefont {Zheng},
  \citenamefont {Chung}, \citenamefont {Corboz}, \citenamefont {Ehlers},
  \citenamefont {Qin}, \citenamefont {Noack}, \citenamefont {Shi},
  \citenamefont {White}, \citenamefont {Zhang},\ and\ \citenamefont
  {Chan}}]{zheng2017stripe}%
  \BibitemOpen
  \bibfield  {author} {\bibinfo {author} {\bibfnamefont {B.-X.}\ \bibnamefont
  {Zheng}}, \bibinfo {author} {\bibfnamefont {C.-M.}\ \bibnamefont {Chung}},
  \bibinfo {author} {\bibfnamefont {P.}~\bibnamefont {Corboz}}, \bibinfo
  {author} {\bibfnamefont {G.}~\bibnamefont {Ehlers}}, \bibinfo {author}
  {\bibfnamefont {M.-P.}\ \bibnamefont {Qin}}, \bibinfo {author} {\bibfnamefont
  {R.~M.}\ \bibnamefont {Noack}}, \bibinfo {author} {\bibfnamefont
  {H.}~\bibnamefont {Shi}}, \bibinfo {author} {\bibfnamefont {S.~R.}\
  \bibnamefont {White}}, \bibinfo {author} {\bibfnamefont {S.}~\bibnamefont
  {Zhang}}, \ and\ \bibinfo {author} {\bibfnamefont {G.~K.-L.}\ \bibnamefont
  {Chan}},\ }\href@noop {} {\bibfield  {journal} {\bibinfo  {journal}
  {Science}\ }\textbf {\bibinfo {volume} {358}},\ \bibinfo {pages} {1155}
  (\bibinfo {year} {2017})}\BibitemShut {NoStop}%
\bibitem [{\citenamefont {Huang}\ \emph {et~al.}(2017)\citenamefont {Huang},
  \citenamefont {Mendl}, \citenamefont {Liu}, \citenamefont {Johnston},
  \citenamefont {Jiang}, \citenamefont {Moritz},\ and\ \citenamefont
  {Devereaux}}]{huang2017numerical}%
  \BibitemOpen
  \bibfield  {author} {\bibinfo {author} {\bibfnamefont {E.~W.}\ \bibnamefont
  {Huang}}, \bibinfo {author} {\bibfnamefont {C.~B.}\ \bibnamefont {Mendl}},
  \bibinfo {author} {\bibfnamefont {S.}~\bibnamefont {Liu}}, \bibinfo {author}
  {\bibfnamefont {S.}~\bibnamefont {Johnston}}, \bibinfo {author}
  {\bibfnamefont {H.-C.}\ \bibnamefont {Jiang}}, \bibinfo {author}
  {\bibfnamefont {B.}~\bibnamefont {Moritz}}, \ and\ \bibinfo {author}
  {\bibfnamefont {T.~P.}\ \bibnamefont {Devereaux}},\ }\href@noop {} {\bibfield
   {journal} {\bibinfo  {journal} {Science}\ }\textbf {\bibinfo {volume}
  {358}},\ \bibinfo {pages} {1161} (\bibinfo {year} {2017})}\BibitemShut
  {NoStop}%
\bibitem [{\citenamefont {Huang}\ \emph {et~al.}(2018)\citenamefont {Huang},
  \citenamefont {Mendl}, \citenamefont {Jiang}, \citenamefont {Moritz},\ and\
  \citenamefont {Devereaux}}]{huang2018stripe}%
  \BibitemOpen
  \bibfield  {author} {\bibinfo {author} {\bibfnamefont {E.~W.}\ \bibnamefont
  {Huang}}, \bibinfo {author} {\bibfnamefont {C.~B.}\ \bibnamefont {Mendl}},
  \bibinfo {author} {\bibfnamefont {H.-C.}\ \bibnamefont {Jiang}}, \bibinfo
  {author} {\bibfnamefont {B.}~\bibnamefont {Moritz}}, \ and\ \bibinfo {author}
  {\bibfnamefont {T.~P.}\ \bibnamefont {Devereaux}},\ }\href@noop {} {\bibfield
   {journal} {\bibinfo  {journal} {npj Quantum Mater.}\ }\textbf {\bibinfo
  {volume} {3}},\ \bibinfo {pages} {22} (\bibinfo {year} {2018})}\BibitemShut
  {NoStop}%
\bibitem [{\citenamefont {Ponsioen}\ \emph {et~al.}(2019)\citenamefont
  {Ponsioen}, \citenamefont {Chung},\ and\ \citenamefont
  {Corboz}}]{ponsioen2019period}%
  \BibitemOpen
  \bibfield  {author} {\bibinfo {author} {\bibfnamefont {B.}~\bibnamefont
  {Ponsioen}}, \bibinfo {author} {\bibfnamefont {S.~S.}\ \bibnamefont {Chung}},
  \ and\ \bibinfo {author} {\bibfnamefont {P.}~\bibnamefont {Corboz}},\
  }\href@noop {} {\bibfield  {journal} {\bibinfo  {journal} {Phys. Rev. B}\
  }\textbf {\bibinfo {volume} {100}},\ \bibinfo {pages} {195141} (\bibinfo
  {year} {2019})}\BibitemShut {NoStop}%
\bibitem [{\citenamefont {Kokalj}(2017)}]{kokalj2017bad}%
  \BibitemOpen
  \bibfield  {author} {\bibinfo {author} {\bibfnamefont {J.}~\bibnamefont
  {Kokalj}},\ }\href@noop {} {\bibfield  {journal} {\bibinfo  {journal} {Phys.
  Rev. B}\ }\textbf {\bibinfo {volume} {95}},\ \bibinfo {pages} {041110}
  (\bibinfo {year} {2017})}\BibitemShut {NoStop}%
\bibitem [{\citenamefont {Huang}\ \emph {et~al.}(2019)\citenamefont {Huang},
  \citenamefont {Sheppard}, \citenamefont {Moritz},\ and\ \citenamefont
  {Devereaux}}]{huang2019strange}%
  \BibitemOpen
  \bibfield  {author} {\bibinfo {author} {\bibfnamefont {E.~W.}\ \bibnamefont
  {Huang}}, \bibinfo {author} {\bibfnamefont {R.}~\bibnamefont {Sheppard}},
  \bibinfo {author} {\bibfnamefont {B.}~\bibnamefont {Moritz}}, \ and\ \bibinfo
  {author} {\bibfnamefont {T.~P.}\ \bibnamefont {Devereaux}},\ }\href@noop {}
  {\bibfield  {journal} {\bibinfo  {journal} {Science}\ }\textbf {\bibinfo
  {volume} {366}},\ \bibinfo {pages} {987} (\bibinfo {year}
  {2019})}\BibitemShut {NoStop}%
\bibitem [{\citenamefont {Cha}\ \emph {et~al.}(2020)\citenamefont {Cha},
  \citenamefont {Patel}, \citenamefont {Gull},\ and\ \citenamefont
  {Kim}}]{cha2020slope}%
  \BibitemOpen
  \bibfield  {author} {\bibinfo {author} {\bibfnamefont {P.}~\bibnamefont
  {Cha}}, \bibinfo {author} {\bibfnamefont {A.~A.}\ \bibnamefont {Patel}},
  \bibinfo {author} {\bibfnamefont {E.}~\bibnamefont {Gull}}, \ and\ \bibinfo
  {author} {\bibfnamefont {E.-A.}\ \bibnamefont {Kim}},\ }\href@noop {}
  {\bibfield  {journal} {\bibinfo  {journal} {Phys. Rev. Research}\ }\textbf
  {\bibinfo {volume} {2}},\ \bibinfo {pages} {033434} (\bibinfo {year}
  {2020})}\BibitemShut {NoStop}%
\bibitem [{\citenamefont {Zheng}\ and\ \citenamefont
  {Chan}(2016)}]{zheng2016ground}%
  \BibitemOpen
  \bibfield  {author} {\bibinfo {author} {\bibfnamefont {B.-X.}\ \bibnamefont
  {Zheng}}\ and\ \bibinfo {author} {\bibfnamefont {G.~K.-L.}\ \bibnamefont
  {Chan}},\ }\href@noop {} {\bibfield  {journal} {\bibinfo  {journal} {Phys.
  Rev. B}\ }\textbf {\bibinfo {volume} {93}},\ \bibinfo {pages} {035126}
  (\bibinfo {year} {2016})}\BibitemShut {NoStop}%
\bibitem [{\citenamefont {Ido}\ \emph {et~al.}(2018)\citenamefont {Ido},
  \citenamefont {Ohgoe},\ and\ \citenamefont {Imada}}]{ido2018competition}%
  \BibitemOpen
  \bibfield  {author} {\bibinfo {author} {\bibfnamefont {K.}~\bibnamefont
  {Ido}}, \bibinfo {author} {\bibfnamefont {T.}~\bibnamefont {Ohgoe}}, \ and\
  \bibinfo {author} {\bibfnamefont {M.}~\bibnamefont {Imada}},\ }\href@noop {}
  {\bibfield  {journal} {\bibinfo  {journal} {Phys. Rev. B}\ }\textbf {\bibinfo
  {volume} {97}},\ \bibinfo {pages} {045138} (\bibinfo {year}
  {2018})}\BibitemShut {NoStop}%
\bibitem [{\citenamefont {Jiang}\ and\ \citenamefont
  {Devereaux}(2019)}]{jiang2019superconductivity}%
  \BibitemOpen
  \bibfield  {author} {\bibinfo {author} {\bibfnamefont {H.-C.}\ \bibnamefont
  {Jiang}}\ and\ \bibinfo {author} {\bibfnamefont {T.~P.}\ \bibnamefont
  {Devereaux}},\ }\href@noop {} {\bibfield  {journal} {\bibinfo  {journal}
  {Science}\ }\textbf {\bibinfo {volume} {365}},\ \bibinfo {pages} {1424}
  (\bibinfo {year} {2019})}\BibitemShut {NoStop}%
\bibitem [{\citenamefont {Shen}\ \emph {et~al.}(2004)\citenamefont {Shen},
  \citenamefont {Ronning}, \citenamefont {Lu}, \citenamefont {Lee},
  \citenamefont {Ingle}, \citenamefont {Meevasana}, \citenamefont {Baumberger},
  \citenamefont {Damascelli}, \citenamefont {Armitage}, \citenamefont {Miller}
  \emph {et~al.}}]{shen2004missing}%
  \BibitemOpen
  \bibfield  {author} {\bibinfo {author} {\bibfnamefont {K.}~\bibnamefont
  {Shen}}, \bibinfo {author} {\bibfnamefont {F.}~\bibnamefont {Ronning}},
  \bibinfo {author} {\bibfnamefont {D.}~\bibnamefont {Lu}}, \bibinfo {author}
  {\bibfnamefont {W.}~\bibnamefont {Lee}}, \bibinfo {author} {\bibfnamefont
  {N.}~\bibnamefont {Ingle}}, \bibinfo {author} {\bibfnamefont
  {W.}~\bibnamefont {Meevasana}}, \bibinfo {author} {\bibfnamefont
  {F.}~\bibnamefont {Baumberger}}, \bibinfo {author} {\bibfnamefont
  {A.}~\bibnamefont {Damascelli}}, \bibinfo {author} {\bibfnamefont
  {N.}~\bibnamefont {Armitage}}, \bibinfo {author} {\bibfnamefont
  {L.}~\bibnamefont {Miller}},  \emph {et~al.},\ }\href@noop {} {\bibfield
  {journal} {\bibinfo  {journal} {Phys. Rev. Lett.}\ }\textbf {\bibinfo
  {volume} {93}},\ \bibinfo {pages} {267002} (\bibinfo {year}
  {2004})}\BibitemShut {NoStop}%
\bibitem [{\citenamefont {Lanzara}\ \emph {et~al.}(2001)\citenamefont
  {Lanzara}, \citenamefont {Bogdanov}, \citenamefont {Zhou}, \citenamefont
  {Kellar}, \citenamefont {Feng}, \citenamefont {Lu}, \citenamefont {Yoshida},
  \citenamefont {Eisaki}, \citenamefont {Fujimori}, \citenamefont {Kishio}
  \emph {et~al.}}]{lanzara2001evidence}%
  \BibitemOpen
  \bibfield  {author} {\bibinfo {author} {\bibfnamefont {A.}~\bibnamefont
  {Lanzara}}, \bibinfo {author} {\bibfnamefont {P.}~\bibnamefont {Bogdanov}},
  \bibinfo {author} {\bibfnamefont {X.}~\bibnamefont {Zhou}}, \bibinfo {author}
  {\bibfnamefont {S.}~\bibnamefont {Kellar}}, \bibinfo {author} {\bibfnamefont
  {D.}~\bibnamefont {Feng}}, \bibinfo {author} {\bibfnamefont {E.}~\bibnamefont
  {Lu}}, \bibinfo {author} {\bibfnamefont {T.}~\bibnamefont {Yoshida}},
  \bibinfo {author} {\bibfnamefont {H.}~\bibnamefont {Eisaki}}, \bibinfo
  {author} {\bibfnamefont {A.}~\bibnamefont {Fujimori}}, \bibinfo {author}
  {\bibfnamefont {K.}~\bibnamefont {Kishio}},  \emph {et~al.},\ }\href@noop {}
  {\bibfield  {journal} {\bibinfo  {journal} {Nature}\ }\textbf {\bibinfo
  {volume} {412}},\ \bibinfo {pages} {510} (\bibinfo {year}
  {2001})}\BibitemShut {NoStop}%
\bibitem [{\citenamefont {Reznik}\ \emph {et~al.}(2006)\citenamefont {Reznik},
  \citenamefont {Pintschovius}, \citenamefont {Ito}, \citenamefont {Iikubo},
  \citenamefont {Sato}, \citenamefont {Goka}, \citenamefont {Fujita},
  \citenamefont {Yamada}, \citenamefont {Gu},\ and\ \citenamefont
  {Tranquada}}]{reznik2006electron}%
  \BibitemOpen
  \bibfield  {author} {\bibinfo {author} {\bibfnamefont {D.}~\bibnamefont
  {Reznik}}, \bibinfo {author} {\bibfnamefont {L.}~\bibnamefont
  {Pintschovius}}, \bibinfo {author} {\bibfnamefont {M.}~\bibnamefont {Ito}},
  \bibinfo {author} {\bibfnamefont {S.}~\bibnamefont {Iikubo}}, \bibinfo
  {author} {\bibfnamefont {M.}~\bibnamefont {Sato}}, \bibinfo {author}
  {\bibfnamefont {H.}~\bibnamefont {Goka}}, \bibinfo {author} {\bibfnamefont
  {M.}~\bibnamefont {Fujita}}, \bibinfo {author} {\bibfnamefont
  {K.}~\bibnamefont {Yamada}}, \bibinfo {author} {\bibfnamefont
  {G.}~\bibnamefont {Gu}}, \ and\ \bibinfo {author} {\bibfnamefont
  {J.}~\bibnamefont {Tranquada}},\ }\href@noop {} {\bibfield  {journal}
  {\bibinfo  {journal} {Nature}\ }\textbf {\bibinfo {volume} {440}},\ \bibinfo
  {pages} {1170} (\bibinfo {year} {2006})}\BibitemShut {NoStop}%
\bibitem [{\citenamefont {Devereaux}\ and\ \citenamefont
  {Hackl}(2007)}]{devereaux2007inelastic}%
  \BibitemOpen
  \bibfield  {author} {\bibinfo {author} {\bibfnamefont {T.~P.}\ \bibnamefont
  {Devereaux}}\ and\ \bibinfo {author} {\bibfnamefont {R.}~\bibnamefont
  {Hackl}},\ }\href@noop {} {\bibfield  {journal} {\bibinfo  {journal} {Rev.
  Mod. Phys.}\ }\textbf {\bibinfo {volume} {79}},\ \bibinfo {pages} {175}
  (\bibinfo {year} {2007})}\BibitemShut {NoStop}%
\bibitem [{\citenamefont {He}\ \emph {et~al.}(2018)\citenamefont {He},
  \citenamefont {Hashimoto}, \citenamefont {Song}, \citenamefont {Chen},
  \citenamefont {He}, \citenamefont {Vishik}, \citenamefont {Moritz},
  \citenamefont {Lee}, \citenamefont {Nagaosa}, \citenamefont {Zaanen} \emph
  {et~al.}}]{he2018rapid}%
  \BibitemOpen
  \bibfield  {author} {\bibinfo {author} {\bibfnamefont {Y.}~\bibnamefont
  {He}}, \bibinfo {author} {\bibfnamefont {M.}~\bibnamefont {Hashimoto}},
  \bibinfo {author} {\bibfnamefont {D.}~\bibnamefont {Song}}, \bibinfo {author}
  {\bibfnamefont {S.-D.}\ \bibnamefont {Chen}}, \bibinfo {author}
  {\bibfnamefont {J.}~\bibnamefont {He}}, \bibinfo {author} {\bibfnamefont
  {I.}~\bibnamefont {Vishik}}, \bibinfo {author} {\bibfnamefont
  {B.}~\bibnamefont {Moritz}}, \bibinfo {author} {\bibfnamefont {D.-H.}\
  \bibnamefont {Lee}}, \bibinfo {author} {\bibfnamefont {N.}~\bibnamefont
  {Nagaosa}}, \bibinfo {author} {\bibfnamefont {J.}~\bibnamefont {Zaanen}},
  \emph {et~al.},\ }\href@noop {} {\bibfield  {journal} {\bibinfo  {journal}
  {Science}\ }\textbf {\bibinfo {volume} {362}},\ \bibinfo {pages} {62}
  (\bibinfo {year} {2018})}\BibitemShut {NoStop}%
\bibitem [{\citenamefont {Hubbard}(1963)}]{hubbard1963week}%
  \BibitemOpen
  \bibfield  {author} {\bibinfo {author} {\bibfnamefont {J.}~\bibnamefont
  {Hubbard}},\ }\href@noop {} {\bibfield  {journal} {\bibinfo  {journal}
  {Theoretical Physics}\ }\textbf {\bibinfo {volume} {276}},\ \bibinfo {pages}
  {238} (\bibinfo {year} {1963})}\BibitemShut {NoStop}%
\bibitem [{\citenamefont {Holstein}(1959)}]{holstein1959studies}%
  \BibitemOpen
  \bibfield  {author} {\bibinfo {author} {\bibfnamefont {T.}~\bibnamefont
  {Holstein}},\ }\href@noop {} {\bibfield  {journal} {\bibinfo  {journal}
  {Annals of physics}\ }\textbf {\bibinfo {volume} {8}},\ \bibinfo {pages}
  {325} (\bibinfo {year} {1959})}\BibitemShut {NoStop}%
\bibitem [{\citenamefont {Clay}\ and\ \citenamefont
  {Hardikar}(2005)}]{clay2005intermediate}%
  \BibitemOpen
  \bibfield  {author} {\bibinfo {author} {\bibfnamefont {R.}~\bibnamefont
  {Clay}}\ and\ \bibinfo {author} {\bibfnamefont {R.}~\bibnamefont
  {Hardikar}},\ }\href@noop {} {\bibfield  {journal} {\bibinfo  {journal}
  {Phys. Rev. Lett.}\ }\textbf {\bibinfo {volume} {95}},\ \bibinfo {pages}
  {096401} (\bibinfo {year} {2005})}\BibitemShut {NoStop}%
\bibitem [{\citenamefont {Nowadnick}\ \emph {et~al.}(2012)\citenamefont
  {Nowadnick}, \citenamefont {Johnston}, \citenamefont {Moritz}, \citenamefont
  {Scalettar},\ and\ \citenamefont {Devereaux}}]{nowadnick2012competition}%
  \BibitemOpen
  \bibfield  {author} {\bibinfo {author} {\bibfnamefont {E.}~\bibnamefont
  {Nowadnick}}, \bibinfo {author} {\bibfnamefont {S.}~\bibnamefont {Johnston}},
  \bibinfo {author} {\bibfnamefont {B.}~\bibnamefont {Moritz}}, \bibinfo
  {author} {\bibfnamefont {R.}~\bibnamefont {Scalettar}}, \ and\ \bibinfo
  {author} {\bibfnamefont {T.}~\bibnamefont {Devereaux}},\ }\href@noop {}
  {\bibfield  {journal} {\bibinfo  {journal} {Phys. Rev. Lett.}\ }\textbf
  {\bibinfo {volume} {109}},\ \bibinfo {pages} {246404} (\bibinfo {year}
  {2012})}\BibitemShut {NoStop}%
\bibitem [{\citenamefont {Nowadnick}\ \emph {et~al.}(2015)\citenamefont
  {Nowadnick}, \citenamefont {Johnston}, \citenamefont {Moritz},\ and\
  \citenamefont {Devereaux}}]{nowadnick2015renormalization}%
  \BibitemOpen
  \bibfield  {author} {\bibinfo {author} {\bibfnamefont {E.}~\bibnamefont
  {Nowadnick}}, \bibinfo {author} {\bibfnamefont {S.}~\bibnamefont {Johnston}},
  \bibinfo {author} {\bibfnamefont {B.}~\bibnamefont {Moritz}}, \ and\ \bibinfo
  {author} {\bibfnamefont {T.}~\bibnamefont {Devereaux}},\ }\href@noop {}
  {\bibfield  {journal} {\bibinfo  {journal} {Phys. Rev. B}\ }\textbf {\bibinfo
  {volume} {91}},\ \bibinfo {pages} {165127} (\bibinfo {year}
  {2015})}\BibitemShut {NoStop}%
\bibitem [{\citenamefont {Johnston}\ \emph {et~al.}(2013)\citenamefont
  {Johnston}, \citenamefont {Nowadnick}, \citenamefont {Kung}, \citenamefont
  {Moritz}, \citenamefont {Scalettar},\ and\ \citenamefont
  {Devereaux}}]{johnston2013determinant}%
  \BibitemOpen
  \bibfield  {author} {\bibinfo {author} {\bibfnamefont {S.}~\bibnamefont
  {Johnston}}, \bibinfo {author} {\bibfnamefont {E.}~\bibnamefont {Nowadnick}},
  \bibinfo {author} {\bibfnamefont {Y.}~\bibnamefont {Kung}}, \bibinfo {author}
  {\bibfnamefont {B.}~\bibnamefont {Moritz}}, \bibinfo {author} {\bibfnamefont
  {R.}~\bibnamefont {Scalettar}}, \ and\ \bibinfo {author} {\bibfnamefont
  {T.}~\bibnamefont {Devereaux}},\ }\href@noop {} {\bibfield  {journal}
  {\bibinfo  {journal} {Phys. Rev. B}\ }\textbf {\bibinfo {volume} {87}},\
  \bibinfo {pages} {235133} (\bibinfo {year} {2013})}\BibitemShut {NoStop}%
\bibitem [{\citenamefont {Wang}\ \emph {et~al.}(2016)\citenamefont {Wang},
  \citenamefont {Moritz}, \citenamefont {Chen}, \citenamefont {Jia},
  \citenamefont {van Veenendaal},\ and\ \citenamefont
  {Devereaux}}]{wang2016using}%
  \BibitemOpen
  \bibfield  {author} {\bibinfo {author} {\bibfnamefont {Y.}~\bibnamefont
  {Wang}}, \bibinfo {author} {\bibfnamefont {B.}~\bibnamefont {Moritz}},
  \bibinfo {author} {\bibfnamefont {C.-C.}\ \bibnamefont {Chen}}, \bibinfo
  {author} {\bibfnamefont {C.}~\bibnamefont {Jia}}, \bibinfo {author}
  {\bibfnamefont {M.}~\bibnamefont {van Veenendaal}}, \ and\ \bibinfo {author}
  {\bibfnamefont {T.~P.}\ \bibnamefont {Devereaux}},\ }\href@noop {} {\bibfield
   {journal} {\bibinfo  {journal} {Phys. Rev. Lett.}\ }\textbf {\bibinfo
  {volume} {116}},\ \bibinfo {pages} {086401} (\bibinfo {year}
  {2016})}\BibitemShut {NoStop}%
\bibitem [{\citenamefont {Mendl}\ \emph {et~al.}(2017)\citenamefont {Mendl},
  \citenamefont {Nowadnick}, \citenamefont {Huang}, \citenamefont {Johnston},
  \citenamefont {Moritz},\ and\ \citenamefont {Devereaux}}]{mendl2017doping}%
  \BibitemOpen
  \bibfield  {author} {\bibinfo {author} {\bibfnamefont {C.~B.}\ \bibnamefont
  {Mendl}}, \bibinfo {author} {\bibfnamefont {E.~A.}\ \bibnamefont
  {Nowadnick}}, \bibinfo {author} {\bibfnamefont {E.~W.}\ \bibnamefont
  {Huang}}, \bibinfo {author} {\bibfnamefont {S.}~\bibnamefont {Johnston}},
  \bibinfo {author} {\bibfnamefont {B.}~\bibnamefont {Moritz}}, \ and\ \bibinfo
  {author} {\bibfnamefont {T.~P.}\ \bibnamefont {Devereaux}},\ }\href@noop {}
  {\bibfield  {journal} {\bibinfo  {journal} {Phys. Rev. B}\ }\textbf {\bibinfo
  {volume} {96}},\ \bibinfo {pages} {205141} (\bibinfo {year}
  {2017})}\BibitemShut {NoStop}%
\bibitem [{\citenamefont {Zhang}\ \emph {et~al.}(2018)\citenamefont {Zhang},
  \citenamefont {Wang}, \citenamefont {Shi}, \citenamefont {Lin}, \citenamefont
  {Zhang}, \citenamefont {Gu}, \citenamefont {Dong},\ and\ \citenamefont
  {Wang}}]{zhang2018light}%
  \BibitemOpen
  \bibfield  {author} {\bibinfo {author} {\bibfnamefont {S.}~\bibnamefont
  {Zhang}}, \bibinfo {author} {\bibfnamefont {Z.}~\bibnamefont {Wang}},
  \bibinfo {author} {\bibfnamefont {L.}~\bibnamefont {Shi}}, \bibinfo {author}
  {\bibfnamefont {T.}~\bibnamefont {Lin}}, \bibinfo {author} {\bibfnamefont
  {M.}~\bibnamefont {Zhang}}, \bibinfo {author} {\bibfnamefont
  {G.}~\bibnamefont {Gu}}, \bibinfo {author} {\bibfnamefont {T.}~\bibnamefont
  {Dong}}, \ and\ \bibinfo {author} {\bibfnamefont {N.}~\bibnamefont {Wang}},\
  }\href@noop {} {\bibfield  {journal} {\bibinfo  {journal} {Phys. Rev. B}\
  }\textbf {\bibinfo {volume} {98}},\ \bibinfo {pages} {020506} (\bibinfo
  {year} {2018})}\BibitemShut {NoStop}%
\bibitem [{\citenamefont {Lu}\ \emph {et~al.}(2012)\citenamefont {Lu},
  \citenamefont {Sota}, \citenamefont {Matsueda}, \citenamefont {Bon{\v{c}}a},\
  and\ \citenamefont {Tohyama}}]{lu2012enhanced}%
  \BibitemOpen
  \bibfield  {author} {\bibinfo {author} {\bibfnamefont {H.}~\bibnamefont
  {Lu}}, \bibinfo {author} {\bibfnamefont {S.}~\bibnamefont {Sota}}, \bibinfo
  {author} {\bibfnamefont {H.}~\bibnamefont {Matsueda}}, \bibinfo {author}
  {\bibfnamefont {J.}~\bibnamefont {Bon{\v{c}}a}}, \ and\ \bibinfo {author}
  {\bibfnamefont {T.}~\bibnamefont {Tohyama}},\ }\href@noop {} {\bibfield
  {journal} {\bibinfo  {journal} {Phys. Rev. Lett.}\ }\textbf {\bibinfo
  {volume} {109}},\ \bibinfo {pages} {197401} (\bibinfo {year}
  {2012})}\BibitemShut {NoStop}%
\bibitem [{\citenamefont {Al-Hassanieh}\ \emph {et~al.}(2013)\citenamefont
  {Al-Hassanieh}, \citenamefont {Rinc{\'o}n}, \citenamefont {Dagotto},\ and\
  \citenamefont {Alvarez}}]{al2013wave}%
  \BibitemOpen
  \bibfield  {author} {\bibinfo {author} {\bibfnamefont {K.~A.}\ \bibnamefont
  {Al-Hassanieh}}, \bibinfo {author} {\bibfnamefont {J.}~\bibnamefont
  {Rinc{\'o}n}}, \bibinfo {author} {\bibfnamefont {E.}~\bibnamefont {Dagotto}},
  \ and\ \bibinfo {author} {\bibfnamefont {G.}~\bibnamefont {Alvarez}},\
  }\href@noop {} {\bibfield  {journal} {\bibinfo  {journal} {Phys. Rev. B}\
  }\textbf {\bibinfo {volume} {88}},\ \bibinfo {pages} {045107} (\bibinfo
  {year} {2013})}\BibitemShut {NoStop}%
\bibitem [{\citenamefont {Rinc{\'o}n}\ \emph {et~al.}(2014)\citenamefont
  {Rinc{\'o}n}, \citenamefont {Al-Hassanieh}, \citenamefont {Feiguin},\ and\
  \citenamefont {Dagotto}}]{rincon2014photoexcitation}%
  \BibitemOpen
  \bibfield  {author} {\bibinfo {author} {\bibfnamefont {J.}~\bibnamefont
  {Rinc{\'o}n}}, \bibinfo {author} {\bibfnamefont {K.}~\bibnamefont
  {Al-Hassanieh}}, \bibinfo {author} {\bibfnamefont {A.~E.}\ \bibnamefont
  {Feiguin}}, \ and\ \bibinfo {author} {\bibfnamefont {E.}~\bibnamefont
  {Dagotto}},\ }\href@noop {} {\bibfield  {journal} {\bibinfo  {journal} {Phys.
  Rev. B}\ }\textbf {\bibinfo {volume} {90}},\ \bibinfo {pages} {155112}
  (\bibinfo {year} {2014})}\BibitemShut {NoStop}%
\bibitem [{\citenamefont {Lu}\ \emph {et~al.}(2015)\citenamefont {Lu},
  \citenamefont {Shao}, \citenamefont {Bon\ifmmode~\check{c}\else \v{c}\fi{}a},
  \citenamefont {Manske},\ and\ \citenamefont {Tohyama}}]{lu2015photoinduced}%
  \BibitemOpen
  \bibfield  {author} {\bibinfo {author} {\bibfnamefont {H.}~\bibnamefont
  {Lu}}, \bibinfo {author} {\bibfnamefont {C.}~\bibnamefont {Shao}}, \bibinfo
  {author} {\bibfnamefont {J.}~\bibnamefont {Bon\ifmmode~\check{c}\else
  \v{c}\fi{}a}}, \bibinfo {author} {\bibfnamefont {D.}~\bibnamefont {Manske}},
  \ and\ \bibinfo {author} {\bibfnamefont {T.}~\bibnamefont {Tohyama}},\
  }\href@noop {} {\bibfield  {journal} {\bibinfo  {journal} {Phys. Rev. B}\
  }\textbf {\bibinfo {volume} {91}},\ \bibinfo {pages} {245117} (\bibinfo
  {year} {2015})}\BibitemShut {NoStop}%
\bibitem [{\citenamefont {Denny}\ \emph {et~al.}(2015)\citenamefont {Denny},
  \citenamefont {Clark}, \citenamefont {Laplace}, \citenamefont {Cavalleri},\
  and\ \citenamefont {Jaksch}}]{denny2015proposed}%
  \BibitemOpen
  \bibfield  {author} {\bibinfo {author} {\bibfnamefont {S.}~\bibnamefont
  {Denny}}, \bibinfo {author} {\bibfnamefont {S.}~\bibnamefont {Clark}},
  \bibinfo {author} {\bibfnamefont {Y.}~\bibnamefont {Laplace}}, \bibinfo
  {author} {\bibfnamefont {A.}~\bibnamefont {Cavalleri}}, \ and\ \bibinfo
  {author} {\bibfnamefont {D.}~\bibnamefont {Jaksch}},\ }\href@noop {}
  {\bibfield  {journal} {\bibinfo  {journal} {Phys. Rev. Lett.}\ }\textbf
  {\bibinfo {volume} {114}},\ \bibinfo {pages} {137001} (\bibinfo {year}
  {2015})}\BibitemShut {NoStop}%
\bibitem [{\citenamefont {Raines}\ \emph {et~al.}(2015)\citenamefont {Raines},
  \citenamefont {Stanev},\ and\ \citenamefont
  {Galitski}}]{raines2015enhancement}%
  \BibitemOpen
  \bibfield  {author} {\bibinfo {author} {\bibfnamefont {Z.~M.}\ \bibnamefont
  {Raines}}, \bibinfo {author} {\bibfnamefont {V.}~\bibnamefont {Stanev}}, \
  and\ \bibinfo {author} {\bibfnamefont {V.~M.}\ \bibnamefont {Galitski}},\
  }\href@noop {} {\bibfield  {journal} {\bibinfo  {journal} {Phys. Rev. B}\
  }\textbf {\bibinfo {volume} {91}},\ \bibinfo {pages} {184506} (\bibinfo
  {year} {2015})}\BibitemShut {NoStop}%
\bibitem [{\citenamefont {Okamoto}\ \emph {et~al.}(2017)\citenamefont
  {Okamoto}, \citenamefont {Hu}, \citenamefont {Cavalleri},\ and\ \citenamefont
  {Mathey}}]{okamoto2017transiently}%
  \BibitemOpen
  \bibfield  {author} {\bibinfo {author} {\bibfnamefont {J.-i.}\ \bibnamefont
  {Okamoto}}, \bibinfo {author} {\bibfnamefont {W.}~\bibnamefont {Hu}},
  \bibinfo {author} {\bibfnamefont {A.}~\bibnamefont {Cavalleri}}, \ and\
  \bibinfo {author} {\bibfnamefont {L.}~\bibnamefont {Mathey}},\ }\href@noop {}
  {\bibfield  {journal} {\bibinfo  {journal} {Phys. Rev. B}\ }\textbf {\bibinfo
  {volume} {96}},\ \bibinfo {pages} {144505} (\bibinfo {year}
  {2017})}\BibitemShut {NoStop}%
\bibitem [{\citenamefont {Freericks}\ \emph {et~al.}(2006)\citenamefont
  {Freericks}, \citenamefont {Turkowski},\ and\ \citenamefont
  {Zlati{\'c}}}]{freericks2006nonequilibrium}%
  \BibitemOpen
  \bibfield  {author} {\bibinfo {author} {\bibfnamefont {J.}~\bibnamefont
  {Freericks}}, \bibinfo {author} {\bibfnamefont {V.}~\bibnamefont
  {Turkowski}}, \ and\ \bibinfo {author} {\bibfnamefont {V.}~\bibnamefont
  {Zlati{\'c}}},\ }\href@noop {} {\bibfield  {journal} {\bibinfo  {journal}
  {Phys. Rev. Lett.}\ }\textbf {\bibinfo {volume} {97}},\ \bibinfo {pages}
  {266408} (\bibinfo {year} {2006})}\BibitemShut {NoStop}%
\bibitem [{\citenamefont {Aoki}\ \emph {et~al.}(2014)\citenamefont {Aoki},
  \citenamefont {Tsuji}, \citenamefont {Eckstein}, \citenamefont {Kollar},
  \citenamefont {Oka},\ and\ \citenamefont {Werner}}]{aoki2014nonequilibrium}%
  \BibitemOpen
  \bibfield  {author} {\bibinfo {author} {\bibfnamefont {H.}~\bibnamefont
  {Aoki}}, \bibinfo {author} {\bibfnamefont {N.}~\bibnamefont {Tsuji}},
  \bibinfo {author} {\bibfnamefont {M.}~\bibnamefont {Eckstein}}, \bibinfo
  {author} {\bibfnamefont {M.}~\bibnamefont {Kollar}}, \bibinfo {author}
  {\bibfnamefont {T.}~\bibnamefont {Oka}}, \ and\ \bibinfo {author}
  {\bibfnamefont {P.}~\bibnamefont {Werner}},\ }\href@noop {} {\bibfield
  {journal} {\bibinfo  {journal} {Rev. Mod. Phys.}\ }\textbf {\bibinfo {volume}
  {86}},\ \bibinfo {pages} {779} (\bibinfo {year} {2014})}\BibitemShut
  {NoStop}%
\bibitem [{\citenamefont {Werner}\ and\ \citenamefont
  {Millis}(2007)}]{werner2007efficient}%
  \BibitemOpen
  \bibfield  {author} {\bibinfo {author} {\bibfnamefont {P.}~\bibnamefont
  {Werner}}\ and\ \bibinfo {author} {\bibfnamefont {A.~J.}\ \bibnamefont
  {Millis}},\ }\href@noop {} {\bibfield  {journal} {\bibinfo  {journal} {Phys.
  Rev. Lett.}\ }\textbf {\bibinfo {volume} {99}},\ \bibinfo {pages} {146404}
  (\bibinfo {year} {2007})}\BibitemShut {NoStop}%
\bibitem [{\citenamefont {Eckstein}\ \emph {et~al.}(2010)\citenamefont
  {Eckstein}, \citenamefont {Kollar},\ and\ \citenamefont
  {Werner}}]{eckstein2010interaction}%
  \BibitemOpen
  \bibfield  {author} {\bibinfo {author} {\bibfnamefont {M.}~\bibnamefont
  {Eckstein}}, \bibinfo {author} {\bibfnamefont {M.}~\bibnamefont {Kollar}}, \
  and\ \bibinfo {author} {\bibfnamefont {P.}~\bibnamefont {Werner}},\
  }\href@noop {} {\bibfield  {journal} {\bibinfo  {journal} {Phys. Rev. B}\
  }\textbf {\bibinfo {volume} {81}},\ \bibinfo {pages} {115131} (\bibinfo
  {year} {2010})}\BibitemShut {NoStop}%
\bibitem [{\citenamefont {Dagotto}(1994)}]{dagotto1994correlated}%
  \BibitemOpen
  \bibfield  {author} {\bibinfo {author} {\bibfnamefont {E.}~\bibnamefont
  {Dagotto}},\ }\href@noop {} {\bibfield  {journal} {\bibinfo  {journal} {Rev.
  Mod. Phys.}\ }\textbf {\bibinfo {volume} {66}},\ \bibinfo {pages} {763}
  (\bibinfo {year} {1994})}\BibitemShut {NoStop}%
\bibitem [{\citenamefont {White}(1992)}]{white1992density}%
  \BibitemOpen
  \bibfield  {author} {\bibinfo {author} {\bibfnamefont {S.~R.}\ \bibnamefont
  {White}},\ }\href@noop {} {\bibfield  {journal} {\bibinfo  {journal} {Phys.
  Rev. Lett.}\ }\textbf {\bibinfo {volume} {69}},\ \bibinfo {pages} {2863}
  (\bibinfo {year} {1992})}\BibitemShut {NoStop}%
\bibitem [{\citenamefont {Zhang}\ \emph {et~al.}(1998)\citenamefont {Zhang},
  \citenamefont {Jeckelmann},\ and\ \citenamefont {White}}]{zhang1998density}%
  \BibitemOpen
  \bibfield  {author} {\bibinfo {author} {\bibfnamefont {C.}~\bibnamefont
  {Zhang}}, \bibinfo {author} {\bibfnamefont {E.}~\bibnamefont {Jeckelmann}}, \
  and\ \bibinfo {author} {\bibfnamefont {S.~R.}\ \bibnamefont {White}},\
  }\href@noop {} {\bibfield  {journal} {\bibinfo  {journal} {Phys. Rev. Lett.}\
  }\textbf {\bibinfo {volume} {80}},\ \bibinfo {pages} {2661} (\bibinfo {year}
  {1998})}\BibitemShut {NoStop}%
\bibitem [{\citenamefont {Brockt}\ \emph {et~al.}(2015)\citenamefont {Brockt},
  \citenamefont {Dorfner}, \citenamefont {Vidmar}, \citenamefont
  {Heidrich-Meisner},\ and\ \citenamefont {Jeckelmann}}]{brockt2015matrix}%
  \BibitemOpen
  \bibfield  {author} {\bibinfo {author} {\bibfnamefont {C.}~\bibnamefont
  {Brockt}}, \bibinfo {author} {\bibfnamefont {F.}~\bibnamefont {Dorfner}},
  \bibinfo {author} {\bibfnamefont {L.}~\bibnamefont {Vidmar}}, \bibinfo
  {author} {\bibfnamefont {F.}~\bibnamefont {Heidrich-Meisner}}, \ and\
  \bibinfo {author} {\bibfnamefont {E.}~\bibnamefont {Jeckelmann}},\
  }\href@noop {} {\bibfield  {journal} {\bibinfo  {journal} {Phys. Rev. B}\
  }\textbf {\bibinfo {volume} {92}},\ \bibinfo {pages} {241106} (\bibinfo
  {year} {2015})}\BibitemShut {NoStop}%
\bibitem [{\citenamefont {Lang}\ and\ \citenamefont {Firsov}(1962)}]{lang1962}%
  \BibitemOpen
  \bibfield  {author} {\bibinfo {author} {\bibfnamefont {I.~G.}\ \bibnamefont
  {Lang}}\ and\ \bibinfo {author} {\bibfnamefont {Y.~A.}\ \bibnamefont
  {Firsov}},\ }\href@noop {} {\bibfield  {journal} {\bibinfo  {journal} {Sov.
  Phys. JETP}\ }\textbf {\bibinfo {volume} {16}},\ \bibinfo {pages} {1301}
  (\bibinfo {year} {1962})}\BibitemShut {NoStop}%
\bibitem [{\citenamefont {Karakuzu}\ \emph {et~al.}(2017)\citenamefont
  {Karakuzu}, \citenamefont {Tocchio}, \citenamefont {Sorella},\ and\
  \citenamefont {Becca}}]{karakuzu2017superconductivity}%
  \BibitemOpen
  \bibfield  {author} {\bibinfo {author} {\bibfnamefont {S.}~\bibnamefont
  {Karakuzu}}, \bibinfo {author} {\bibfnamefont {L.~F.}\ \bibnamefont
  {Tocchio}}, \bibinfo {author} {\bibfnamefont {S.}~\bibnamefont {Sorella}}, \
  and\ \bibinfo {author} {\bibfnamefont {F.}~\bibnamefont {Becca}},\
  }\href@noop {} {\bibfield  {journal} {\bibinfo  {journal} {Phys. Rev. B}\
  }\textbf {\bibinfo {volume} {96}},\ \bibinfo {pages} {205145} (\bibinfo
  {year} {2017})}\BibitemShut {NoStop}%
\bibitem [{\citenamefont {Ohgoe}\ and\ \citenamefont
  {Imada}(2017)}]{ohgoe2017competition}%
  \BibitemOpen
  \bibfield  {author} {\bibinfo {author} {\bibfnamefont {T.}~\bibnamefont
  {Ohgoe}}\ and\ \bibinfo {author} {\bibfnamefont {M.}~\bibnamefont {Imada}},\
  }\href@noop {} {\bibfield  {journal} {\bibinfo  {journal} {Phys. Rev. Lett.}\
  }\textbf {\bibinfo {volume} {119}},\ \bibinfo {pages} {197001} (\bibinfo
  {year} {2017})}\BibitemShut {NoStop}%
\bibitem [{\citenamefont {Costa}\ \emph {et~al.}(2020)\citenamefont {Costa},
  \citenamefont {Seki}, \citenamefont {Yunoki},\ and\ \citenamefont
  {Sorella}}]{costa2020phase}%
  \BibitemOpen
  \bibfield  {author} {\bibinfo {author} {\bibfnamefont {N.~C.}\ \bibnamefont
  {Costa}}, \bibinfo {author} {\bibfnamefont {K.}~\bibnamefont {Seki}},
  \bibinfo {author} {\bibfnamefont {S.}~\bibnamefont {Yunoki}}, \ and\ \bibinfo
  {author} {\bibfnamefont {S.}~\bibnamefont {Sorella}},\ }\href@noop {}
  {\bibfield  {journal} {\bibinfo  {journal} {Commun. Phys.}\ }\textbf
  {\bibinfo {volume} {3}},\ \bibinfo {pages} {80} (\bibinfo {year}
  {2020})}\BibitemShut {NoStop}%
\bibitem [{\citenamefont {Chung}\ \emph {et~al.}(2020)\citenamefont {Chung},
  \citenamefont {Qin}, \citenamefont {Zhang}, \citenamefont {Schollw\"ock},\
  and\ \citenamefont {White}}]{chung2020plaquette}%
  \BibitemOpen
  \bibfield  {author} {\bibinfo {author} {\bibfnamefont {C.-M.}\ \bibnamefont
  {Chung}}, \bibinfo {author} {\bibfnamefont {M.}~\bibnamefont {Qin}}, \bibinfo
  {author} {\bibfnamefont {S.}~\bibnamefont {Zhang}}, \bibinfo {author}
  {\bibfnamefont {U.}~\bibnamefont {Schollw\"ock}}, \ and\ \bibinfo {author}
  {\bibfnamefont {S.~R.}\ \bibnamefont {White}},\ }\href@noop {} {\bibfield
  {journal} {\bibinfo  {journal} {Phys. Rev. B}\ }\textbf {\bibinfo {volume}
  {102}},\ \bibinfo {pages} {041106} (\bibinfo {year} {2020})}\BibitemShut
  {NoStop}%
\bibitem [{\citenamefont {Maier}\ \emph {et~al.}(2005)\citenamefont {Maier},
  \citenamefont {Jarrell}, \citenamefont {Schulthess}, \citenamefont {Kent},\
  and\ \citenamefont {White}}]{maier2005systematic}%
  \BibitemOpen
  \bibfield  {author} {\bibinfo {author} {\bibfnamefont {T.~A.}\ \bibnamefont
  {Maier}}, \bibinfo {author} {\bibfnamefont {M.}~\bibnamefont {Jarrell}},
  \bibinfo {author} {\bibfnamefont {T.}~\bibnamefont {Schulthess}}, \bibinfo
  {author} {\bibfnamefont {P.}~\bibnamefont {Kent}}, \ and\ \bibinfo {author}
  {\bibfnamefont {J.}~\bibnamefont {White}},\ }\href@noop {} {\bibfield
  {journal} {\bibinfo  {journal} {Phys. Rev. Lett.}\ }\textbf {\bibinfo
  {volume} {95}},\ \bibinfo {pages} {237001} (\bibinfo {year}
  {2005})}\BibitemShut {NoStop}%
\bibitem [{\citenamefont {Fulde}\ and\ \citenamefont
  {Ferrell}(1964)}]{fulde1964superconductivity}%
  \BibitemOpen
  \bibfield  {author} {\bibinfo {author} {\bibfnamefont {P.}~\bibnamefont
  {Fulde}}\ and\ \bibinfo {author} {\bibfnamefont {R.~A.}\ \bibnamefont
  {Ferrell}},\ }\href@noop {} {\bibfield  {journal} {\bibinfo  {journal}
  {Physical Review}\ }\textbf {\bibinfo {volume} {135}},\ \bibinfo {pages}
  {A550} (\bibinfo {year} {1964})}\BibitemShut {NoStop}%
\bibitem [{\citenamefont {Larkin}\ and\ \citenamefont
  {Ovchinnikov}(1965)}]{larkin1965nonuniform}%
  \BibitemOpen
  \bibfield  {author} {\bibinfo {author} {\bibfnamefont {A.}~\bibnamefont
  {Larkin}}\ and\ \bibinfo {author} {\bibfnamefont {Y.~N.}\ \bibnamefont
  {Ovchinnikov}},\ }\href@noop {} {\bibfield  {journal} {\bibinfo  {journal}
  {Soviet Physics-JETP}\ }\textbf {\bibinfo {volume} {20}},\ \bibinfo {pages}
  {762} (\bibinfo {year} {1965})}\BibitemShut {NoStop}%
\bibitem [{def()}]{definitionOfSC}%
  \BibitemOpen
  \href@noop {} {}\bibinfo {note} {{It is debatable whether a short-range
  superconductivity correlation can be conceptually classified as a
  light-induced superconducting state. Apart from the abstract definitions, a
  practical statement relies on the experimental consequences. Here, the
  short-range correlation can be detected by the onset of superconductivity
  gap.}}\BibitemShut {Stop}%
\bibitem [{\citenamefont {Knap}\ \emph {et~al.}(2016)\citenamefont {Knap},
  \citenamefont {Babadi}, \citenamefont {Refael}, \citenamefont {Martin},\ and\
  \citenamefont {Demler}}]{knap2016dynamical}%
  \BibitemOpen
  \bibfield  {author} {\bibinfo {author} {\bibfnamefont {M.}~\bibnamefont
  {Knap}}, \bibinfo {author} {\bibfnamefont {M.}~\bibnamefont {Babadi}},
  \bibinfo {author} {\bibfnamefont {G.}~\bibnamefont {Refael}}, \bibinfo
  {author} {\bibfnamefont {I.}~\bibnamefont {Martin}}, \ and\ \bibinfo {author}
  {\bibfnamefont {E.}~\bibnamefont {Demler}},\ }\href@noop {} {\bibfield
  {journal} {\bibinfo  {journal} {Phys. Rev. B}\ }\textbf {\bibinfo {volume}
  {94}},\ \bibinfo {pages} {214504} (\bibinfo {year} {2016})}\BibitemShut
  {NoStop}%
\bibitem [{\citenamefont {Macridin}\ \emph {et~al.}(2006)\citenamefont
  {Macridin}, \citenamefont {Moritz}, \citenamefont {Jarrell},\ and\
  \citenamefont {Maier}}]{macridin2006synergistic}%
  \BibitemOpen
  \bibfield  {author} {\bibinfo {author} {\bibfnamefont {A.}~\bibnamefont
  {Macridin}}, \bibinfo {author} {\bibfnamefont {B.}~\bibnamefont {Moritz}},
  \bibinfo {author} {\bibfnamefont {M.}~\bibnamefont {Jarrell}}, \ and\
  \bibinfo {author} {\bibfnamefont {T.}~\bibnamefont {Maier}},\ }\href@noop {}
  {\bibfield  {journal} {\bibinfo  {journal} {Physical review letters}\
  }\textbf {\bibinfo {volume} {97}},\ \bibinfo {pages} {056402} (\bibinfo
  {year} {2006})}\BibitemShut {NoStop}%
\bibitem [{\citenamefont {Wang}\ \emph {et~al.}(2017)\citenamefont {Wang},
  \citenamefont {Claassen}, \citenamefont {Moritz},\ and\ \citenamefont
  {Devereaux}}]{wang2017producing}%
  \BibitemOpen
  \bibfield  {author} {\bibinfo {author} {\bibfnamefont {Y.}~\bibnamefont
  {Wang}}, \bibinfo {author} {\bibfnamefont {M.}~\bibnamefont {Claassen}},
  \bibinfo {author} {\bibfnamefont {B.}~\bibnamefont {Moritz}}, \ and\ \bibinfo
  {author} {\bibfnamefont {T.}~\bibnamefont {Devereaux}},\ }\href@noop {}
  {\bibfield  {journal} {\bibinfo  {journal} {Phys. Rev. B}\ }\textbf {\bibinfo
  {volume} {96}},\ \bibinfo {pages} {235142} (\bibinfo {year}
  {2017})}\BibitemShut {NoStop}%
\bibitem [{\citenamefont {Shvaika}\ \emph {et~al.}(2018)\citenamefont
  {Shvaika}, \citenamefont {Matveev}, \citenamefont {Devereaux},\ and\
  \citenamefont {Freericks}}]{shvaika2018interpreting}%
  \BibitemOpen
  \bibfield  {author} {\bibinfo {author} {\bibfnamefont {A.}~\bibnamefont
  {Shvaika}}, \bibinfo {author} {\bibfnamefont {O.}~\bibnamefont {Matveev}},
  \bibinfo {author} {\bibfnamefont {T.}~\bibnamefont {Devereaux}}, \ and\
  \bibinfo {author} {\bibfnamefont {J.}~\bibnamefont {Freericks}},\ }\href@noop
  {} {\bibfield  {journal} {\bibinfo  {journal} {Condensed Matter Physics}\
  }\textbf {\bibinfo {volume} {21}},\ \bibinfo {pages} {33707} (\bibinfo {year}
  {2018})}\BibitemShut {NoStop}%
\bibitem [{\citenamefont {Wang}\ \emph
  {et~al.}(2018{\natexlab{c}})\citenamefont {Wang}, \citenamefont {Devereaux},\
  and\ \citenamefont {Chen}}]{wang2018theory}%
  \BibitemOpen
  \bibfield  {author} {\bibinfo {author} {\bibfnamefont {Y.}~\bibnamefont
  {Wang}}, \bibinfo {author} {\bibfnamefont {T.~P.}\ \bibnamefont {Devereaux}},
  \ and\ \bibinfo {author} {\bibfnamefont {C.-C.}\ \bibnamefont {Chen}},\
  }\href@noop {} {\bibfield  {journal} {\bibinfo  {journal} {Phys. Rev. B}\
  }\textbf {\bibinfo {volume} {98}},\ \bibinfo {pages} {245106} (\bibinfo
  {year} {2018}{\natexlab{c}})}\BibitemShut {NoStop}%
\bibitem [{\citenamefont {Matveev}\ \emph {et~al.}(2019)\citenamefont
  {Matveev}, \citenamefont {Shvaika}, \citenamefont {Devereaux},\ and\
  \citenamefont {Freericks}}]{matveev2019stroboscopic}%
  \BibitemOpen
  \bibfield  {author} {\bibinfo {author} {\bibfnamefont {O.}~\bibnamefont
  {Matveev}}, \bibinfo {author} {\bibfnamefont {A.}~\bibnamefont {Shvaika}},
  \bibinfo {author} {\bibfnamefont {T.}~\bibnamefont {Devereaux}}, \ and\
  \bibinfo {author} {\bibfnamefont {J.}~\bibnamefont {Freericks}},\ }\href@noop
  {} {\bibfield  {journal} {\bibinfo  {journal} {Phys. Rev. Lett.}\ }\textbf
  {\bibinfo {volume} {122}},\ \bibinfo {pages} {247402} (\bibinfo {year}
  {2019})}\BibitemShut {NoStop}%
\bibitem [{\citenamefont {Budden}\ \emph {et~al.}(2021)\citenamefont {Budden},
  \citenamefont {Gebert}, \citenamefont {Buzzi}, \citenamefont {Jotzu},
  \citenamefont {Wang}, \citenamefont {Matsuyama}, \citenamefont {Meier},
  \citenamefont {Laplace}, \citenamefont {Pontiroli}, \citenamefont {Ricc{\`o}}
  \emph {et~al.}}]{budden2021evidence}%
  \BibitemOpen
  \bibfield  {author} {\bibinfo {author} {\bibfnamefont {M.}~\bibnamefont
  {Budden}}, \bibinfo {author} {\bibfnamefont {T.}~\bibnamefont {Gebert}},
  \bibinfo {author} {\bibfnamefont {M.}~\bibnamefont {Buzzi}}, \bibinfo
  {author} {\bibfnamefont {G.}~\bibnamefont {Jotzu}}, \bibinfo {author}
  {\bibfnamefont {E.}~\bibnamefont {Wang}}, \bibinfo {author} {\bibfnamefont
  {T.}~\bibnamefont {Matsuyama}}, \bibinfo {author} {\bibfnamefont
  {G.}~\bibnamefont {Meier}}, \bibinfo {author} {\bibfnamefont
  {Y.}~\bibnamefont {Laplace}}, \bibinfo {author} {\bibfnamefont
  {D.}~\bibnamefont {Pontiroli}}, \bibinfo {author} {\bibfnamefont
  {M.}~\bibnamefont {Ricc{\`o}}},  \emph {et~al.},\ }\href@noop {} {\bibfield
  {journal} {\bibinfo  {journal} {Nature Physics}\ }\textbf {\bibinfo {volume}
  {17}},\ \bibinfo {pages} {611} (\bibinfo {year} {2021})}\BibitemShut
  {NoStop}%
\bibitem [{\citenamefont {Su}\ and\ \citenamefont
  {Zhang}(2020)}]{su2020coincidence}%
  \BibitemOpen
  \bibfield  {author} {\bibinfo {author} {\bibfnamefont {Y.}~\bibnamefont
  {Su}}\ and\ \bibinfo {author} {\bibfnamefont {C.}~\bibnamefont {Zhang}},\
  }\href@noop {} {\bibfield  {journal} {\bibinfo  {journal} {Phys. Rev. B}\
  }\textbf {\bibinfo {volume} {101}},\ \bibinfo {pages} {205110} (\bibinfo
  {year} {2020})}\BibitemShut {NoStop}%
\bibitem [{\citenamefont {Mahmood}\ \emph {et~al.}(2021)\citenamefont
  {Mahmood}, \citenamefont {Devereaux}, \citenamefont {Abbamonte},\ and\
  \citenamefont {Morr}}]{mahmood2021distinguishing}%
  \BibitemOpen
  \bibfield  {author} {\bibinfo {author} {\bibfnamefont {F.}~\bibnamefont
  {Mahmood}}, \bibinfo {author} {\bibfnamefont {T.}~\bibnamefont {Devereaux}},
  \bibinfo {author} {\bibfnamefont {P.}~\bibnamefont {Abbamonte}}, \ and\
  \bibinfo {author} {\bibfnamefont {D.~K.}\ \bibnamefont {Morr}},\ }\href@noop
  {} {\bibfield  {journal} {\bibinfo  {journal} {arXiv:2108.04260}\ } (\bibinfo
  {year} {2021})}\BibitemShut {NoStop}%
\bibitem [{\citenamefont {Wandel}\ \emph {et~al.}(2020)\citenamefont {Wandel},
  \citenamefont {Boschini}, \citenamefont {Neto}, \citenamefont {Shen},
  \citenamefont {Na}, \citenamefont {Zohar}, \citenamefont {Wang},
  \citenamefont {Welch}, \citenamefont {Seaberg}, \citenamefont {Koralek} \emph
  {et~al.}}]{wandel2020light}%
  \BibitemOpen
  \bibfield  {author} {\bibinfo {author} {\bibfnamefont {S.}~\bibnamefont
  {Wandel}}, \bibinfo {author} {\bibfnamefont {F.}~\bibnamefont {Boschini}},
  \bibinfo {author} {\bibfnamefont {E.}~\bibnamefont {Neto}}, \bibinfo {author}
  {\bibfnamefont {L.}~\bibnamefont {Shen}}, \bibinfo {author} {\bibfnamefont
  {M.}~\bibnamefont {Na}}, \bibinfo {author} {\bibfnamefont {S.}~\bibnamefont
  {Zohar}}, \bibinfo {author} {\bibfnamefont {Y.}~\bibnamefont {Wang}},
  \bibinfo {author} {\bibfnamefont {G.}~\bibnamefont {Welch}}, \bibinfo
  {author} {\bibfnamefont {M.}~\bibnamefont {Seaberg}}, \bibinfo {author}
  {\bibfnamefont {J.}~\bibnamefont {Koralek}},  \emph {et~al.},\ }\href@noop {}
  {\bibfield  {journal} {\bibinfo  {journal} {arXiv:2003.04224}\ } (\bibinfo
  {year} {2020})}\BibitemShut {NoStop}%
\bibitem [{\citenamefont {Kogar}\ \emph {et~al.}(2020)\citenamefont {Kogar},
  \citenamefont {Zong}, \citenamefont {Dolgirev}, \citenamefont {Shen},
  \citenamefont {Straquadine}, \citenamefont {Bie}, \citenamefont {Wang},
  \citenamefont {Rohwer}, \citenamefont {Tung}, \citenamefont {Yang} \emph
  {et~al.}}]{kogar2020light}%
  \BibitemOpen
  \bibfield  {author} {\bibinfo {author} {\bibfnamefont {A.}~\bibnamefont
  {Kogar}}, \bibinfo {author} {\bibfnamefont {A.}~\bibnamefont {Zong}},
  \bibinfo {author} {\bibfnamefont {P.~E.}\ \bibnamefont {Dolgirev}}, \bibinfo
  {author} {\bibfnamefont {X.}~\bibnamefont {Shen}}, \bibinfo {author}
  {\bibfnamefont {J.}~\bibnamefont {Straquadine}}, \bibinfo {author}
  {\bibfnamefont {Y.-Q.}\ \bibnamefont {Bie}}, \bibinfo {author} {\bibfnamefont
  {X.}~\bibnamefont {Wang}}, \bibinfo {author} {\bibfnamefont {T.}~\bibnamefont
  {Rohwer}}, \bibinfo {author} {\bibfnamefont {I.-C.}\ \bibnamefont {Tung}},
  \bibinfo {author} {\bibfnamefont {Y.}~\bibnamefont {Yang}},  \emph {et~al.},\
  }\href@noop {} {\bibfield  {journal} {\bibinfo  {journal} {Nat. Phys.}\
  }\textbf {\bibinfo {volume} {16}},\ \bibinfo {pages} {159} (\bibinfo {year}
  {2020})}\BibitemShut {NoStop}%
\bibitem [{\citenamefont {Demler}\ and\ \citenamefont
  {Zhang}(1995)}]{demler1995theory}%
  \BibitemOpen
  \bibfield  {author} {\bibinfo {author} {\bibfnamefont {E.}~\bibnamefont
  {Demler}}\ and\ \bibinfo {author} {\bibfnamefont {S.-C.}\ \bibnamefont
  {Zhang}},\ }\href@noop {} {\bibfield  {journal} {\bibinfo  {journal} {Phys.
  Rev. Lett.}\ }\textbf {\bibinfo {volume} {75}},\ \bibinfo {pages} {4126}
  (\bibinfo {year} {1995})}\BibitemShut {NoStop}%
\bibitem [{\citenamefont {Demler}\ \emph {et~al.}(1998)\citenamefont {Demler},
  \citenamefont {Kohno},\ and\ \citenamefont {Zhang}}]{demler1998pi}%
  \BibitemOpen
  \bibfield  {author} {\bibinfo {author} {\bibfnamefont {E.}~\bibnamefont
  {Demler}}, \bibinfo {author} {\bibfnamefont {H.}~\bibnamefont {Kohno}}, \
  and\ \bibinfo {author} {\bibfnamefont {S.-C.}\ \bibnamefont {Zhang}},\
  }\href@noop {} {\bibfield  {journal} {\bibinfo  {journal} {Phys. Rev. B}\
  }\textbf {\bibinfo {volume} {58}},\ \bibinfo {pages} {5719} (\bibinfo {year}
  {1998})}\BibitemShut {NoStop}%
\bibitem [{\citenamefont {Freericks}\ \emph {et~al.}(2009)\citenamefont
  {Freericks}, \citenamefont {Krishnamurthy},\ and\ \citenamefont
  {Pruschke}}]{freericks2009theoretical}%
  \BibitemOpen
  \bibfield  {author} {\bibinfo {author} {\bibfnamefont {J.}~\bibnamefont
  {Freericks}}, \bibinfo {author} {\bibfnamefont {H.}~\bibnamefont
  {Krishnamurthy}}, \ and\ \bibinfo {author} {\bibfnamefont {T.}~\bibnamefont
  {Pruschke}},\ }\href@noop {} {\bibfield  {journal} {\bibinfo  {journal}
  {Phys. Rev. Lett.}\ }\textbf {\bibinfo {volume} {102}},\ \bibinfo {pages}
  {136401} (\bibinfo {year} {2009})}\BibitemShut {NoStop}%
\bibitem [{\citenamefont {Lenar{\v{c}}i{\v{c}}}\ \emph
  {et~al.}(2014)\citenamefont {Lenar{\v{c}}i{\v{c}}}, \citenamefont
  {Gole{\v{z}}}, \citenamefont {Bon{\v{c}}a},\ and\ \citenamefont
  {Prelov{\v{s}}ek}}]{lenarvcivc2014optical}%
  \BibitemOpen
  \bibfield  {author} {\bibinfo {author} {\bibfnamefont {Z.}~\bibnamefont
  {Lenar{\v{c}}i{\v{c}}}}, \bibinfo {author} {\bibfnamefont {D.}~\bibnamefont
  {Gole{\v{z}}}}, \bibinfo {author} {\bibfnamefont {J.}~\bibnamefont
  {Bon{\v{c}}a}}, \ and\ \bibinfo {author} {\bibfnamefont {P.}~\bibnamefont
  {Prelov{\v{s}}ek}},\ }\href@noop {} {\bibfield  {journal} {\bibinfo
  {journal} {Phys. Rev. B}\ }\textbf {\bibinfo {volume} {89}},\ \bibinfo
  {pages} {125123} (\bibinfo {year} {2014})}\BibitemShut {NoStop}%
\bibitem [{\citenamefont {Shimizu}\ and\ \citenamefont
  {Yuge}(2011)}]{shimizu2011sum}%
  \BibitemOpen
  \bibfield  {author} {\bibinfo {author} {\bibfnamefont {A.}~\bibnamefont
  {Shimizu}}\ and\ \bibinfo {author} {\bibfnamefont {T.}~\bibnamefont {Yuge}},\
  }\href@noop {} {\bibfield  {journal} {\bibinfo  {journal} {Journal of the
  Physical Society of Japan}\ }\textbf {\bibinfo {volume} {80}},\ \bibinfo
  {pages} {093706} (\bibinfo {year} {2011})}\BibitemShut {NoStop}%
\bibitem [{\citenamefont {Hackl}\ \emph {et~al.}(2020)\citenamefont {Hackl},
  \citenamefont {Guaita}, \citenamefont {Shi}, \citenamefont {Haegeman},
  \citenamefont {Demler},\ and\ \citenamefont {Cirac}}]{hackl2020geometry}%
  \BibitemOpen
  \bibfield  {author} {\bibinfo {author} {\bibfnamefont {L.}~\bibnamefont
  {Hackl}}, \bibinfo {author} {\bibfnamefont {T.}~\bibnamefont {Guaita}},
  \bibinfo {author} {\bibfnamefont {T.}~\bibnamefont {Shi}}, \bibinfo {author}
  {\bibfnamefont {J.}~\bibnamefont {Haegeman}}, \bibinfo {author}
  {\bibfnamefont {E.}~\bibnamefont {Demler}}, \ and\ \bibinfo {author}
  {\bibfnamefont {J.~I.}\ \bibnamefont {Cirac}},\ }\href@noop {} {\bibfield
  {journal} {\bibinfo  {journal} {SciPost}\ }\textbf {\bibinfo {volume} {9}},\
  \bibinfo {pages} {048} (\bibinfo {year} {2020})}\BibitemShut {NoStop}%
\bibitem [{\citenamefont {Manmana}\ \emph {et~al.}(2005)\citenamefont
  {Manmana}, \citenamefont {Muramatsu},\ and\ \citenamefont
  {Noack}}]{manmana2007strongly}%
  \BibitemOpen
  \bibfield  {author} {\bibinfo {author} {\bibfnamefont {S.~R.}\ \bibnamefont
  {Manmana}}, \bibinfo {author} {\bibfnamefont {A.}~\bibnamefont {Muramatsu}},
  \ and\ \bibinfo {author} {\bibfnamefont {R.~M.}\ \bibnamefont {Noack}},\
  }\href@noop {} {\bibfield  {journal} {\bibinfo  {journal} {AIP Conf. Proc.}\
  }\textbf {\bibinfo {volume} {789}},\ \bibinfo {pages} {269} (\bibinfo {year}
  {2005})}\BibitemShut {NoStop}%
\bibitem [{\citenamefont {Balzer}\ \emph {et~al.}(2012)\citenamefont {Balzer},
  \citenamefont {Gdaniec},\ and\ \citenamefont {Potthoff}}]{balzer2012krylov}%
  \BibitemOpen
  \bibfield  {author} {\bibinfo {author} {\bibfnamefont {M.}~\bibnamefont
  {Balzer}}, \bibinfo {author} {\bibfnamefont {N.}~\bibnamefont {Gdaniec}}, \
  and\ \bibinfo {author} {\bibfnamefont {M.}~\bibnamefont {Potthoff}},\
  }\href@noop {} {\bibfield  {journal} {\bibinfo  {journal} {J. Phys. Condens.
  Matter}\ }\textbf {\bibinfo {volume} {24}},\ \bibinfo {pages} {035603}
  (\bibinfo {year} {2012})}\BibitemShut {NoStop}%
\bibitem [{\citenamefont {Hochbruck}\ and\ \citenamefont
  {Lubich}(1997)}]{hochbruck1997krylov}%
  \BibitemOpen
  \bibfield  {author} {\bibinfo {author} {\bibfnamefont {M.}~\bibnamefont
  {Hochbruck}}\ and\ \bibinfo {author} {\bibfnamefont {C.}~\bibnamefont
  {Lubich}},\ }\href@noop {} {\bibfield  {journal} {\bibinfo  {journal} {SIAM
  J. Numer. Anal.}\ }\textbf {\bibinfo {volume} {34}},\ \bibinfo {pages} {1911}
  (\bibinfo {year} {1997})}\BibitemShut {NoStop}%
\bibitem [{\citenamefont {Innerberger}\ \emph {et~al.}(2020)\citenamefont
  {Innerberger}, \citenamefont {Worm}, \citenamefont {Prauhart},\ and\
  \citenamefont {Kauch}}]{innerberger2020electron}%
  \BibitemOpen
  \bibfield  {author} {\bibinfo {author} {\bibfnamefont {M.}~\bibnamefont
  {Innerberger}}, \bibinfo {author} {\bibfnamefont {P.}~\bibnamefont {Worm}},
  \bibinfo {author} {\bibfnamefont {P.}~\bibnamefont {Prauhart}}, \ and\
  \bibinfo {author} {\bibfnamefont {A.}~\bibnamefont {Kauch}},\ }\href@noop {}
  {\bibfield  {journal} {\bibinfo  {journal} {The European Physical Journal
  Plus}\ }\textbf {\bibinfo {volume} {135}},\ \bibinfo {pages} {1} (\bibinfo
  {year} {2020})}\BibitemShut {NoStop}%
\end{thebibliography}%
\end{document}